\documentclass[12pt]{iopart}

\usepackage{iopams}

\usepackage{mathrsfs}
\usepackage{bm}
\usepackage{hyperref}
\usepackage{graphicx}
\usepackage{float}
\usepackage{verbatim}
\usepackage{subcaption}
\usepackage{color}
\usepackage{amsopn}

\newcommand{\R}{\mathbb{R}}
\newcommand{\Z}{\mathbb{Z}}
\renewcommand{\leq}{\leqslant}
\renewcommand{\geq}{\geqslant}
\newcommand{\grad}{\nabla} 
\newcommand{\dv}{\operatorname{div}} 
\newcommand{\sgn}{\operatorname{sign}} 
\DeclareMathOperator{\curl}{\operatorname{curl}} 
 
\DeclareMathOperator{\tpr}{\otimes} 
\newcommand{\dd}{\operatorname{d}\!}
\newcommand{\DD}{\mathsf{D}}
\newcommand{\ccurl}{\mathsf{curl}\,}
\newcommand{\n}{\bm{n}}
\newcommand{\dframe}{(\n_1,\n_2,\n)}
\newcommand{\cframe}{(\vx,\vy,\vz)}
\newcommand{\vx}{\bm{e}_{x}}
\newcommand{\vy}{\bm{e}_{y}}
\newcommand{\vz}{\bm{e}_{z}}

\newcommand{\er}{\bm{e}_{r}}
\newcommand{\et}{\bm{e}_{\vartheta}}
\newcommand{\x}{\bm{x}}

\newcommand{\vb}{\bm{b}}
\newcommand{\tW}{\mathbf{W}}
\newcommand{\tP}{\mathbf{P}}
\newcommand{\tD}{\mathbf{D}}
\newcommand{\tI}{\mathbf{I}}
\newcommand{\cc}{\bm{c}}
\newcommand{\cd}{\bm{d}}
\newcommand{\normal}{\bm{\nu}}
\newcommand{\surface}{\mathscr{S}}
\newcommand{\nablas}{\nabla\!_\mathrm{s}}

\newcommand{\sloh}{M_{\rm l}}
\newcommand{\slod}{M_{\rm c}}

\newcommand{\f}{f}
\newcommand{\gradf}{\grad\f}
\newcommand{\qS}{S^*}
\newcommand{\qb}{b^*}
\newcommand{\qT}{T^*}
\newcommand{\qq}{q^*}

\newcommand{\dchar}{(S,T,b_1,b_2,q)}
\newcommand{\fchar}{(fS,fT,fb_1,fb_2,fq)}
\newcommand{\dstar}{(fS^\ast,fT^\ast,fb_1^\ast,fb_2^\ast,fq^\ast)}

\newcommand{\gradn}{\grad\n}
\newcommand{\Fdensity}{W_\mathrm{OF}}

\newcommand{\disk}{\mathbb{S}^1}
\newcommand{\tanset}{\mathcal{T}(\alpha_0)}
\newcommand{\trans}{^\mathsf{T}}

\newcommand{\nalert}[1]{{\color{black}{#1}}}
\newcommand{\curve}{\mathscr{C}}

\begin{document}
\title[Relieving nematic geometric frustration]{Relieving nematic geometric frustration in the  plane}

\author{Andrea Pedrini$^1$, Epifanio G Virga$^1$}
\address{$^1$ Dipartimento di Matematica, Universit\`a di Pavia, Via Ferrata 5, I-27100 Pavia, Italy}

\eads{\mailto{andrea.pedrini@unipv.it}, \mailto{eg.virga@unipv.it}}

\begin{abstract}
 Frustration in \nalert{nematic-ordered} media (endowed with a director field) is treated in a purely geometric fashion in a flat, two-dimensional space. We recall the definition of \emph{quasi-uniform} distortions and envision these as viable ways to relieve director fields prescribed on either a straight line or the unit circle. We prove that \nalert{using a \emph{planar spiral} is  the only way to fill the whole plane with a quasi-uniform distortion}. Apart from that, all relieving quasi-uniform distortions can at most be defined in a half-plane; however, in a generic sense, they are all asymptotically spirals. 
\end{abstract}
\vspace{2pc}


\section{Introduction}\label{sec:intro}
Nematic liquid crystals are perhaps the most typical example of a soft matter system whose order parameter is a unit vector $\n$ enjoying the head-tail symmetry, which requires all their physical properties to be invariant under the transformation $\n\mapsto-\n$. Such an order parameter is often called a \emph{director} to emphasize that only the direction of $\n$ has physical significance, whereas in a \emph{spin} also the sense of orientation is meaningful. On occasion, a director is said to generate a \emph{line field}, as opposed to a vector field.

The microscopic origin of $\n$ is to be retraced in the orientation of the elongated molecules that constitute these fascinating, ordered phases: on average, these molecules tend to be oriented alike, with equally likely distributions of heads and tails, whenever these can be distinguished from one another.

Properly speaking, the target of the director mapping $\n$ is the real projective plane $\mathbb{RP}^2$, although in a number of practical cases (but not all), $\n$ is \emph{orientable}, meaning that it can be lifted into the unit sphere $\mathbb{S}^2$ \cite{ball:orientable}.

Our approach here will be more general than our original liquid crystal motivation might suggest. The director field $\n$ will have any possible interpretation compatible with nematic symmetry. Our focus will be on \emph{geometric frustration} and ways of relieving it in two space dimensions. By geometric frustration we refer to any means capable of preventing a director field from filling space \emph{uniformly}. Beyond its intuitive meaning, the latter property is properly defined by identifying scalar measures of distortion, which we call the \emph{distortion characteristics}, and requiring them to be \emph{constant} throughout space.

This notion was introduced in \cite{virga:uniform} in three-dimensional Euclidean space and further extended to the non-Euclidean case in \cite{dasilva:moving,pollard:intrinsic}. In flat two-dimensional space, the request of uniformity confines $\n$ to be a trivial constant field \cite{niv:geometric}, and the question then arises as to whether prescribing $\n$ on a line in the plane, so as \emph{not} to be constant there, induces a geometric frustration that can be relieved via a purely geometric mechanism, with no reference to any energy consideration.

The purely geometric mechanism we envision was introduced in \cite{pedrini:liquid}; it appears to be the most natural extension of the notion of uniform distortion, one for which the distortion characteristics instead of being constant are in constant ratios to one another, so that a single function of position in space is left free. We call these distortions \emph{quasi-uniform}.

This paper explores the possibility that such an avenue can indeed be taken to relieve frustration in flat, two-dimensional space. The material presented here is organized as follows.

In \sref{sec:characteristiscs}, we recall the definition of distortion characteristics and their role in defining uniformity. Section~\ref{sec:quasiuniform} is concerned with the definition of quasi-uniform distortions and their representation in the plane. In \sref{sec:halfplane}, we show how to fill half a plane with a quasi-uniform distortion relieving a geometric frustration concentrated on a straight line. It will emerge there that some frustrations can be relieved globally, but others only locally. Incidentally, our analysis will prove that no genuine quasi-uniform distortion can exist in the whole plane. Geometric distortions enforced on a circle will be considered in \sref{sec:defect}. The possibility of relieving them in the whole plane outside the circle will be related to the topological charge of the frustration. In \sref{sec:universal}, we extract from the families of quasi-uniform distortions constructed in this paper some features that they have in common. \nalert{The notion of frustration adopted in this paper admittedly bears a rather extended meaning; a more restricted one, which also entails a definition of \emph{one-dimensional} uniformity is introduced and discussed in \sref{sec:1d_uniformity}.} Finally, in \sref{sec:conclusions}, we collect our main conclusions and comment on the general nature of the asymptotic distortion field relieving the two classes of geometric frustration considered in this work. 

A number of appendices close this paper; there we display the mathematical details of our development that could have hampered our presentation in the main text. 

\section{Distortion characteristics}\label{sec:characteristiscs}
A natural measure of distortion for a director field $\n$ is its spatial gradient $\nabla\n$. Were $\nabla\n\equiv\mathbf{0}$, $\n$ would have the same orientation everywhere in the domain in space under consideration. Being a tensor, $\nabla\n$ is (covariantly) affected by a change of frame (observer). We want to extract from it  invariant measures of distortions that bear an intrinsic meaning (independent of observers).

Here this goal is achieved by means of a 
decomposition of $\gradn$ first proposed in \cite{machon:umbilic} and then reprised and reinterpreted in  \cite{selinger:interpretation}, where the main players are  the \emph{splay} scalar $S:=\dv\n$, the \emph{twist} pseudoscalar $T:=\n\cdot\curl\n$ and the \emph{bend} vector $\vb:=\n\times\curl\n$:
\begin{equation}\label{eq:grad_n}
\grad\n = -\vb\tpr \n + \frac{1}{2}T\tW(\n) + \frac{1}{2}S\tP(\n) + \tD.
\end{equation}
In \eref{eq:grad_n}, $\tW(\n)$ denotes the skew-symmetric tensor with axial vector $\n$, $\tP(\n):=\tI-\n\tpr\n$ is the projector on the plane orthogonal to $\n$ and $\tD$ is a traceless symmetric tensor for which $\tD\n=\bm{0}$. Whenever $\tD\neq\bm{0}$, its properties ensure that it can be represented as
\begin{equation}\label{eq:D}
\tD = q(\n_1\tpr\n_1 - \n_2\tpr\n_2)
\end{equation}
by choosing an orthonormal basis of eigenvectors $(\n_1,\n_2)$  in the plane orthogonal to $\n$, such that the positive scalar $q$ is the eigenvalue of $\n_1$, and this is oriented so that $\n=\n_1\times\n_2$. We call $(\n_1,\n_2,\n)$ the \emph{distortion frame} and $q$ the \emph{octupolar splay}.\footnote{We prefer the adjective \emph{octupolar} to  \emph{biaxial}  and \emph{thetrahedral}, used in \cite{selinger:interpretation} and  \cite{selinger:director}, respectively, in order to stress the connection with an \emph{octupolar tensor} employed in  \cite{pedrini:liquid} to represent pictorially  $S$, $\vb$, and $q$.}

In the distortion frame, $\tW(\n)$ is represented as
\begin{equation}
	\label{eq:W_representation}
	\tW(\n) = \n_2\tpr\n_1 - \n_1\tpr\n_2
\end{equation}
and, by its own definition, the bend vector can always be decomposed as
\begin{equation}
	\label{eq:b_representation}
\vb:=b_1\n_1+b_2\n_2,	
\end{equation}
for suitable scalars $b_1$ and $b_2$.\footnote{When $\tD=\bm{0}$ and $\vb\neq\bm{0}$, the choice of $\n_1$ and $\n_2$ can be made by requiring that $\vb=b_1\n_1$ and then setting $\n_2:=\n\times\n_1$, while for $\tD=\bm{0}$ and $\vb=\bm{0}$ any pair $(\n_1,\n_2)$ in the plane orthogonal to $\n$ could be employed as constituents of a  suitable distortion frame.}
The following identity, which is a direct consequence of \eref{eq:grad_n}, can be used to determine $q$,\footnote{\nalert{We defer the reader to \ref{app:oseen_frank} for a link between $q^2$ and the classical \emph{saddle-splay} distortion measure in the Oseen-Frank theory of liquid crystals.}}
\begin{equation}\label{eq:q_identity}
2q^2 = \tr(\gradn)^2 + \frac12T^2 - \frac12S^2.
\end{equation}

The decomposition in \eref{eq:grad_n} also suggests a natural question as to the existence of \emph{uniform} distortions, for   which the director field $\n$ is not necessarily constant in space, but $\gradn$ has everywhere the same representation in the intrinsic distortion frame. This notion was made formally precise in \cite{virga:uniform} by prescribing that the \emph{distortion characteristics} $(S,b_1,b_2,T,q)$ remain \emph{constant} in the whole domain where $\n$ is defined. A complete characterization of uniform distortions in the whole three-dimensional Euclidean space was also obtained in \cite{virga:uniform}: they are precisely Meyer's \emph{heliconical} distortions \cite{meyer:structural}, for which either
\begin{equation}\label{eq:uniform_distortions}
	S = 0,\ T = 2q,\ b_1 = b_2 = b
\quad\mathrm{or}\quad
S = 0,\ T = -2q,\ b_1 = -b_2 = b,
\end{equation}
where $q>0$ and $b$ are arbitrary scalars. 

In general, a distortion will be said to be a \emph{twist-bend}, whenever
$S=0$ and $T=\pm2q$, with distortion characteristics \emph{not} necessarily constant. Thus, uniform distortions in three-dimensional space are special twist-bend distortions. 

We find it instructive to comment in \ref{app:oseen_frank} about the bearing that  identity  \eref{eq:q_identity} has on the elastic free energy of nematic liquid crystals.\footnote{These comments have no direct impact on our development, which is purely geometric.}

In two space dimensions, the existence of uniform distortions  depends on the geometry of the surface $\surface$ where the director field $\n$ lies. A  uniform distortion, with both $S\neq0$ and $B\neq0$, where $S$ is to be interpreted as the \emph{covariant} divergence of $\n$ and $B$ as the norm of the \nalert{covariant} bend vector,\footnote{The \emph{covariant} gradient $\DD\n$ of a director field $\n$ on a surface $\surface$ can be defined as the projection onto $\surface$ of the surface gradient $\nablas\n$, $\DD\n:=\tP(\normal)\nablas\n$, where $\tP(\normal):\tI-\normal\otimes\normal$ and $\normal$ is an outer unit normal to $\surface$. Here then $S=\tr\DD\n$ and $B=|\n\times\ccurl\n|$, where $\ccurl\n$ is twice the axial vector associated with the skew-symmetric part of $\DD\n$.} is possible only if the Gaussian curvature $K$ of $\surface$ is negative and such that $B^2 + S^2 = -K$ \cite{niv:geometric}. Thus, only a constant field, for which both $S\equiv0$ and $B\equiv0$, is an admissible uniform distortion  on a flat surface \cite{meyer:structural}.

The notion of uniform distortion in flat three-dimensional space in \cite{virga:uniform} has also been extended  to curved spaces in \cite{dasilva:moving} and \cite{pollard:intrinsic}.

In the following sections we study a broader class of distortions in two space dimensions by relaxing the definition of uniformity; then we present explicit constructions of two families of such distortions in the plane.

\section{Quasi-uniform distortions}\label{sec:quasiuniform}
A \emph{quasi-uniform} distorsion \cite{pedrini:liquid} is represented by a nematic director field $\n$ with distortion characteristics in a constant ratio to one another. More formally, either the distortion characteristics $\dchar$ are all zero,  but one, or they are all proportional through constants to a nonzero scalar function $\f=\f(\x)$ of position $\x$ in space,
\begin{equation}\label{eq:qu_characteristics}
S = \qS\f,\quad b_1=\qb_1\f,\quad b_2=\qb_2\f,\quad T=\qT\f,\quad q=\qq\f,
\end{equation}
with $\qS$, $\qb_1$, $\qb_2$, $\qT\in\R$ and $\qq\geq0$ all constants. For a quasi-uniform distortion, decomposition \eref{eq:grad_n} is therefore factorized by $\f$
and, by explicitly expanding $\gradn$ in the distortion frame $(\n_1,\n_2,\n)$, we give it the form
\begin{equation}\label{eq:qu_gradn}
\eqalign{
 \grad\n 
 = \f\Bigg[&\Bigg(\frac{\qS}{2}+\qq\Bigg)\n_1\otimes\n_1 - \frac{\qT}{2}\n_1\otimes\n_2 - \qb_{1}\n_1\otimes\n \\
 &+ \frac{\qT}{2}\n_2\otimes\n_1 + \Bigg(\frac{\qS}{2}-\qq\Bigg)\n_2\otimes\n_2 - \qb_{2}\n_2\otimes\n\Bigg].}
\end{equation}
To ease our notation, we shall suppress the superscript ${}^*$ wherever the distinction between  distortion characteristics and their associated constants need not be specified; in such a case, the distortion characteristics will be denoted $(\f S, \f T, \f b_1, \f b_2, \f q)$; otherwise, they will be denoted  $(\f \qS, \f \qT, \f \qb_1, \f \qb_2, \f \qq)$. 

\begin{figure}
	\centering
	\begin{subfigure}[b]{0.3\textwidth}
		\centering
		\includegraphics[width=\textwidth]{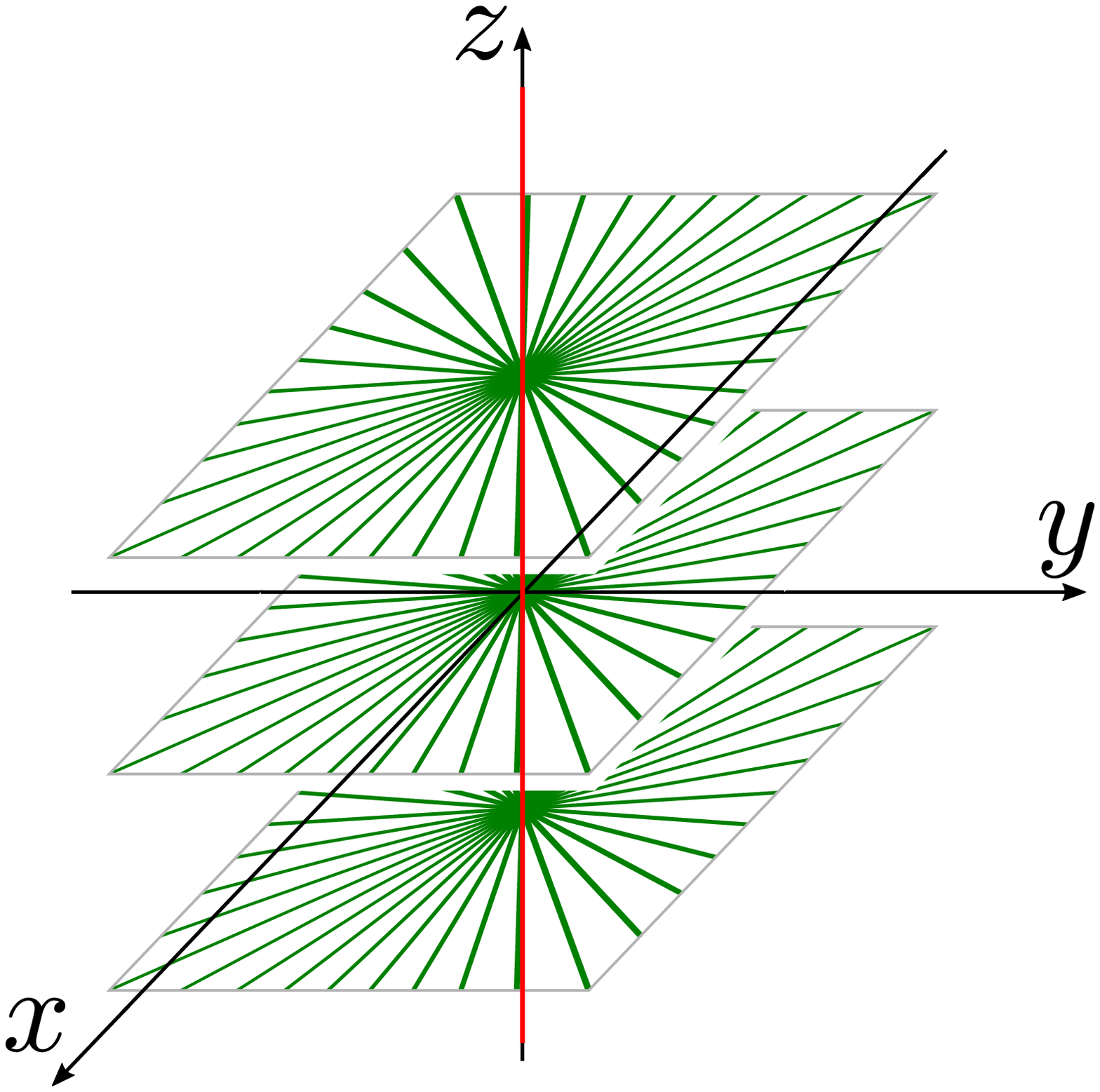}
		\caption{Planar splay.}
	\end{subfigure}
	$\quad$
	\begin{subfigure}[b]{0.3\textwidth}
		\centering
		\includegraphics[width=\textwidth]{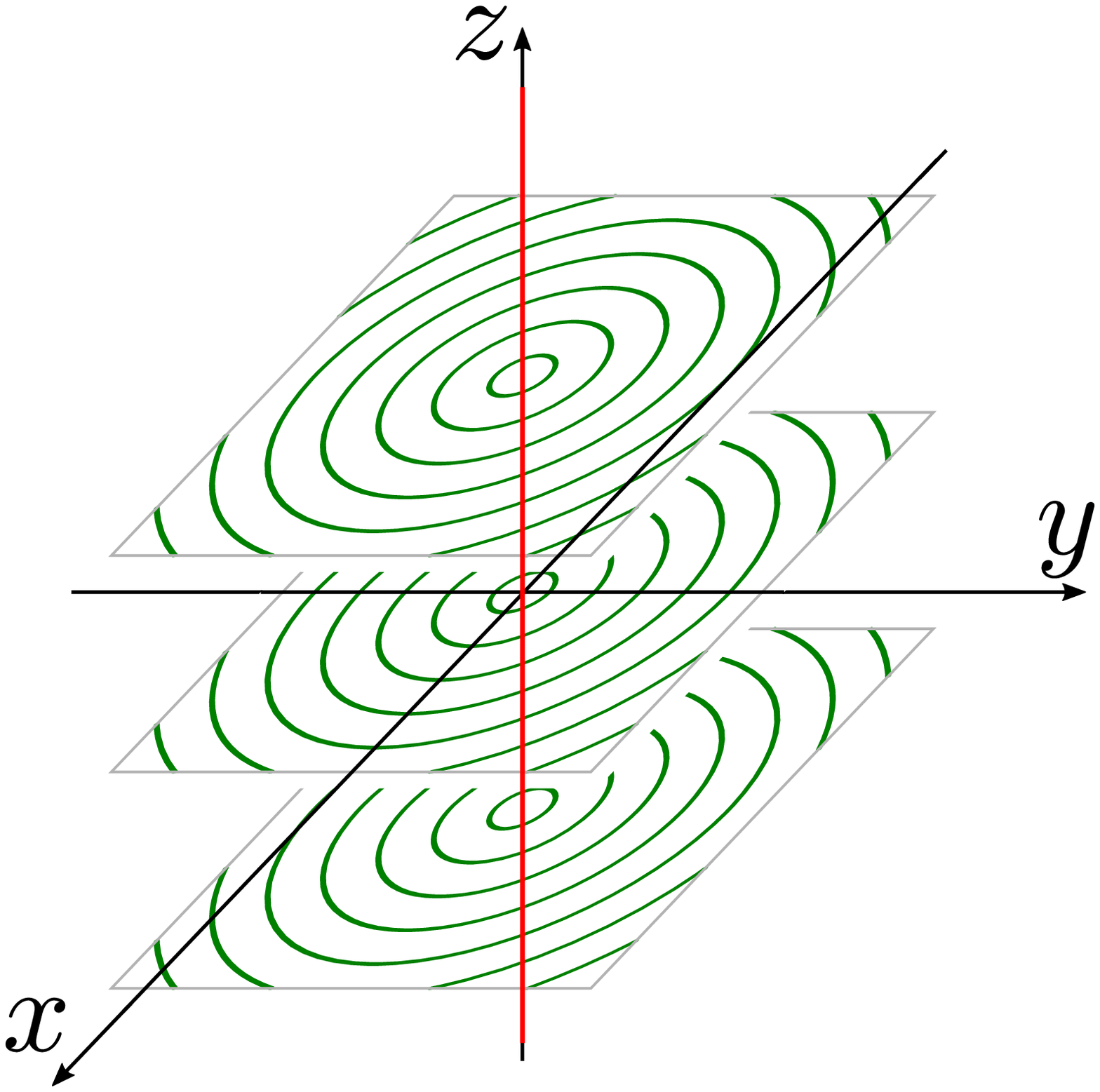}
		\caption{Pure bend.}
	\end{subfigure}
	$\quad$
	\begin{subfigure}[b]{0.3\textwidth}
		\centering
		\includegraphics[width=\textwidth]{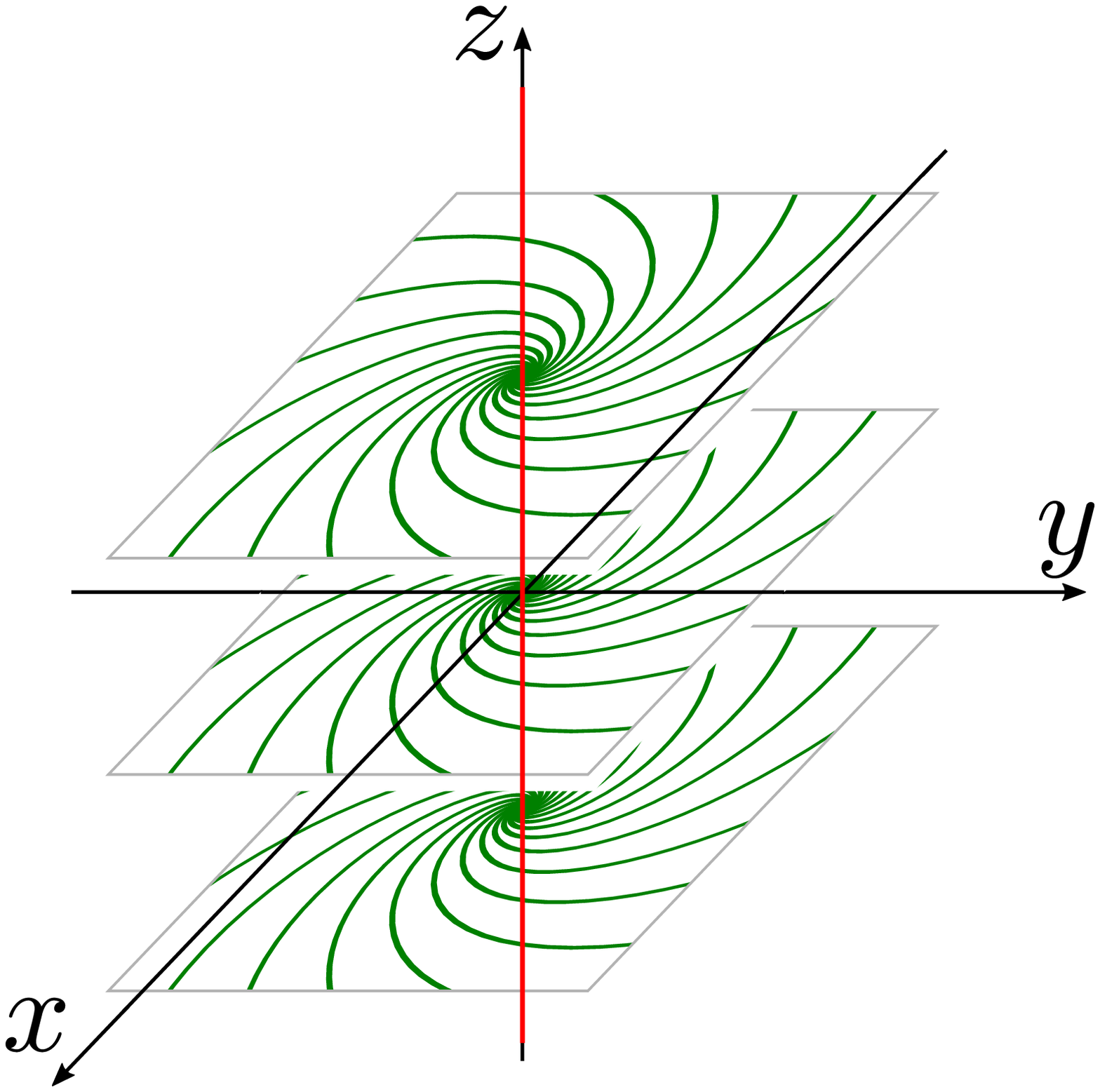}
		\caption{Splay-bend spirals.}
	\end{subfigure}
	\caption{Planar quasi-uniform distortions extended to three-dimensional space. Field lines are depicted in green, while the red $z$-axis is a line defect.}
	\label{fig:quasi-unifom_extended}
\end{figure}

In \cite{pedrini:liquid}, by a direct computation of the distortion characteristics, we showed the existence of genuine quasi-uniform distortions (which are not uniform). They are the following \emph{elementary} fields (see \fref{fig:quasi-unifom_extended}):
\begin{enumerate}
\item the \emph{hedgehog}, represented in a Cartesian frame $\cframe$ as
\begin{equation}
	\label{eq:hedgehog}
	\n=(x^2+y^2+z^2)^{-\frac12}(x\vx+y\vy+z\vz),
\end{equation} 
\item the \emph{pure bend} $\n=(x^2+y^2)^{-\frac12}(-y\vx+x\vy)$,
\item the \emph{planar splay} $\n=(x^2+y^2)^{-\frac12}(x\vx+y\vy)$, and
\item the \emph{planar spiral} 
\begin{equation}\label{eq:spirals}
\n=(x^2+y^2)^{-\frac12}[(x\cos\alpha-y\sin\alpha)\vx+(x\sin\alpha+y\cos\alpha\vy)],
\end{equation}
which interpolates between planar splay and pure bend as the parameter $\alpha$ ranges in the interval $[0,\frac{\pi}{2}]$.
\end{enumerate}

The question as to the existence of quasi-uniform distortions other than these can  be formally rephrased  as the quest for the conditions that a scalar function $\f$ must satisfy for  \eref{eq:qu_gradn} to apply. This formal avenue is undertaken in \ref{app:compatibility}, where we reproduce the approach followed in \cite{virga:uniform}, only requiring quasi-uniformity, instead of uniformity. Here, we rather prefer a more direct, constructive approach and start from  an instructive somewhat negative result.  

\subsection{Planar quasi-uniform distortions}\label{sec:quasiuniform_splaybend}
Consider a planar, non-constant distortion for which 
\begin{equation}\label{eq:splay-bend}
 \n = \cos\varphi\vx + \sin\varphi\vy,
\end{equation}
where $\varphi=\varphi(x,y)$ depends only on  the $(x,y)$ coordinates in the plane, so that  $(\gradn)\vz\equiv\bm{0}$. Moreover, since $\n\cdot\vz\equiv0$, we also have that $(\gradn)\trans\vz\equiv\bm{0}$, and so $\curl\n\parallel\vz$ and $T\equiv0$. By applying both sides of equation \eref{eq:grad_n} to $\vz$, we then obtain that
\begin{equation}
\label{eq:D_planar}
\tD\vz=-\frac12S\vz,
\end{equation}	
which by \eref{eq:D} implies that either
\begin{equation}
	\label{eq:n_dicothomy}
	\n_2\equiv\vz\quad\mathrm{if}\quad S>0\quad\mathrm{or}\quad
		\n_1\equiv\vz\quad\mathrm{if}\quad S<0
\end{equation}
and, correspondingly, 
\begin{equation}
	\label{eq:splay_bend_definition}
	S=\pm2q.
\end{equation}
In general, we  say that a distortion is a \emph{splay-bend}, whenever $T=0$ and \eref{eq:splay_bend_definition} applies. Thus, any planar distortion is a splay-bend, no matter whether quasi-uniform or not.

We now show that there is no non-trivial  function $\varphi=\varphi(x)$ of a single variable $x$ that makes $\n$ in \eref{eq:splay-bend} a \emph{modulated} quasi-uniform field on the whole plane.
This  result is obtained by direct inspection of the distortion characteristics. In fact, by calling $\varphi_{,x}$ the derivative of $\varphi$, the gradient of the director becomes 
\begin{equation}\label{eq:gradn_sp}
 \grad\n = \varphi_{,x}(-\sin\varphi\vx\tpr\vx + \cos\varphi\vy\tpr\vx) = \varphi_{,x}\n_\perp\tpr\vx,
\end{equation}
with $\n_\perp:= \vz\times\n=-\sin\varphi\vx + \cos\varphi\vy$.
Since the skew-symmetric part of $\gradn$ is
\begin{equation}\label{eq:sp_skw}
\grad\n_{\rm skw} = \frac{\varphi_{,x}\cos\varphi}{2}(\vy\tpr\vx - \vx\tpr\vy),
\end{equation}
$\curl\n = \varphi_{,x}\cos\varphi\vz$ and $\vb=-\varphi_{,x}\cos\varphi\n_\perp$. Furthermore $S=-\varphi_{,x}\sin\varphi$ and 
\begin{equation}\label{eq:bsplay_sp}
\tD = \frac{\varphi_{,x}\sin\varphi}2(\vz\tpr\vz - \n_\perp\tpr\n_\perp),
\end{equation}
in agreement with \eref{eq:splay_bend_definition}.
As a result, the distortion characteristics $\dstar$ can be written by making use of one the following expressions:
\begin{eqnarray}\label{eq:sp_char}
\fl
\f = -\frac{\varphi_{,x}\sin\varphi}{2},
\ \qS=2,
\ \qb_1= 2\cot\varphi,
\ \qb_2=\qT=0,
\ \qq=1,
\qquad&\mathrm{if}\ \varphi_{,x}\sin\varphi\leq0, \\
\fl
\f = \frac{\varphi_{,x}\sin\varphi}{2},
\ \qS=-2,
\ \qb_1=\qT=0,
\ \qb_2= -2\cot\varphi,
\ \qq=1,
\qquad&\mathrm{if}\ \varphi_{,x}\sin\varphi\geq0,
\end{eqnarray}
 and it is clear that $\n$  is never quasi-uniform, unless $\varphi$ is constant  (which would make $\n$ constant too). 

We note that all planar elementary quasi-uniform distortions, ranging from pure bend to planar splay, have \emph{defects}: in the plane, the origin is indeed a singular point; it extends to an entire line defect in the third dimention $z$, as illustrated in \fref{fig:quasi-unifom_extended}. 

The lack of modulated quasi-uniform distortions filling the whole plane and the presence of defects in the most elementary ones suggest that in our quest we should restrict the domain of $\n$ and allow for  the existence of defects.  In the next two sections, we present the explicit construction of two families of planar quasi-uniform distortions: one is defined in a half-plane and prescribed  on the boundary,  the other includes fields winding outside a circle with prescribed topological charge. 
These will identify two separate mechanisms for relieving frustration in the plane, from a line and from a circle, respectively. We shall see that not all frustrations can be relieved in this way;   geometric criteria will be given that single out those that can. 

\section{Filling a half-plane}\label{sec:halfplane}
Here we consider the most general family of planar director fields $\n$ in \eref{eq:splay-bend} by letting $\varphi=\varphi(x, y)$  depend  on both coordinates $(x,y)$ in the plane. 
The distortion gradient is then given by
\begin{equation}\label{eq:gradn_half}
\fl
 \grad\n = - \varphi_{,x}\sin\varphi\vx\tpr\vx - \varphi_{,y}\sin\varphi\vx\tpr\vy + \varphi_{,x}\cos\varphi\vy\tpr\vx + \varphi_{,y}\cos\varphi\vy\tpr\vy.
\end{equation}
Taking $S=2q$ (for $S=-2q$, we proceed similarly) and letting the distortion frame comprise the unit vectors (see \ref{app:frustrated_line} for details)
\begin{equation}\label{eq:frame_half}
 \n_1 = - \sin\varphi\vx + \cos\varphi\vy
 \quad\textrm{and}\quad
 \n_2 = \vz,
\end{equation}
we express the distortion characteristics  as 
\begin{equation}\label{eq:modes_half}
\fl
 S = \varphi_{,y}\cos\varphi - \varphi_{,x}\sin\varphi,
 \quad
 b_1 = -(\varphi_{,x}\cos\varphi + \varphi_{,y}\sin\varphi),
 \quad
 b_2 = T = 0,
 \quad
 q = \frac S2.
 \quad
\end{equation}
Then a quasi-uniform distortion with both $\varphi_{,x}\not\equiv0$ and $\varphi_{,y}\not\equiv0$ must satisfy
\begin{equation}\label{eq:condition1a_half}
 -(\varphi_{,x}\cos\varphi + \varphi_{,y}\sin\varphi) = \qb(\varphi_{,y}\cos\varphi - \varphi_{,x}\sin\varphi)
\end{equation}
for some constant $\qb\in\R$, which implies that 
\begin{equation}\label{eq:condition1b_half}
 \varphi_{,x}(\cos\varphi - \qb\sin\varphi) 
 + \varphi_{,y}(\sin\varphi + \qb\cos\varphi) = 0.
\end{equation}
In this way, all distortion characteristics are factorized by the function
\begin{equation}\label{eq:f_line}
\nalert{
\f=\frac12(\varphi_{,y}\cos\varphi - \varphi_{,x}\sin\varphi)}
\end{equation}
via constants
\begin{equation}\label{eq:constants_half}
 \qS = 2,
 \quad
 \qb_1 = 2\qb,
 \quad
 \qb_2 = \qT = 0,
 \quad
 \qq = 1.
 \quad
\end{equation}

To solve the quasi-linear partial differential equation \eref{eq:condition1b_half}, we parameterize $x$, $y$, and $\varphi$ by introducing an auxiliary variable $t\in\R$ and we then follow the propagation of the  values $x=x_0$, $y=y_0$, and $\varphi_0=\varphi(x_0,y_0)$  for $t=0$ along the solutions of \eref{eq:condition1b_half}, as entailed by  the classical \emph{method of characteristics} (see e.g. Sect.\,4.8 of \cite{salsa:partial}, for a fresh didactic exposition\footnote{A classical reference is Chapt.\,2 of \cite{courant:methods}; for the related Lagrange-Charpit method, the reader is also referred to \cite{delgado:lagrange}.}), which reduces \eref{eq:condition1b_half} to the following system of ordinary differential equations (often referred to as the Lagrange-Charpit system):
\begin{equation}\label{eq:chareq_half}
 \frac{\dd x}{\dd t} = \cos\varphi - \qb\sin\varphi,
\qquad
 \frac{\dd y}{\dd t} = \sin\varphi + \qb\cos\varphi,
 \qquad 
 \frac{\dd\varphi}{\dd t} = 0.
\end{equation}
Therefore, on any characteristic curve
\begin{equation}\label{eq:chareq_nonpar_half}
y = \frac{\sin\varphi_0 + \qb\cos\varphi_0}{\cos\varphi_0 - \qb\sin\varphi_0}(x - x_0)
\quad \mathrm{and} \quad 
\varphi(x,y) = \varphi(x_0,0) =:\varphi_0(x_0),
\end{equation}
where $\varphi_0(x_0)$ is the prescribed value of $\varphi$ at $x_0$: all characteristic curves are straight line; they are parallel to $\vy$ (vertical) for $\tan\varphi_0=1/\qb$ and  parallel to $\vx$ (horizontal) for $\tan\varphi_0=-\qb$.

The solution of \eref{eq:condition1b_half} is uniquely defined only at those points in the plane belonging precisely to a single  characteristic line. Since the slope of a characteristic line \eref{eq:chareq_nonpar_half} depends on $\qb$ (which is constant) and $\varphi_0$ (which propagates unchanged on the entire line), the unique solution filling the \emph{whole} plane must have $\varphi$ \emph{constant}, and so it generates thorough \eref{eq:gradn_half} a constant director field $\n$.\footnote{\nalert{In three space dimensions, general compatibility conditions for the function $f$ were given in \cite{pollard:intrinsic} for a \emph{regular} quasi-uniform distortion to fill the whole space (see their equation (48)). Examples were also given to show that this class in not empty.}}

\begin{figure}[h]
\centering
\includegraphics[width=.6\textwidth]{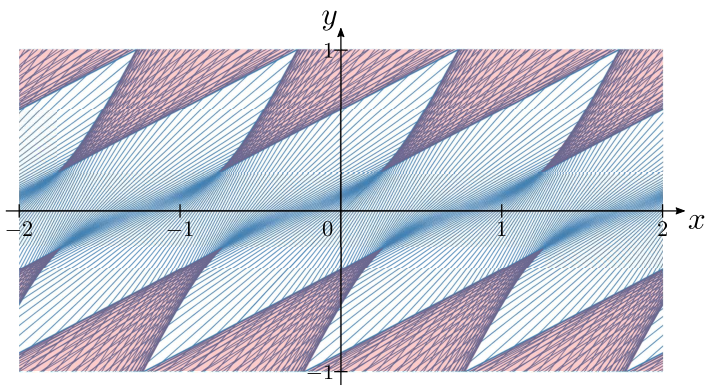}
\caption{Example of a bounded  domain (in white) for a planar and periodic,  quasi-uniform distortion \eref{eq:splay-bend}. Precisely, in \eref{eq:chareq_nonpar_half}, $\qb=1$ and $\varphi_0(x_0)= \pi\sin(\pi x_0)/10$. The  characteristic (straight) lines where $\varphi_0$ propagates unchanged are drawn in blue: they intersect one another in the  (red) region where $\n$  \emph{cannot} be uniquely defined.}
\label{fig:lozenges}
\end{figure}

To construct a genuine planar quasi-uniform distortion, we must therefore prescribe a non-constant $\varphi_0$ on the straight line $y=0$, so  that  there are no intersections of characteristic lines in the half-plane $y>0$. It follows from direct inspection of \eref{eq:chareq_nonpar_half} that the only way to achieve this goal is to ensure that  the slope along the line $y=0$,
\begin{equation}\label{eq:slope_half}
\sloh := \frac{\sin\varphi_0 + \qb\cos\varphi_0}{\cos\varphi_0 - \qb\sin\varphi_0},
\end{equation}
is a monotonic function of $x_0\in\R$. The violation of such a condition produces a planar domain of non-intersecting characteristic lines that does \emph{not} extend to a whole half-plane, as shown in \fref{fig:lozenges} where the boundary condition $\varphi_0(x_0)= \varepsilon\sin(2\pi x_0/\ell)$ represents a periodic  perturbation of the constant field $\varphi_0\equiv0$, with period $\ell$ and amplitude  a (small) parameter $\varepsilon$.

\begin{figure}
	\centering
	\begin{subfigure}[b]{0.3\textwidth}
		\centering
		\includegraphics[width=\textwidth]{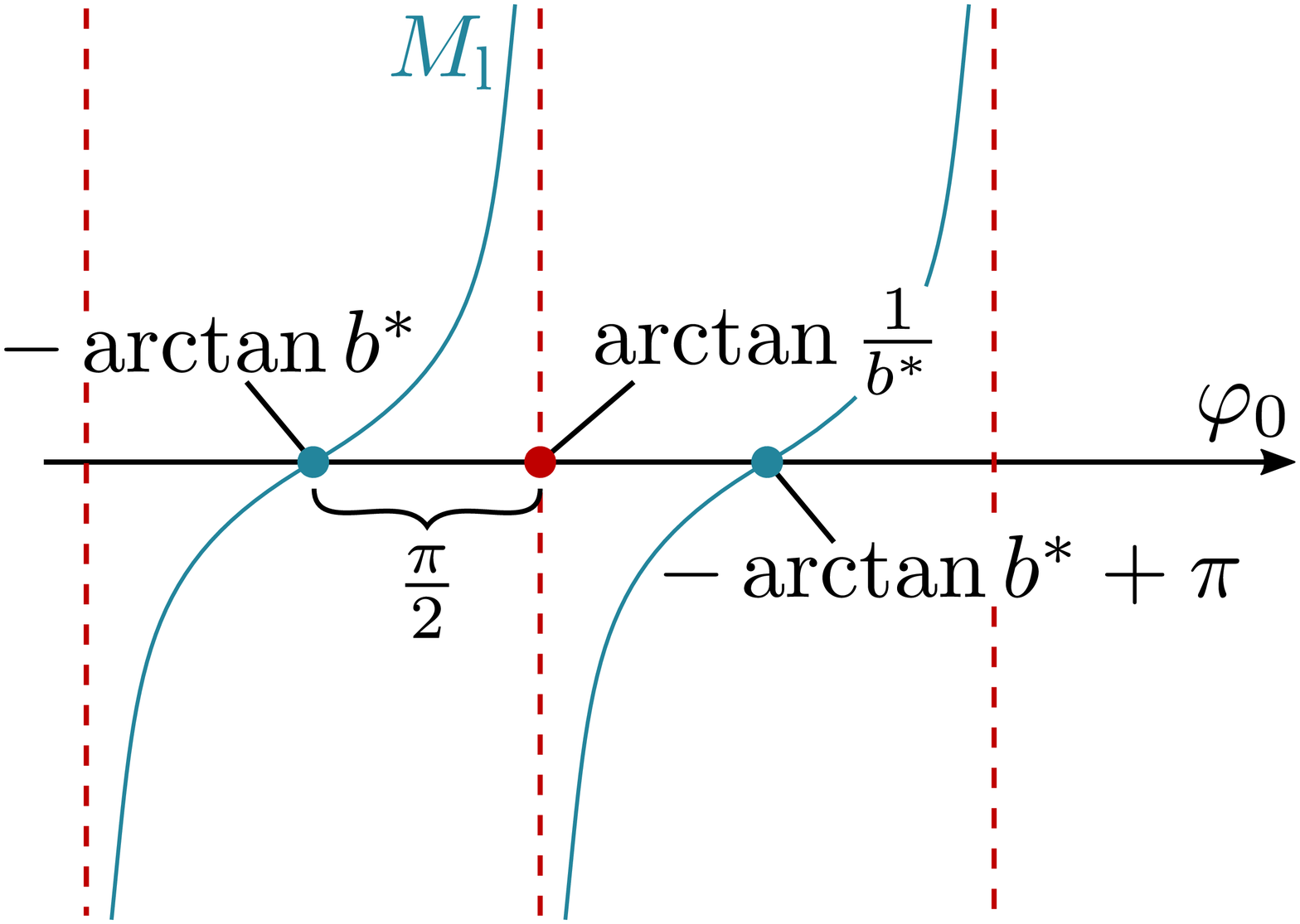}
		\caption{Plot of $\sloh$.}
	\end{subfigure}
	$\qquad$
	\begin{subfigure}[b]{0.6\textwidth}
		\centering
		\includegraphics[width=\textwidth]{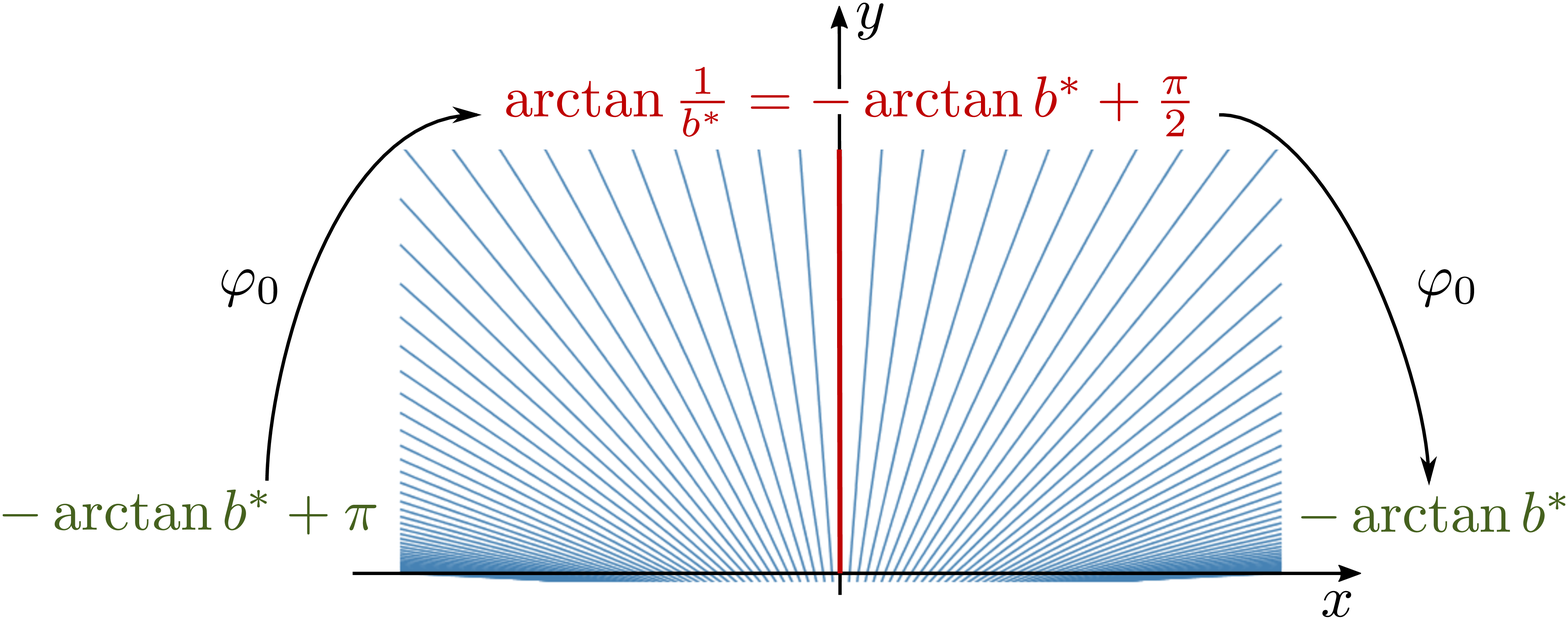}
		\caption{Range of corresponding characteristics.}
	\end{subfigure}
	\caption{Qualitative study of the slope $\sloh$ in \eref{eq:slope_half}; its graph is plotted in panel (a). This function vanishes for $\varphi_0=-\arctan\qb+k\pi$, with $k\in\mathbb{Z}$, giving rise to horizontal characteristic lines, while for $\varphi_0=-\arctan\qb+\pi/2+k\pi$ $\sloh$ diverges, jumping from $-\infty$ to $+\infty$ and producing vertical  characteristic lines (red in panel  (b)).}
	\label{fig:slope}
\end{figure}

\Fref{fig:slope} illustrates a qualitative study of the slope $\sloh$ in   \eref{eq:slope_half}.
It shows how $\varphi_0$ is confined to vary in a range of amplitude at most $\pi$, between $-\arctan\qb + k\pi$ and $-\arctan\qb + (k-1)\pi$ (for $k\in\Z$); $\varphi_0$ must also be a (not necessarily strictly) decreasing function of $x_0$ in order to select the half-plane $y>0$ as domain for $\n$ (an increasing $\varphi_0$ would select the half-plane $y<0$). 

\begin{figure}[h]
	\centering
	\begin{subfigure}[b]{0.45\textwidth}
		\centering
		\includegraphics[width=\textwidth]{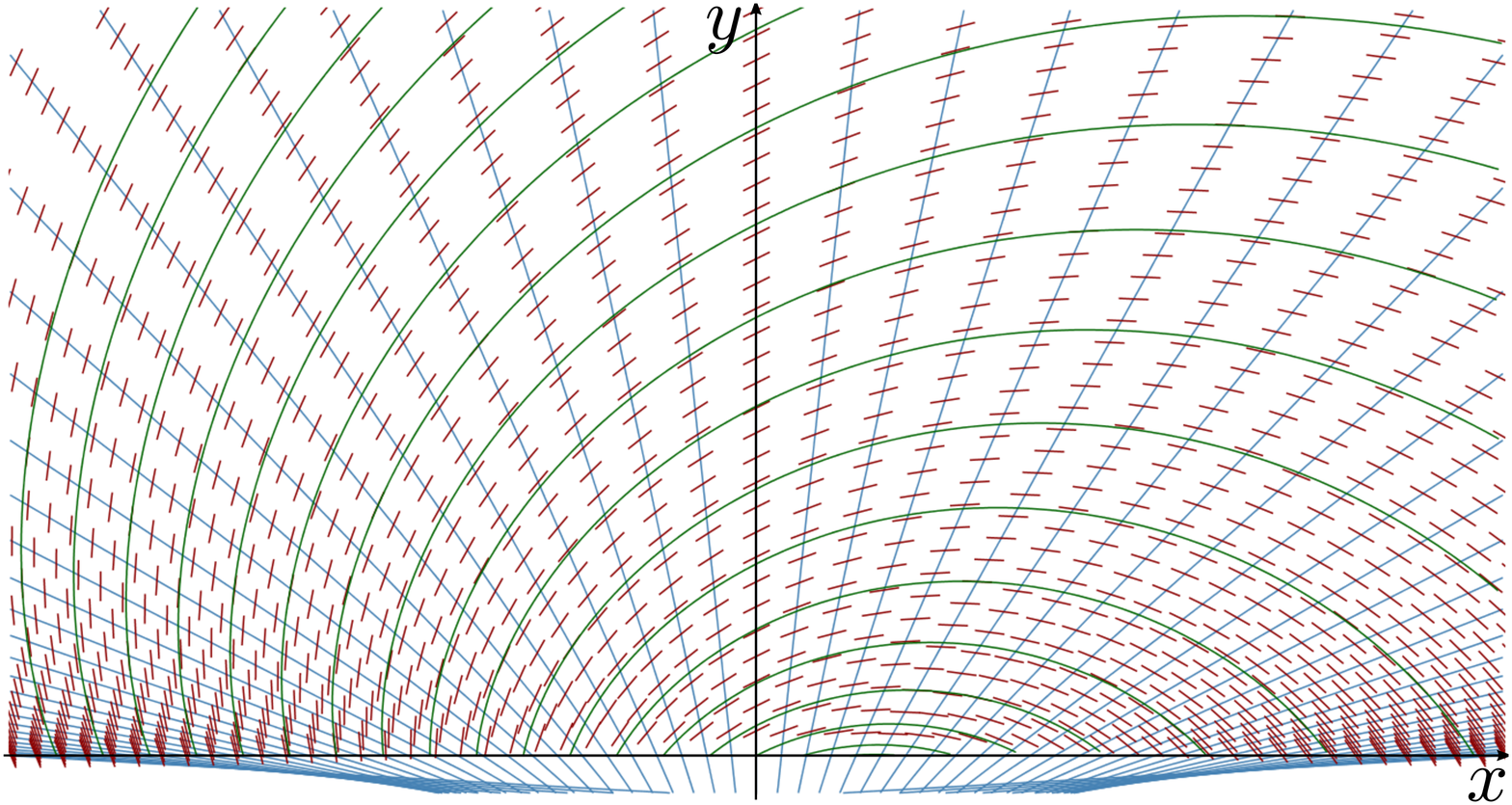}
		\caption{$\varphi_0(x_0)=-\pi(\tanh x_0 + 1)/2 - \arctan2$.}\label{fig:halfplane1}
	\end{subfigure}
	$\qquad$
	\begin{subfigure}[b]{0.45\textwidth}
		\centering
		\includegraphics[width=\textwidth]{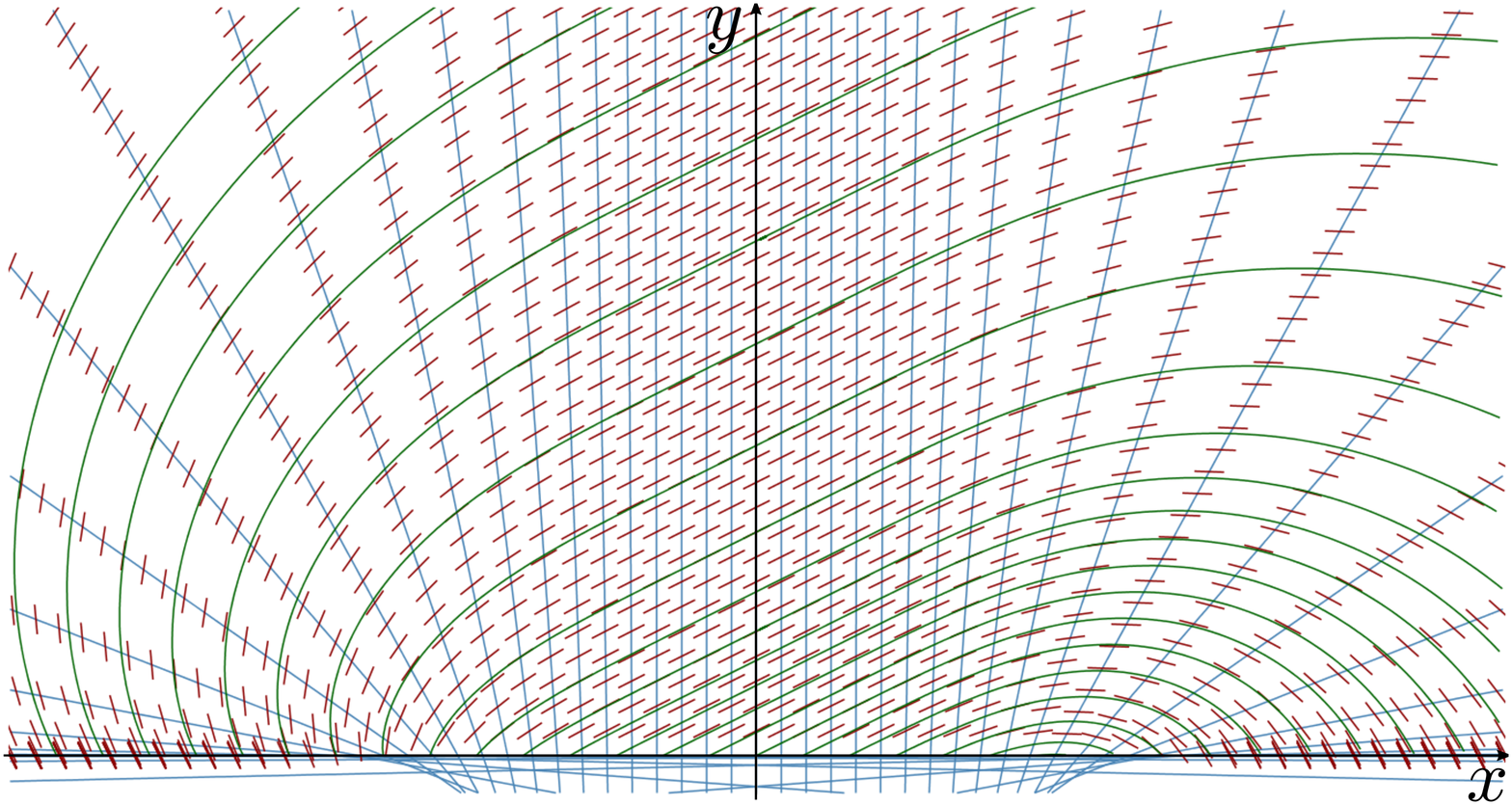}
		\caption{$\varphi_0(x_0)=-\pi( x_0^5 + 1)/2 - \arctan2$.}\label{fig:halfplane2}
	\end{subfigure}
	\caption{Integral lines (in green) of planar quasi-uniform distortions in \eref{eq:splay-bend} with $\qb=2$ and two choices of relivable frustrations $\varphi_0(x_0)$. The blue straight lines are the characteristic lines where $\varphi_0$ propagates unchanged: they are vertical at $x_0=0$ and horizontal for $x_0\to\pm\infty$. The nematic directors are represented by short red segments. The field $\n$ is relieved  in the half-plane $y>0$, as in both panels (a) and (b) $\varphi_0$ is a decreasing function of $x_0$.}
	\label{fig:halfplane}
\end{figure}

These criteria also identify the nematic frustrations localized on the line $y=0$ that can be relieved quasi-uniformly in the half-plane $y>0$.
Thus, a relivable frustration is, for example, 
\begin{equation}\label{eq:phi0_half}
\varphi_0(x_0)=-\frac\pi2(\tanh x_0 + 1) - \arctan\qb,
\end{equation}
which propagates to  a quasi-uniform distortion whose integral lines are shown in \fref{fig:halfplane1}. Any variation in the rate at which $\varphi_0$ varies with  $x_0$ would generate a different, but equally  relivable quasi-uniform field, like  the one illustrated in \fref{fig:halfplane2}.

\nalert{It is interesting to determine  the function $f$ that characterizes a frustration that can be relieved quasi-uniformly in a half-plane. The explicit expression for $f$ is obtained from \eref{eq:f_line} in \ref{app:frustrated_line} by a change of variables that maps $(x,y)$ into $(x_0,s)$, where $s\in\mathbb{R}$ describes the characteristic through $(x_0,0)$, for any given $x_0\in\mathbb{R}$. In the new variables, the line $y=0$ corresponds to $s=0$ and there $f$ is a function of $x_0$ only, which reads as
\begin{equation}
	\label{eq:f_y=0}
	f(x_0,0)=-\frac12\frac{\varphi_0'}{\sin\varphi_0+b^\ast\cos\varphi_0}.
\end{equation}
On the other hand, along each characteristic,
\begin{equation}
	\label{eq:f_line_infinity}
	\lim_{|s|\to\infty}f(x_0,s)=0\quad\forall\ x_0\in\mathbb{R}.
\end{equation}

Level sets of $f$ in the plane $(x,y)$ have a clear geometric interpretation: they represent  \emph{restricted loci} of uniformity, where all distortion characteristics are constant, although their values can be different on different level sets. Equation \ref{eq:f_y=0} suffices to show that the line of frustration is \emph{not} a restricted locus of uniformity. We shall return to this issue in \sref{sec:1d_uniformity}, in connection with a notion of lower-dimensional uniformity.}

In \sref{sec:universal}, we shall provide a further example of relivable frustration  and we shall investigate the behaviour of the corresponding quasi-uniform field far away from the origin.

\section{Winding outside a circle}\label{sec:defect}
In this section, we present a construction for a class of planar quasi-uniform  distortions winding outside a circle upon which $\n$ is prescribed. We consider the $(x,y)$-plane deprived of the region enclosed by the (unit) circle $\disk$ with center at the origin of a Cartesian frame $(\vx,\vy)$.\footnote{A uniform scaling would map $\disk$ onto a circle of any selected radius $r_0$ with no qualitative consequences on our development.} We imagine $\n$ prescribed on $\disk$ with a given \emph{topological charge} $m$ (also known as the \emph{winding number}) and we ask whether it can be relieved quasi-uniformly outside $\disk$. Again by use of the method of characteristics, depending on the value of $m$, we shall find either the elementary, planar quasi-uniform distortions in \fref{fig:quasi-unifom_extended} or new families of planar quasi-uniform distortions defined outside $\disk$, some filling the rest of the plane, others living only in a half-plane.

In the present geometric setting, we find it convenient to  rewrite the nematic field \eref{eq:splay-bend} in the local frame $(\er,\et)$ of polar coordinates $(r,\vartheta)$ as
\begin{equation}\label{eq:defect}
 \n = \cos\alpha\er + \sin\alpha\et,
 \quad\mathrm{with}\quad \alpha=\alpha(r,\vartheta),\quad\mathrm{for}\quad r>1,
\end{equation}
so that the azimuthal angle $\varphi$ between $\n$ and $\vx$ is given by $\varphi=\alpha + \vartheta$, where $\vartheta\in[0,2\pi)$. Here $\alpha$ represents the \emph{local} orientation angle of $\n$ relative to the radial direction $\er$.
For the gradient of $\n$ we then have
\begin{equation}\label{eq:gradn_defect}
\eqalign{
 \grad\n &=
  - \alpha_{,r}\sin\alpha\er\tpr\er - \frac{1+\alpha_{,\vartheta}}{r}\sin\alpha\er\tpr\et \\
 &\qquad
 + \alpha_{,r}\cos\alpha\et\tpr\er + \frac{1+\alpha_{,\vartheta}}{r}\cos\alpha\et\tpr\et}
\end{equation}
and the distortion characteristics become (see \ref{app:sb_defect} for details)
\begin{equation}\label{eq:characteristic_defect}
\fl
 S = \frac{1+\alpha_{,\vartheta}}{r}\cos\alpha - \alpha_{,r}\sin\alpha,\ 
 b_1 = \alpha_{,r}\cos\alpha+\frac{1+\alpha_{,\vartheta}}{r}\sin\alpha,\  
 b_2 = T = 0,\ 
 q = \frac S2,
\end{equation}
valid for $S\geq0$. The natural choice for the factorizing function is then $f=q$, so that
\begin{equation}\label{eq:constants_defect}
\qS=2,
\quad \qb_2=\qT=0,
\quad \qq=1.
\end{equation} 
For a genuine quasi-uniform distortion we also require $\f$ not to vanish identically. It follows from \eref{eq:characteristic_defect} that if $f\neq0$ then
\begin{equation}\label{eq:defect_cond}
 \frac{1+\alpha_{,\vartheta}}{r}\cos\alpha \neq \alpha_{,r}\sin\alpha
\end{equation} 
and 
\begin{equation}\label{eq:b1_defect}
\quad \qb_1= 2\frac{(1 + \alpha_{,\vartheta})\sin\alpha + r\alpha_{,r}\cos\alpha}{(1 + \alpha_{,\vartheta})\cos\alpha - r\alpha_{,r}\sin\alpha}
= 2\frac{(1 + \alpha_{,\vartheta})\tan\alpha + r\alpha_{,r}}{1 + \alpha_{,\vartheta} - r\alpha_{,r}\tan\alpha}.
\end{equation} 
Without loss of generality, we also assume that $\qb_1\geq0$.\footnote[1]{Were this not the case, we could take $(-\n_1,-\n_2,\n)$ as  distorsion frame  instead of $\dframe$, thus recovering the case $\qb_1\geq0$.} 

Quasi-uniformity is then achieved when $\qb_1 = 2\qb \geq 0$ is constant, that is, when
\begin{equation}\label{eq:defect_pde}
 r\alpha_{,r}(\qb\sin\alpha+\cos\alpha) - \alpha_{,\vartheta}(\qb\cos\alpha-\sin\alpha) = \qb\cos\alpha - \sin\alpha.
\end{equation}
By using the method of characteristics, we parametrize $r$, $\vartheta$ and $\alpha$ in $t$ and we follow the propagation of the  values $r=r_0$, $\vartheta=\vartheta_0$, and $\alpha_0=\alpha(1,\vartheta_0)$ for $t=0$ along the solutions of \eref{eq:defect_pde}. The associated Lagrange–Charpit equations are
\begin{equation}\label{eq:defect_chareq}
 \fl
 \frac{\dd r}{\dd t} = r(\qb\sin\alpha+\cos\alpha),
 \quad 
 \frac{\dd\vartheta}{\dd t} = -(\qb\cos\alpha-\sin\alpha),
\quad
\frac{\dd\alpha}{\dd t} = \qb\cos\alpha - \sin\alpha.
\end{equation}

In the following we focus on the main properties of the characteristic curves for \eref{eq:defect_chareq}: more details and full computations are collected in \ref{app:sb_char_curves}.

By solving the nonparametric equations
\begin{equation}\label{eq:defect_chareq_nonpar}
\frac{\dd\alpha}{\dd\vartheta} = -1
\qquad \mathrm{and} \qquad
\frac{\dd r}{\dd\vartheta} = -\frac{r(\qb\sin\alpha+\cos\alpha)}{\qb\cos\alpha-\sin\alpha}
\end{equation}
obtained from \eref{eq:defect_chareq}, we arrive at
\begin{equation}\label{eq:defect_dalphadtheta}
 \alpha = \varphi_0 - \vartheta, \quad \mathrm{with}\quad\varphi_0(\vartheta_0):=\alpha_0 + \vartheta_0,\quad\alpha_0:=\alpha(1,\vartheta_0)
\end{equation}
and
\begin{equation}\label{eq:defect_drdtheta}
\fl
 r = R_0|\qb\cos(\varphi_0 - \vartheta)-\sin(\varphi_0 - \vartheta)|^{-1},
 \quad \mathrm{with }\quad R_0:=|\qb\cos\alpha_0-\sin\alpha_0|>0 .
\end{equation}
We note that both $\varphi=\alpha+\vartheta$ and $R:=r|\qb\cos\alpha-\sin\alpha|$ are constant along the characteristic curves. Therefore, by recalling that $x=r\cos\vartheta$ and $y=r\sin\vartheta$, with the aid of \eref{eq:defect_dalphadtheta}, we also see that the quantity
\begin{equation}\label{eq:characteristic_circle}
 |(\qb \cos\varphi_0 - \sin\varphi_0)x + (\qb \sin\varphi_0 + \cos\varphi_0)y| = R \\
\end{equation}
is constant along a characteristic curve. Thus, any such curve is a  straight line with slope 
\begin{equation}\label{eq:defect_slope}
\slod:=-\frac{\qb \cos\varphi_0 - \sin\varphi_0}{\qb \sin\varphi_0 + \cos\varphi_0}
\end{equation}
through the point  $(\cos\vartheta_0,\sin\vartheta_0)$ of $\disk$. \nalert{Moreover, by \eref{eq:characteristic_defect}, the function $f$ can also be written in terms of $\varphi$ and $\alpha$ as
\begin{equation}
	\label{eq:f_circle}
	f=\frac12\left(\frac{1}{r}\varphi_{,\vartheta}\cos\alpha-\varphi_{,r}\sin\alpha\right).
\end{equation}
}

The domain $\mathcal{D}$ outside $\disk$ where the characteristic lines do  not cross each other is bounded by the characteristic lines tangent to $\disk$. The latter touch $\disk$  at points $(\cos\vartheta_0^*, \sin\vartheta_0^*)$, where $\vartheta_0^*$ solves the equation
\begin{equation}
	\label{eq:tangent_characteristic_lines}
\tan\alpha_0(\vartheta_0^*)=-\frac{1}{\qb}.	
\end{equation}
Letting $\tanset$ denote the set of roots of \eref{eq:tangent_characteristic_lines} in $[0,2\pi)$, we can represent $\mathcal{D}$ as 
\begin{equation}\label{eq:defect_domain}
\fl
 \mathcal{D} =  \Big\{(x,y)\in\R^2 \,|\, x^2+y^2>1\Big\}
\bigcap_{\vartheta_0\in\tanset}\Big\{(x,y)\in\R^2 \,|\, x\cos\vartheta_0+y\sin\vartheta_0<1\Big\}
 .
\end{equation}

To provide concrete examples of this construction, we prescribe $\n$ on $\disk$ with a given \emph{topological charge} $m$. Since $\n$ enjoys the nematic symmetry, which identifies $\n$ and $-\n$, $2m$ is the integer that counts the number of times the azimuthal angle $\varphi_0$ winds around $\disk$  while $\vartheta_0$ runs once along it. Thus, we have that   $\varphi_0(\vartheta_0+2\pi) = \varphi_0(\vartheta_0) + 2m\pi$, and so $\alpha$ is subject to the following condition on its trace $\alpha_0$ on $\disk$,
\begin{equation}\label{eq:alpha_0_equation}
\alpha_0(\vartheta_0 + 2\pi) - \alpha_0(\vartheta_0) = 2(m-1)\pi. 
\end{equation}

\nalert{Introducing coordinates $(\vartheta_0,s)$ in the plane $(x,y)$, with $s\in\mathbb{R}$ describing the characteristic in \eref{eq:characteristic_circle} with given $\vartheta_0$, we can obtain $f$ from \ref{eq:f_circle} (see \ref{app:sb_defect}); its explicit form is in general too complicated to be easily interpreted, but on $\disk$ it simplifies into
\begin{equation}
	\label{eq:f_disk_s=0}
	f(\vartheta_0,0)=\frac12\frac{\varphi_0'}{b^\ast\sin\alpha_0+\cos\alpha_0},
\end{equation}
which parallels \eref{eq:f_y=0}, while 
\begin{equation}
	\label{eq:f_disk_infinity}
	\lim_{|s|\to\infty}f(\vartheta_0,s)=0\quad\forall\ \vartheta_0\in[0,2\pi),
\end{equation}
which parallels \eref{eq:f_line_infinity}. We conclude from \eref{eq:f_disk_s=0} that $\disk$ is not a level set of $f$.}

\begin{figure}[h]
\centering
 \begin{subfigure}[b]{0.3\textwidth}
 \centering
  \includegraphics[width=\textwidth]{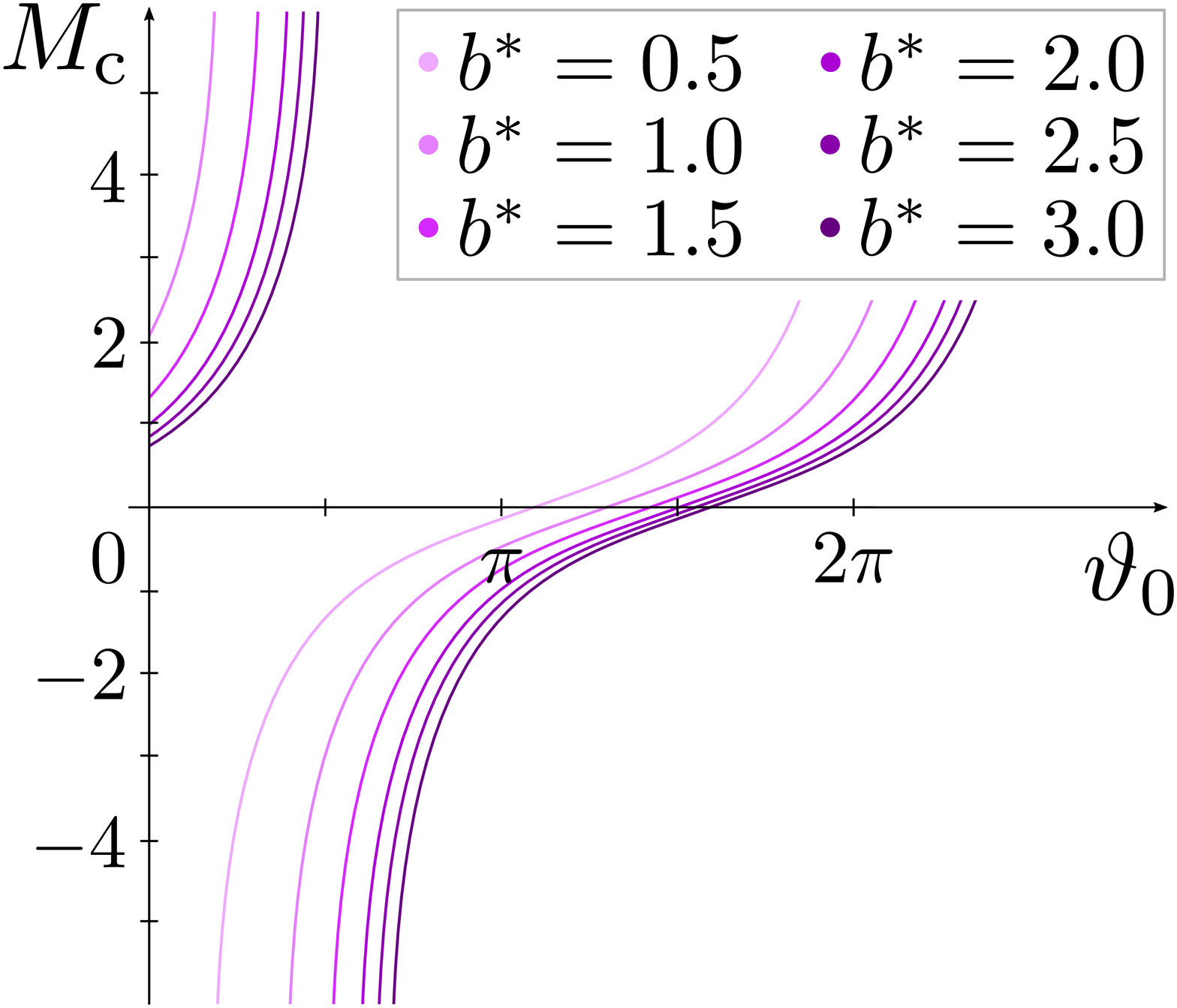}
  \caption{$m=1/2$.}
 \end{subfigure}
 \quad
 \begin{subfigure}[b]{0.3\textwidth}
 \centering
  \includegraphics[width=\textwidth]{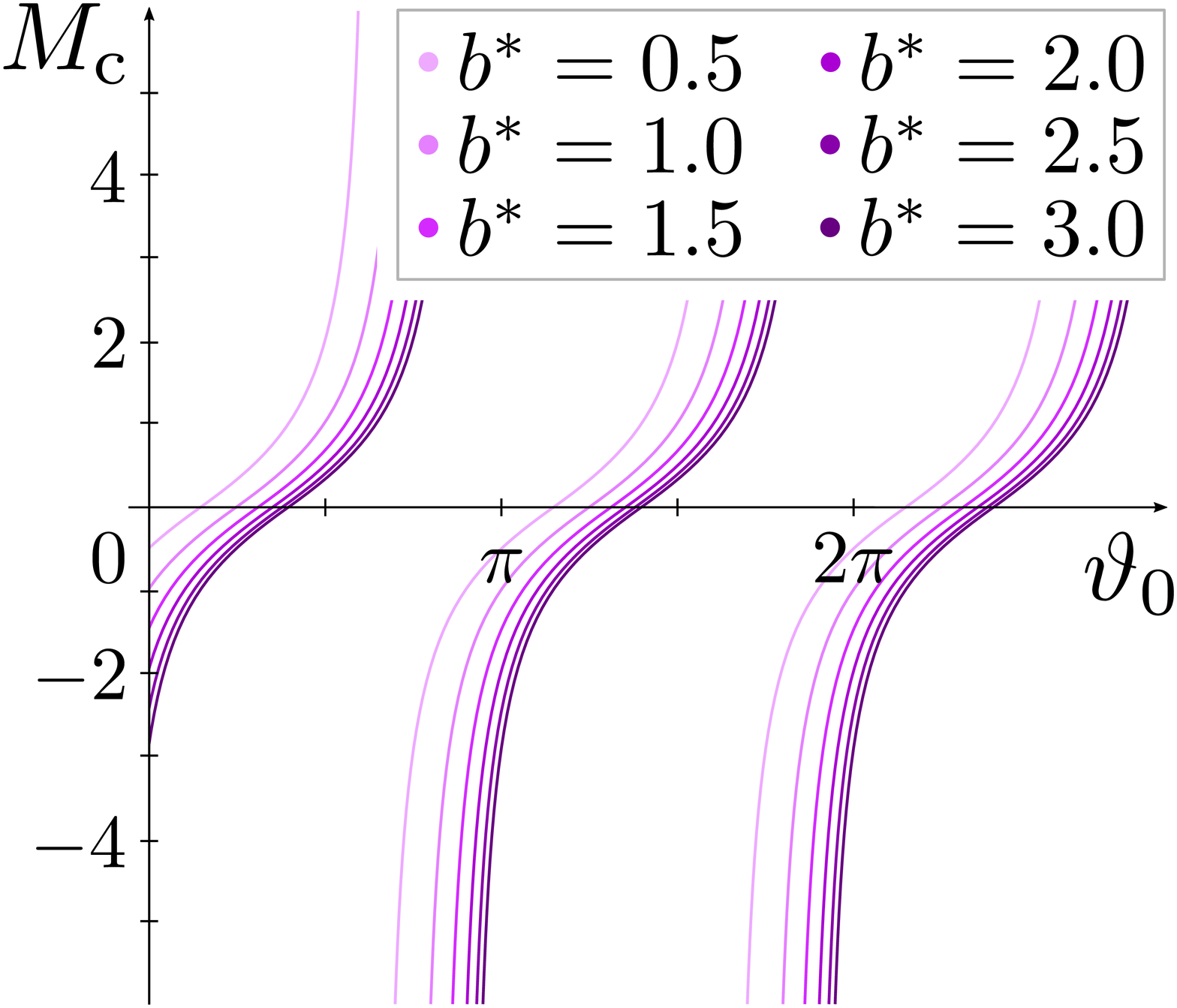}
  \caption{$m=1$.}
 \end{subfigure}
  \quad 
 \begin{subfigure}[b]{0.3\textwidth}
 \centering
  \includegraphics[width=\textwidth]{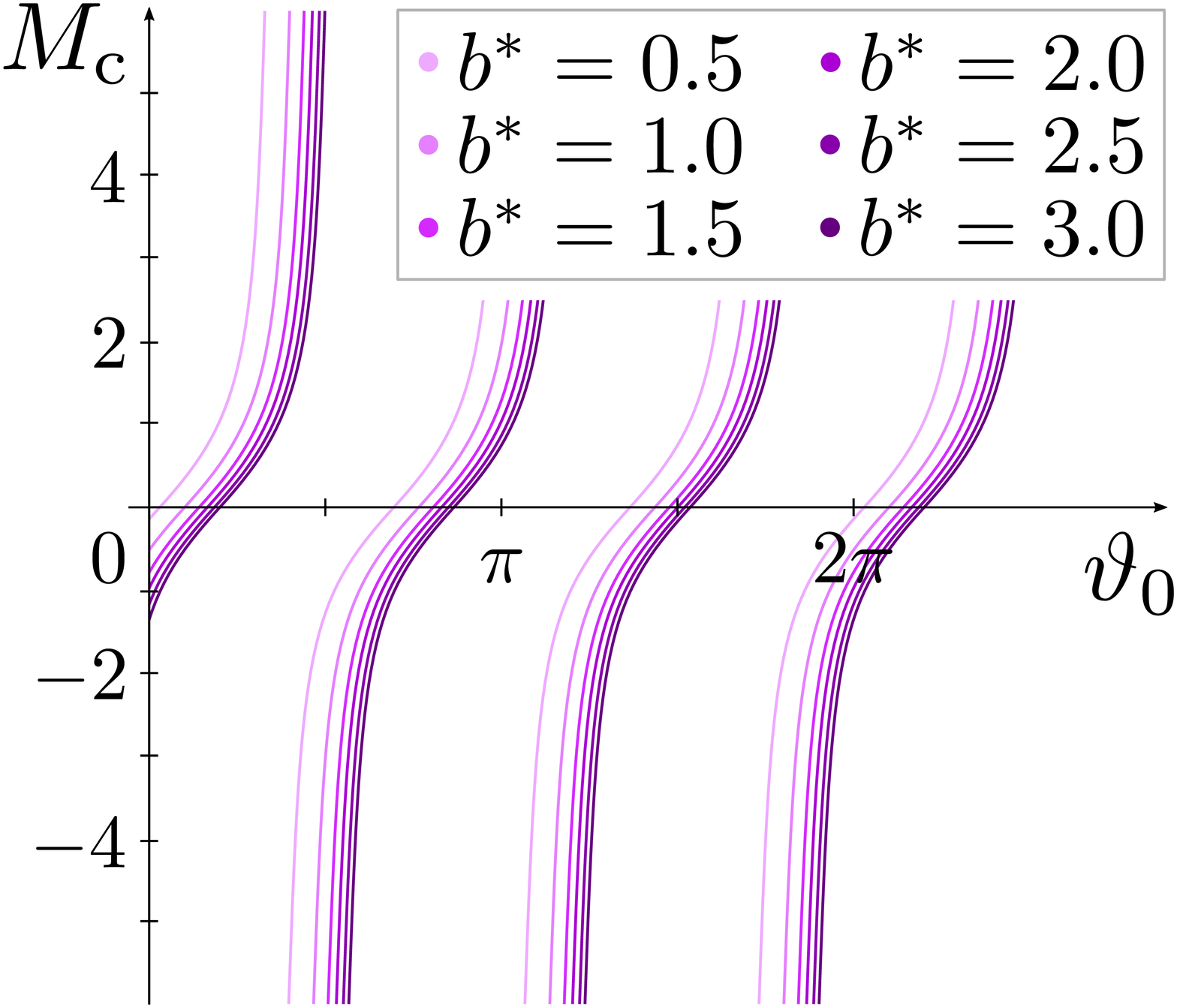}
  \caption{$m=3/2$.}
 \end{subfigure}
\caption{Illustrating how the slope $\slod$ in \eref{eq:defect_slope} depends on $\qb$ when $\vartheta_0\in[0,2\pi)$ for  $\alpha_0$ as in \eref{eq:alpha0}. In panel (b), $c_0=0$, while in panels (a) and (c) $c_0=-3\pi(m-1)/2-\arctan(1/\qb)$, so that $\tanset=\{3\pi/2\}$. In all cases, the period of $\slod$ is  $2m$, independently on $\qb$.}
\label{fig:slope_disk}
\end{figure}

A simple function that satisfies \eref{eq:alpha_0_equation} is 
\begin{equation}\label{eq:alpha0}
 \alpha_0(\vartheta_0) = (m-1)\vartheta_0 + c_0
 \quad\mathrm{for}\quad\vartheta_0\in[0,2\pi),
\end{equation}
where $c_0$ is a constant that for $m\neq1$ merely affects the field $\n$ by a rigid rotation. The corresponding nematic fields $\n$ are also known as \emph{Frank's disclinations} \cite{frank:theory}. For this choice of $\alpha_0$, we have that on each characteristic line,
\begin{equation}\label{eq:phi}
\varphi(r, \vartheta) = \alpha(r, \vartheta) + \vartheta = \alpha_0(\vartheta_0) + \vartheta_0 = \varphi_0(\vartheta_0) = m\vartheta_0 + c_0.
\end{equation}
It should be noticed that the field $\n$ generated by the propagation of $\varphi_0$ along characteristics has by continuity the same topological charge $m$ on any circuit enclosing $\disk$ in $\mathcal{D}$, insensitive to the fact that the slope of characteristics $\slod$ also depends on $\qb$. Indeed, $\slod$ takes on any specified value (for example, $0$) exactly $2m$ times as $\vartheta_0$ covers the interval $[0,2\pi)$, as shown in \fref{fig:slope_disk}.

If $m=0$ then $\varphi\equiv c_0$ and $\n$ is everywhere constant and its domain is the whole plane. If $m=1$ then $\mathcal{D}$ is the whole  plane outside $\disk$, apart from the \emph{resonant} case where $c_0=\arctan\qb $. In this case, characteristics can also be extended inside $\disk$ and the field lines of $\n$ are logarithmic spirals emanating from the origin with constant local angle $\alpha=\arctan\qb$; they range  from straight lines (as for the planar splay) when $\qb \to0$ to concentric circles (as for the pure bend) when $\qb \to+\infty$.

\begin{figure}
\centering
 \begin{subfigure}[b]{0.3\textwidth}
 \centering
  \includegraphics[width=\textwidth]{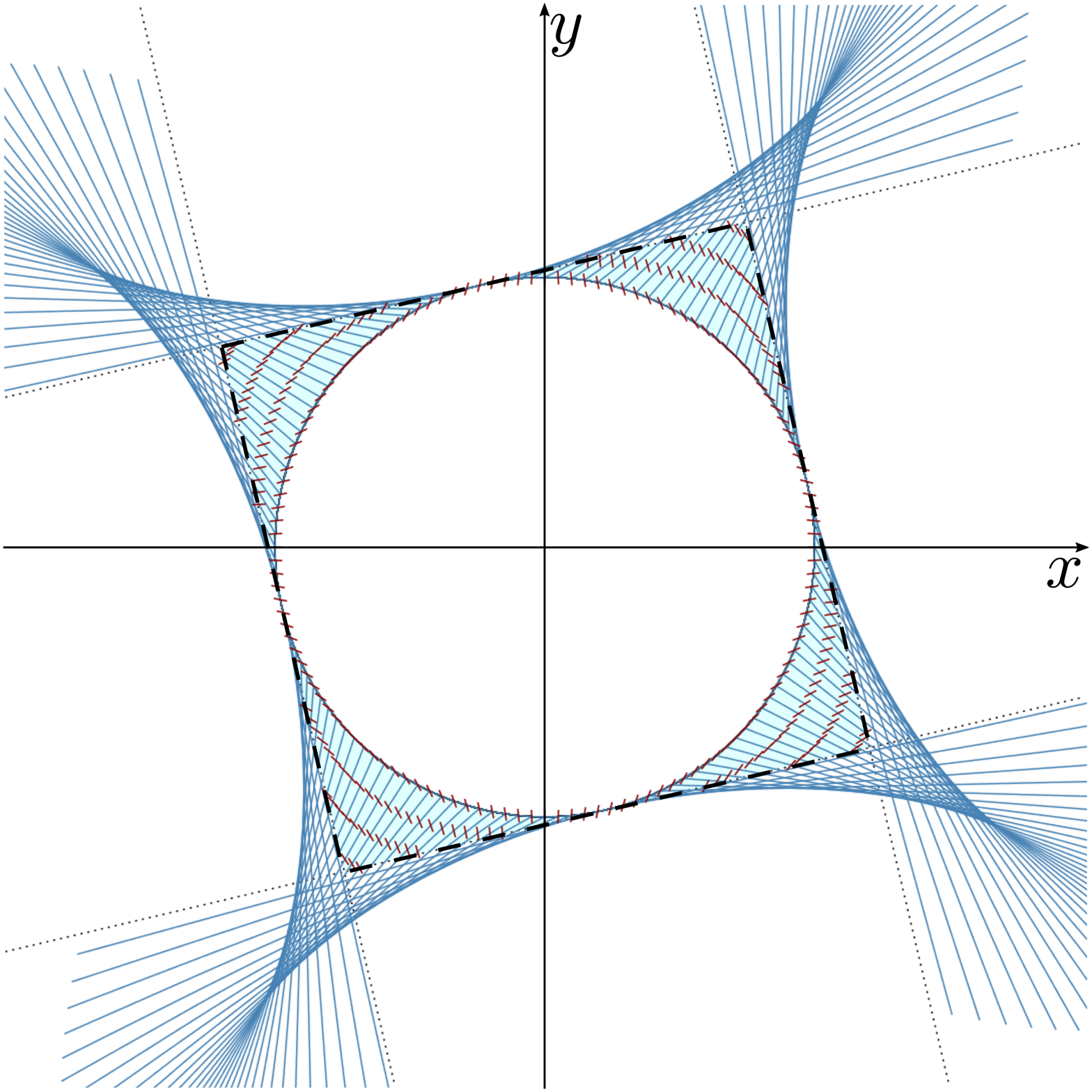}
  \caption{$m=-1$.}
 \end{subfigure}
 \ 
 \begin{subfigure}[b]{0.3\textwidth}
 \centering
  \includegraphics[width=\textwidth]{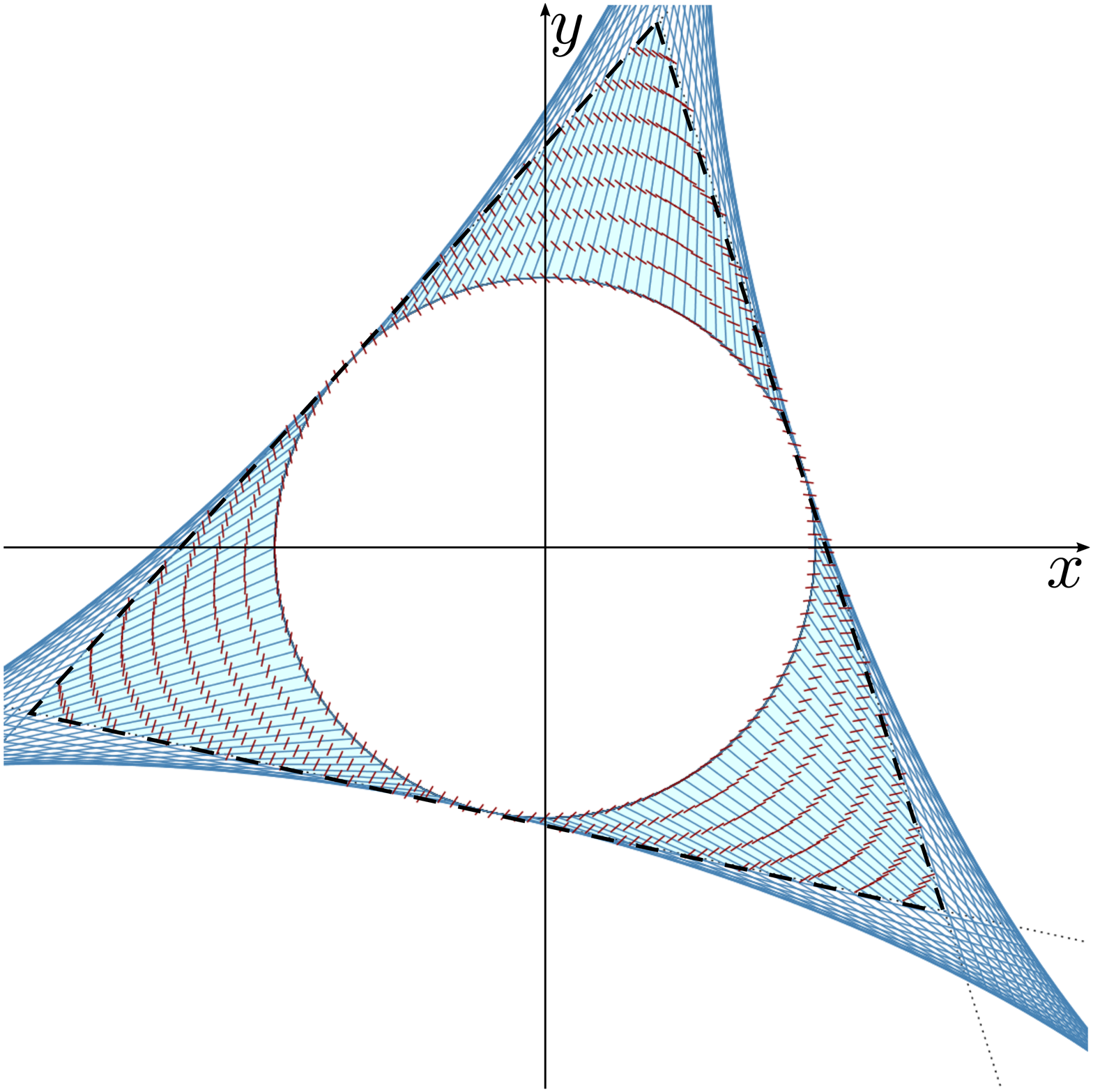}
  \caption{$m=-1/2$.}
 \end{subfigure}
 \ 
 \begin{subfigure}[b]{0.3\textwidth}
 \centering
  \includegraphics[width=\textwidth]{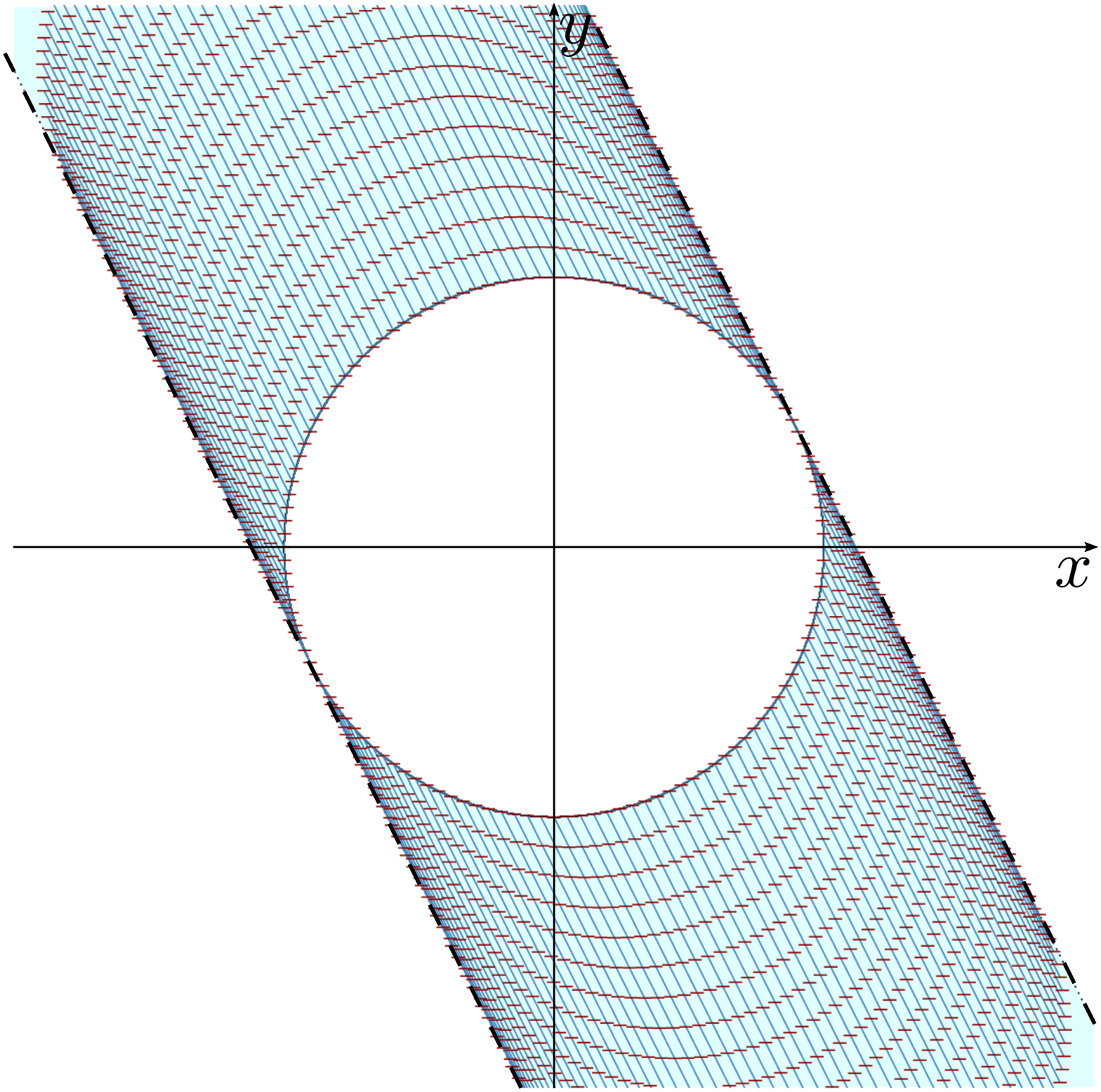}
  \caption{$m=0$.}
 \end{subfigure}
 \\
 \begin{subfigure}[b]{0.3\textwidth}
 \centering
  \includegraphics[width=\textwidth]{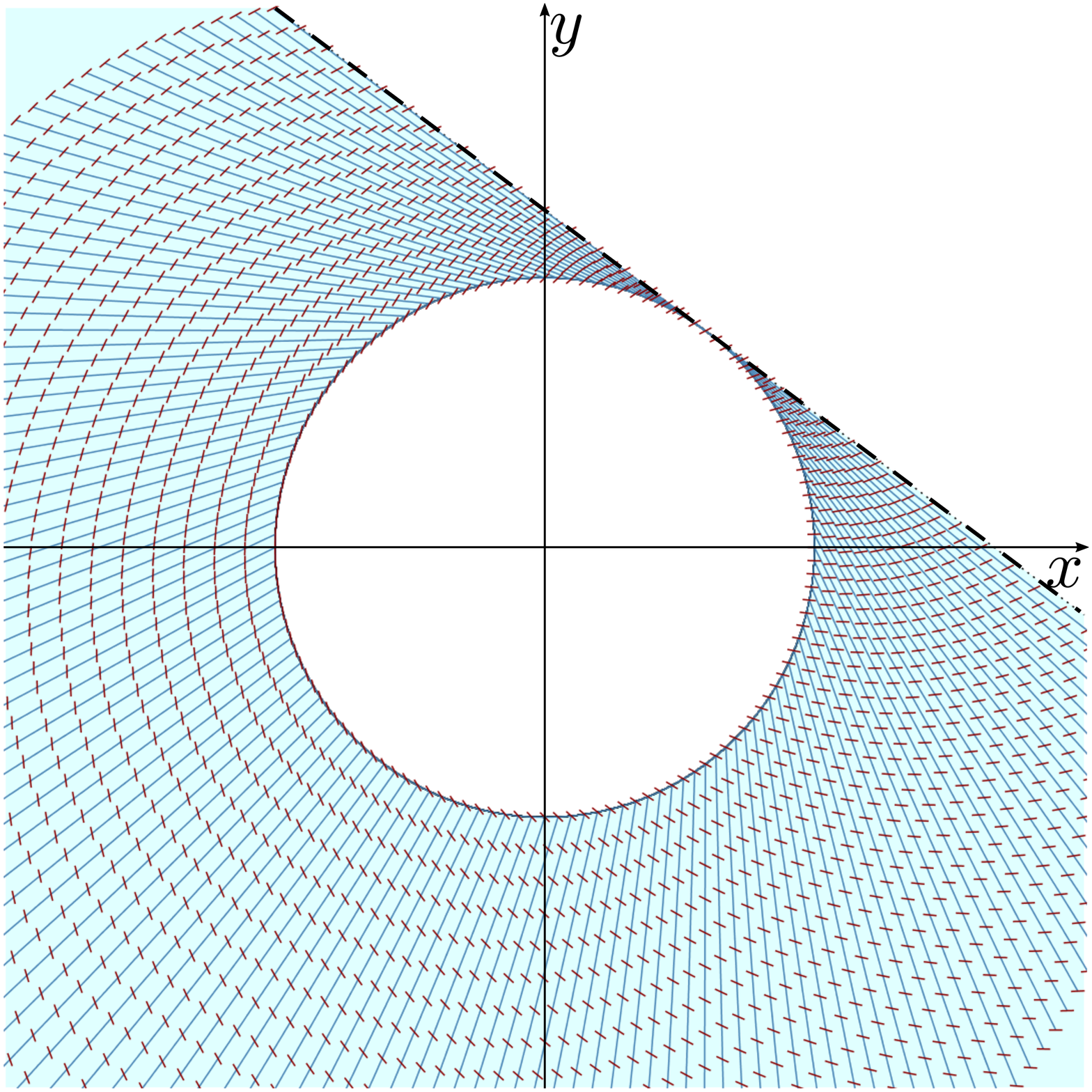}
  \caption{$m=1/2$.}
 \end{subfigure}
 \ 
 \begin{subfigure}[b]{0.3\textwidth}
 \centering
  \includegraphics[width=\textwidth]{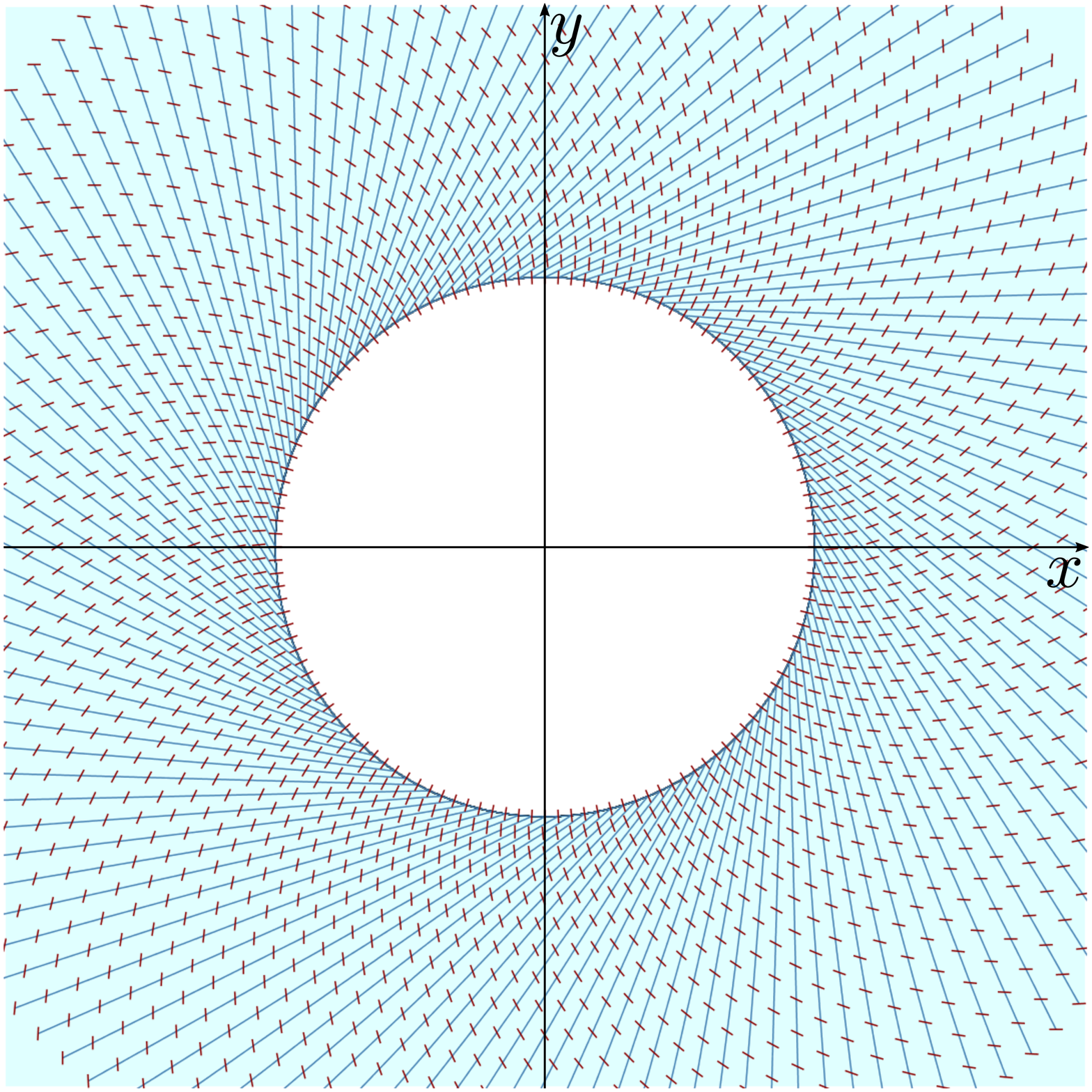}
  \caption{$m=1$.}
 \end{subfigure}
 \ 
 \begin{subfigure}[b]{0.3\textwidth}
 \centering
  \includegraphics[width=\textwidth]{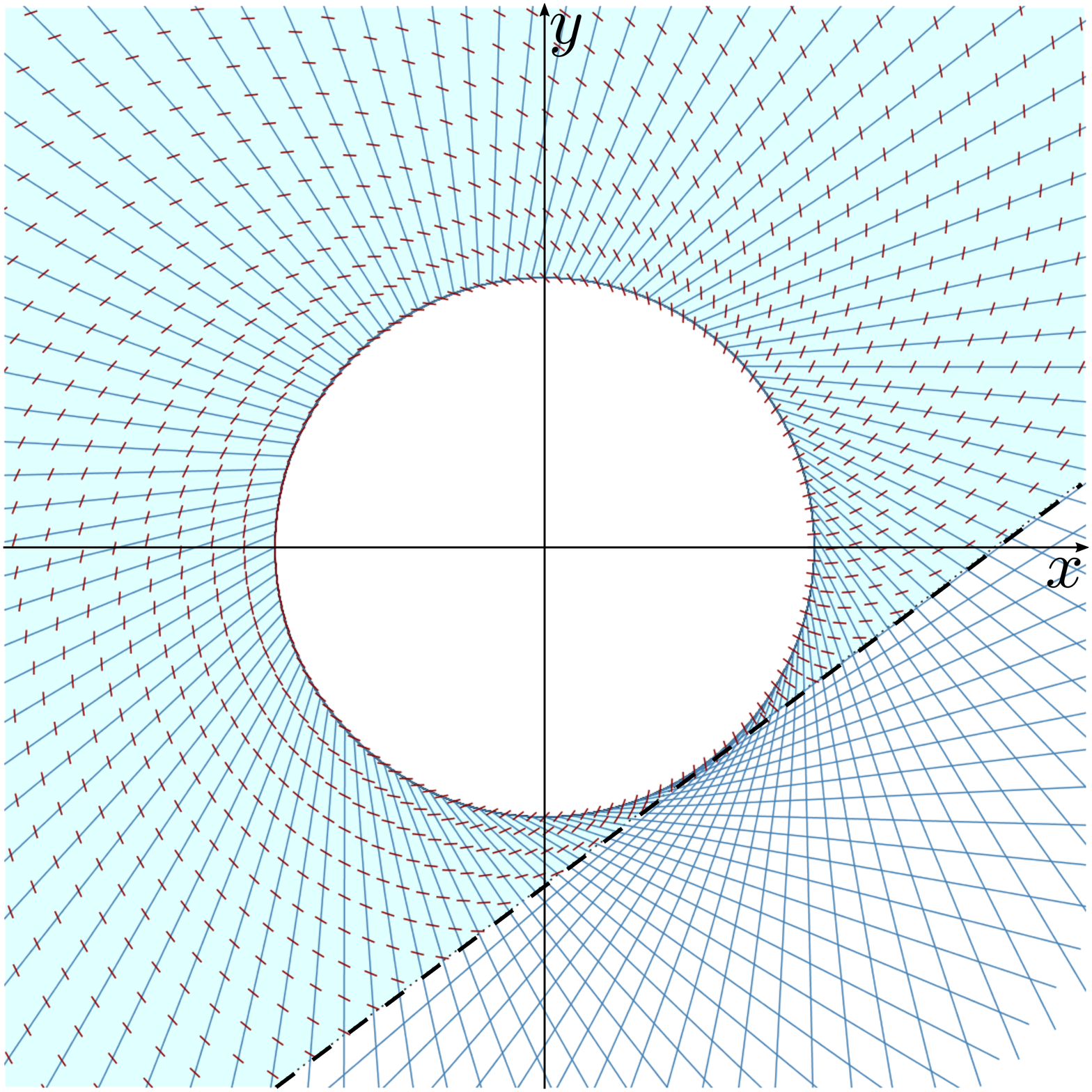}
  \caption{$m=3/2$.}
 \end{subfigure}
  \\
 \begin{subfigure}[b]{0.3\textwidth}
 \centering
  \includegraphics[width=\textwidth]{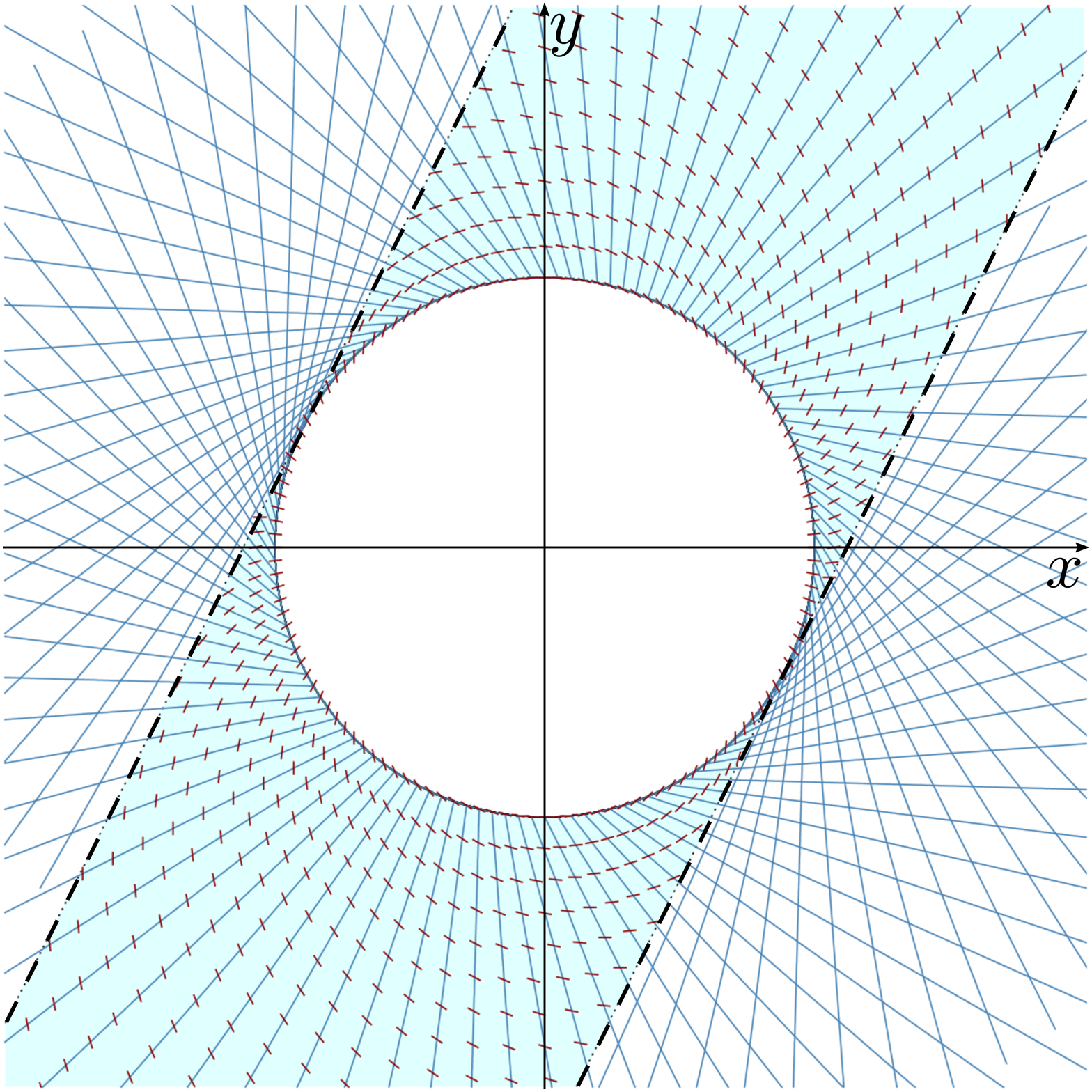}
  \caption{$m=2$.}
 \end{subfigure}
 \ 
 \begin{subfigure}[b]{0.3\textwidth}
 \centering
  \includegraphics[width=\textwidth]{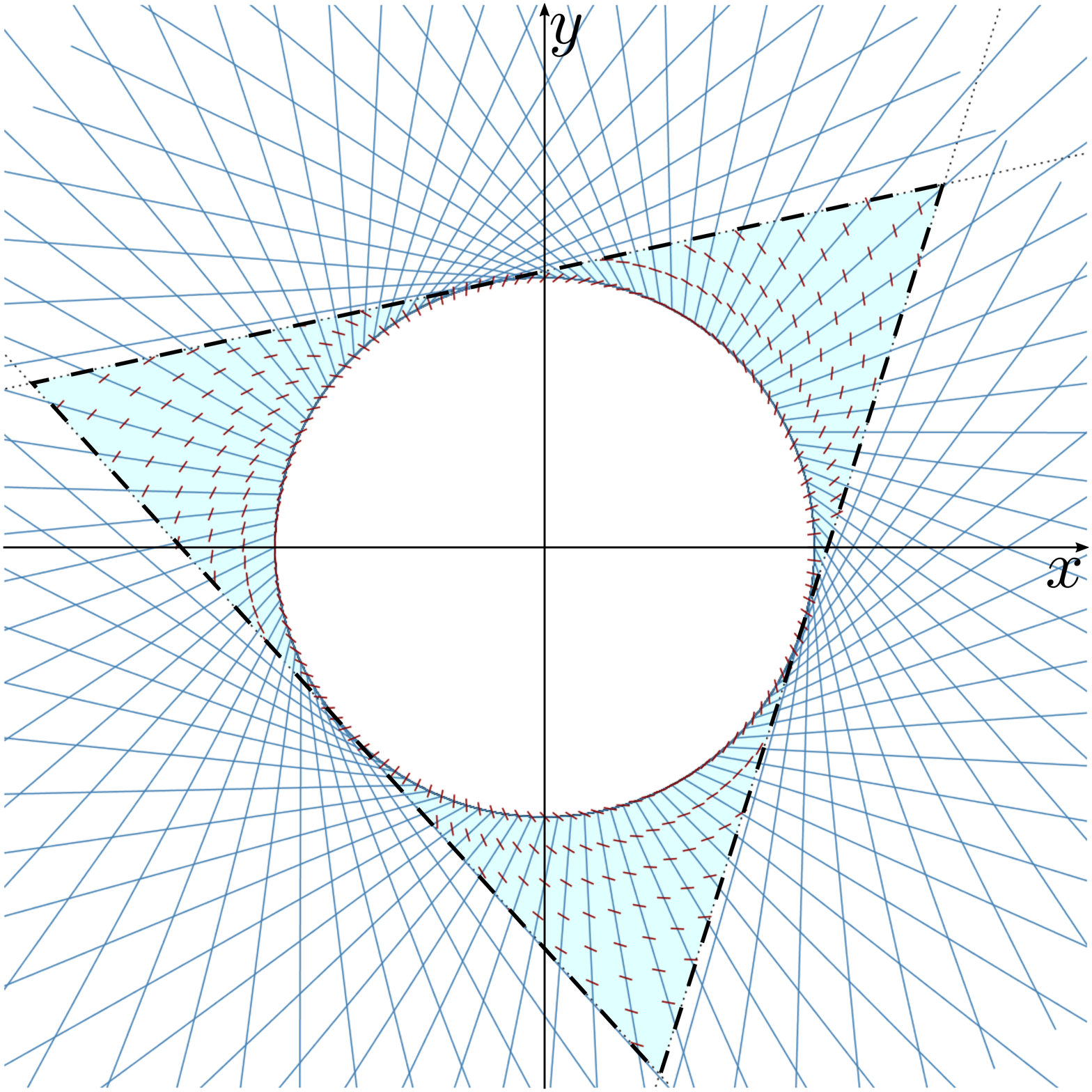}
  \caption{$m=5/2$.}
 \end{subfigure}
 \ 
 \begin{subfigure}[b]{0.3\textwidth}
 \centering
  \includegraphics[width=\textwidth]{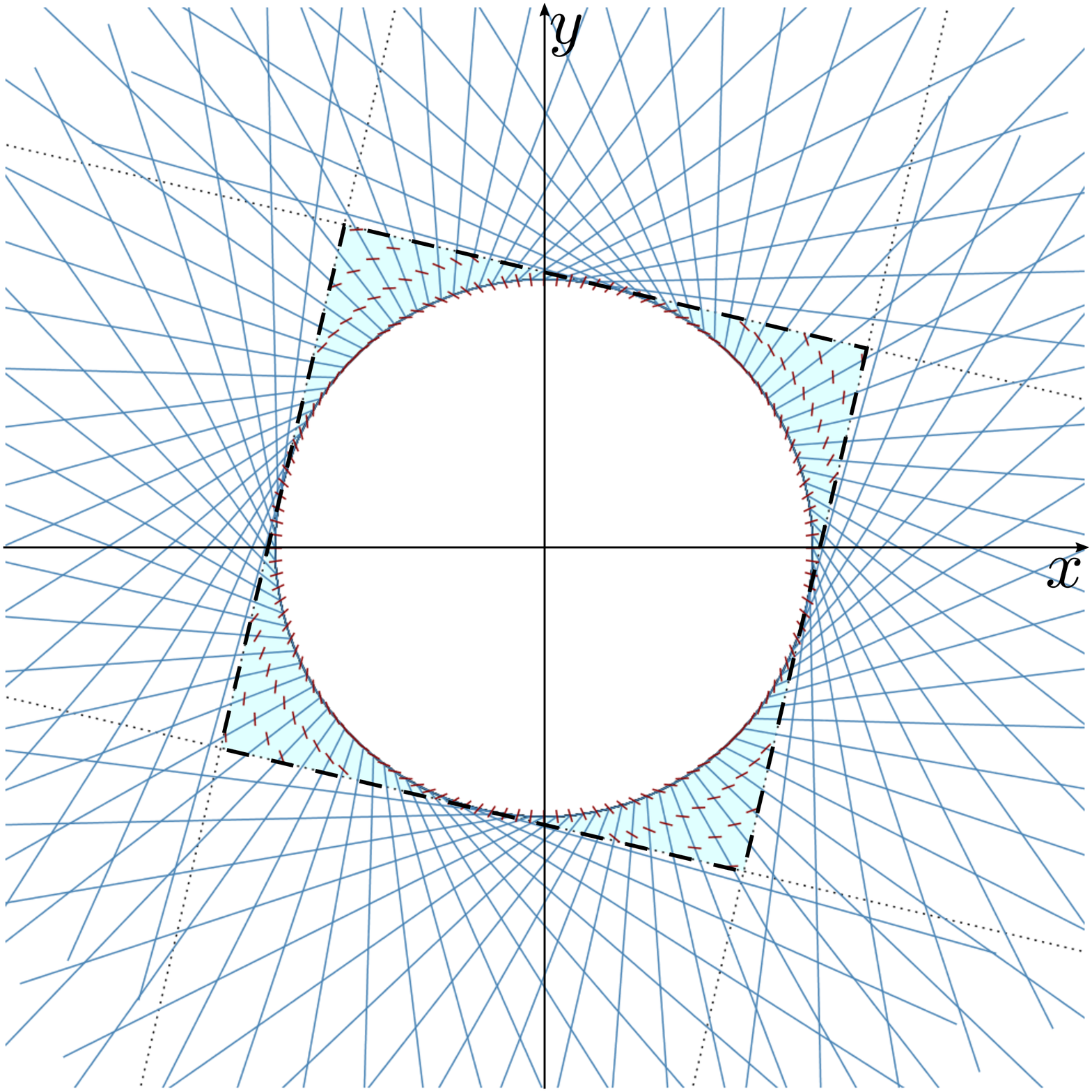}
  \caption{$m=3$.}
 \end{subfigure}
\caption{Domain $\mathcal{D}$ in \eref{eq:defect_domain} for planar quasi-uniform splay-bend distortions winding  outside $\disk$ for different values of the topological charge $m$. Directors are represented by red segments, while  blue straight lines are characteristic lines along which $\n$ propagates unchanged, starting from the frustration prescribed on $\disk$  with angle $\alpha$ as in \eref{eq:alpha0}: here $c_0=0$, while $\qb=2$. $\mathcal{D}$ is the dark blue region  bounded by black dashed lines, corresponding to characteristic lines tangent to $\disk$  at angles $\vartheta_0^*\in\tanset$  in \eref{eq:theta0}. Panels (a), (b), (h), and (i) provide examples where $\mathcal{D}$ is bounded.  In panel (c), $\n$ can actually be extended to the whole plane because it is constant, while in panel (g) this is not possible and the domain is unbounded but in one direction only. The most interesting cases are shown in panels (d), (e), and (f): they realize new families of quasi-uniform splay-bend distortions, other than the well-known quasi-uniform distortions of pure bend, pure splay and spirals.}
\label{fig:domain}
\end{figure}

To obtain $\mathcal{D}$ when  $m\not\in\{0,1\}$, we first construct the set $\tanset$ for $\alpha_0$ as in \eref{eq:alpha0}; it comprises 
$2|m-1|$ elements:
\begin{equation}\label{eq:theta0}
 \tanset=\Big\{ -\frac{\arctan(1/\qb)+c_0}{m-1}+\frac{n\pi}{m-1}\,|\,
 n=1,\dots,2|m-1|\Big\}.
\end{equation}
Thus, $\mathcal{D}$ is the (open) regular polyhedron with $2|m-1|$ vertices circumscribed to $\disk$; \fref{fig:domain} shows examples of either bounded and unbounded domains. 

The most interesting cases other than $m=1$ are those for $m=1/2$ and $m=3/2$, where there is only one element in $\tanset$; the corresponding nematic fields $\n$ fill a whole half-plane outside $\disk$.

\begin{figure}[h]
	\centering
	\begin{subfigure}[b]{0.4\textwidth}
		\centering
		\includegraphics[width=\textwidth]{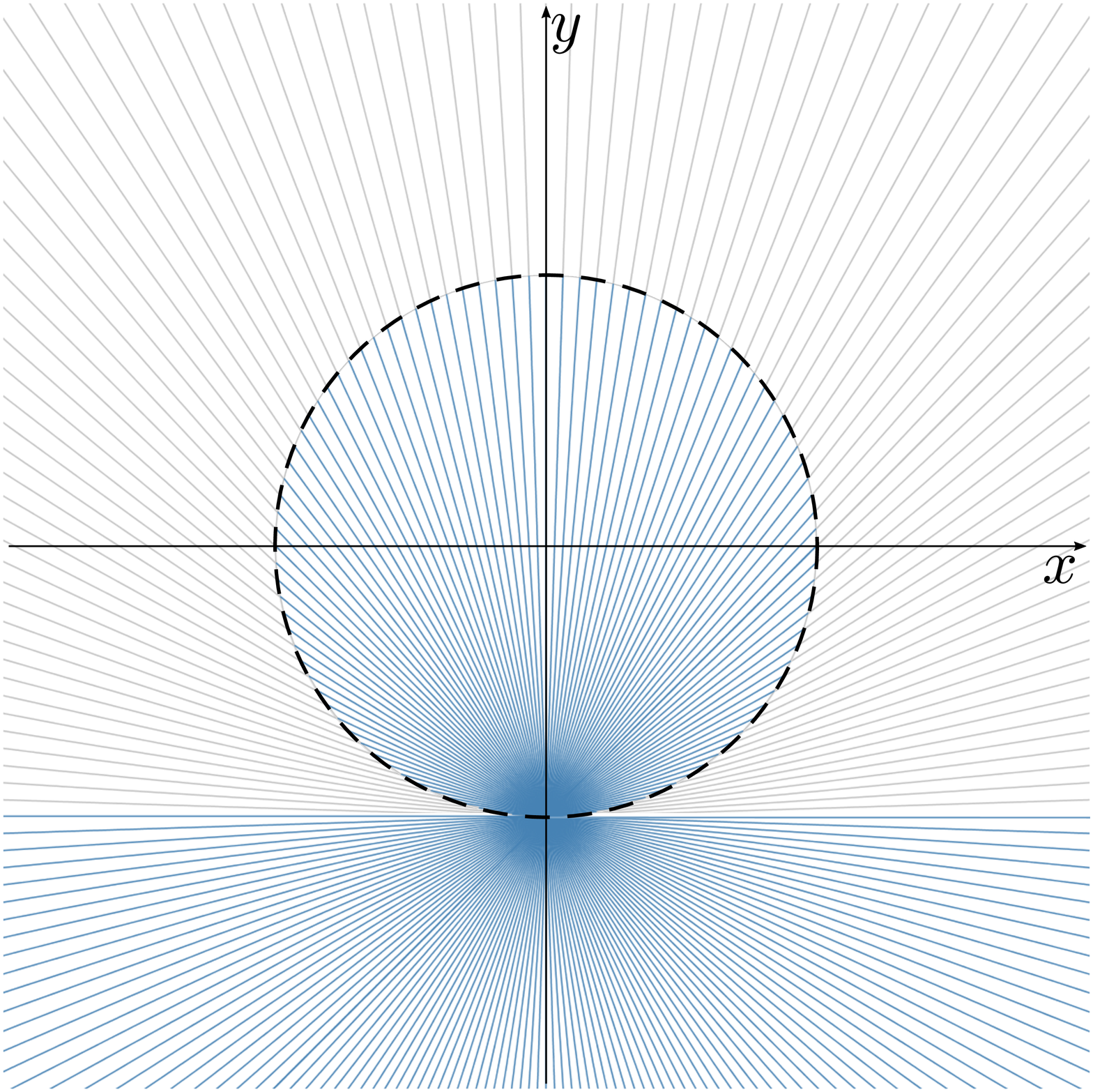}
		\caption{Characteristic lines.}
	\end{subfigure}
	$\qquad$
	\begin{subfigure}[b]{0.4\textwidth}
		\centering
		\includegraphics[width=\textwidth]{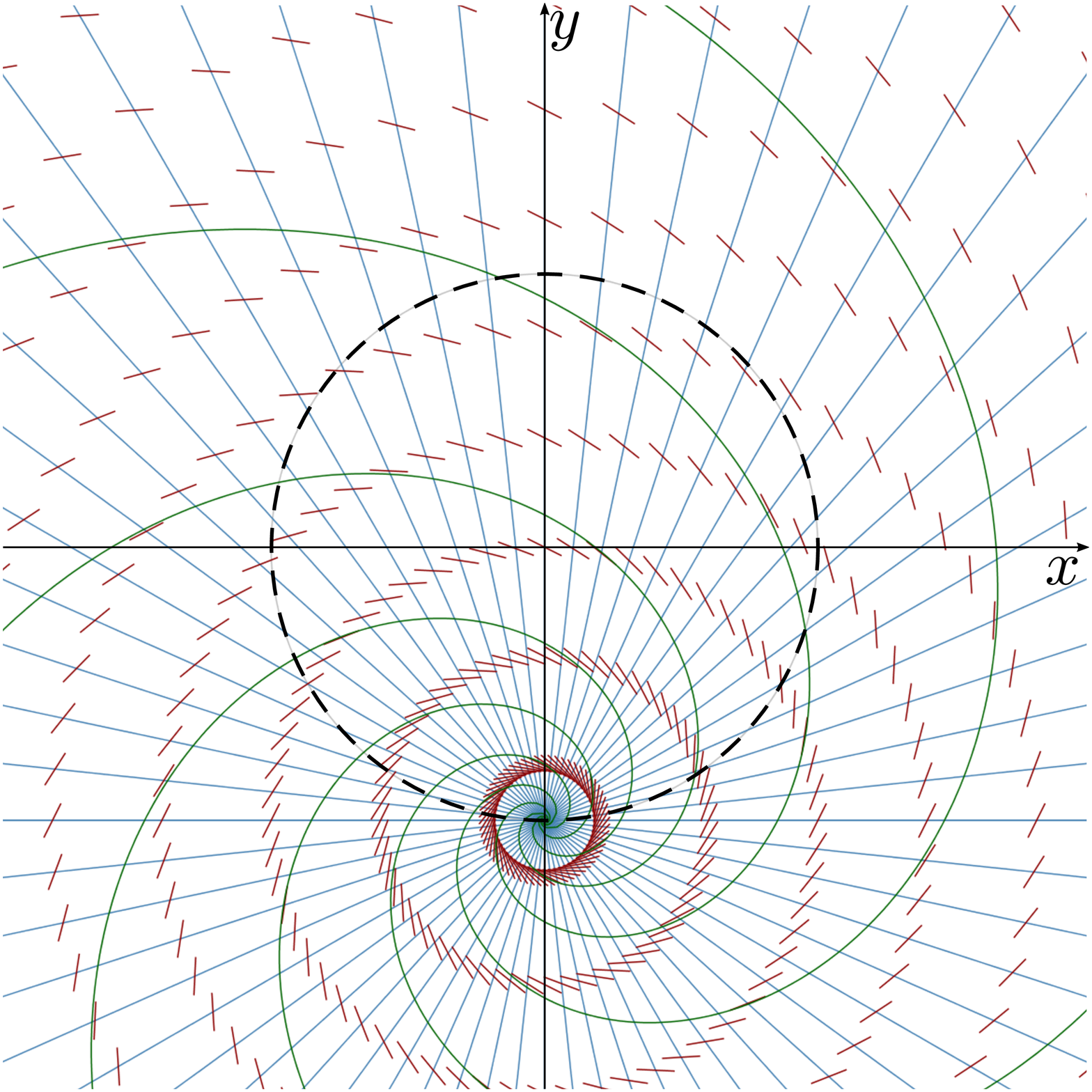}
		\caption{Field lines.}
	\end{subfigure}
	\caption{Planar quasi-uniform distortion in \eref{eq:defect} winding outside $\disk$ (black dashed line) with local angle $\alpha$ prescribed   as in \eref{eq:alpha0} with $m=1/2$. Here $\qb=2$ and $c_0:=-3\pi(m-1)/2-\arctan(1/\qb)$, so that   $\tanset=\{3\pi/2\}$. The straight lines are the characteristics where the azimuthal angle $\varphi$ propagates unchanged. In panel (a), the characteristic (gray) lines starting from $\disk$ are extended inside $\disk$ and beyond (blue lines). In panel (b),  directors are represented by short red segments, while the field lines of $\n$ are drawn in green: they are logarithmic spirals emanating from the point of $\disk$ where all the characteristic lines intersect ($\vartheta_0=3\pi/2$). The nematic field $\n$ is the same as the planar spiral in \eref{eq:spirals}, simply shifted to have its  defect at the point  $(0,-1)$ in the $(x,y)$ plane.}
	\label{fig:m05}
\end{figure}

To understand better  the kind of fields we have thus produced, we study their extendibility along characteristic lines inside $\disk$ (and possibly beyond). This construction reveals full extendibility in the case $m=1/2$, but the extended field $\n$ turns out to be nothing but a planar spiral as in \eref{eq:spirals}, only  shifted with its defect at the single point on $\disk$ where a characteristic line is tangent (see \fref{fig:m05}).

\begin{figure}[h]
	\centering
	\begin{subfigure}[b]{0.4\textwidth}
		\centering
		\includegraphics[width=\textwidth]{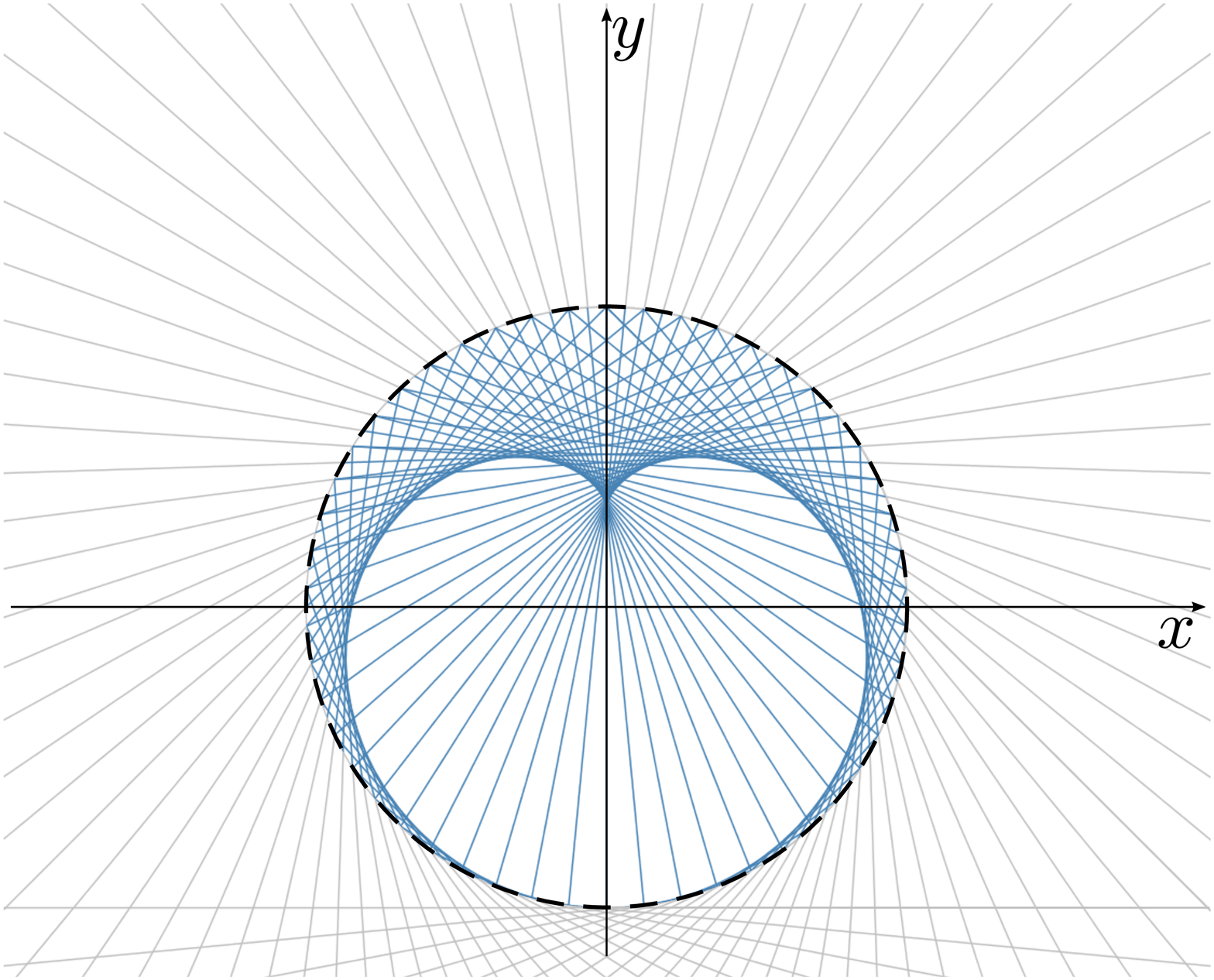}
		\caption{Characteristic lines.}
	\end{subfigure}
	$\qquad$
	\begin{subfigure}[b]{0.4\textwidth}
		\centering
		\includegraphics[width=\textwidth]{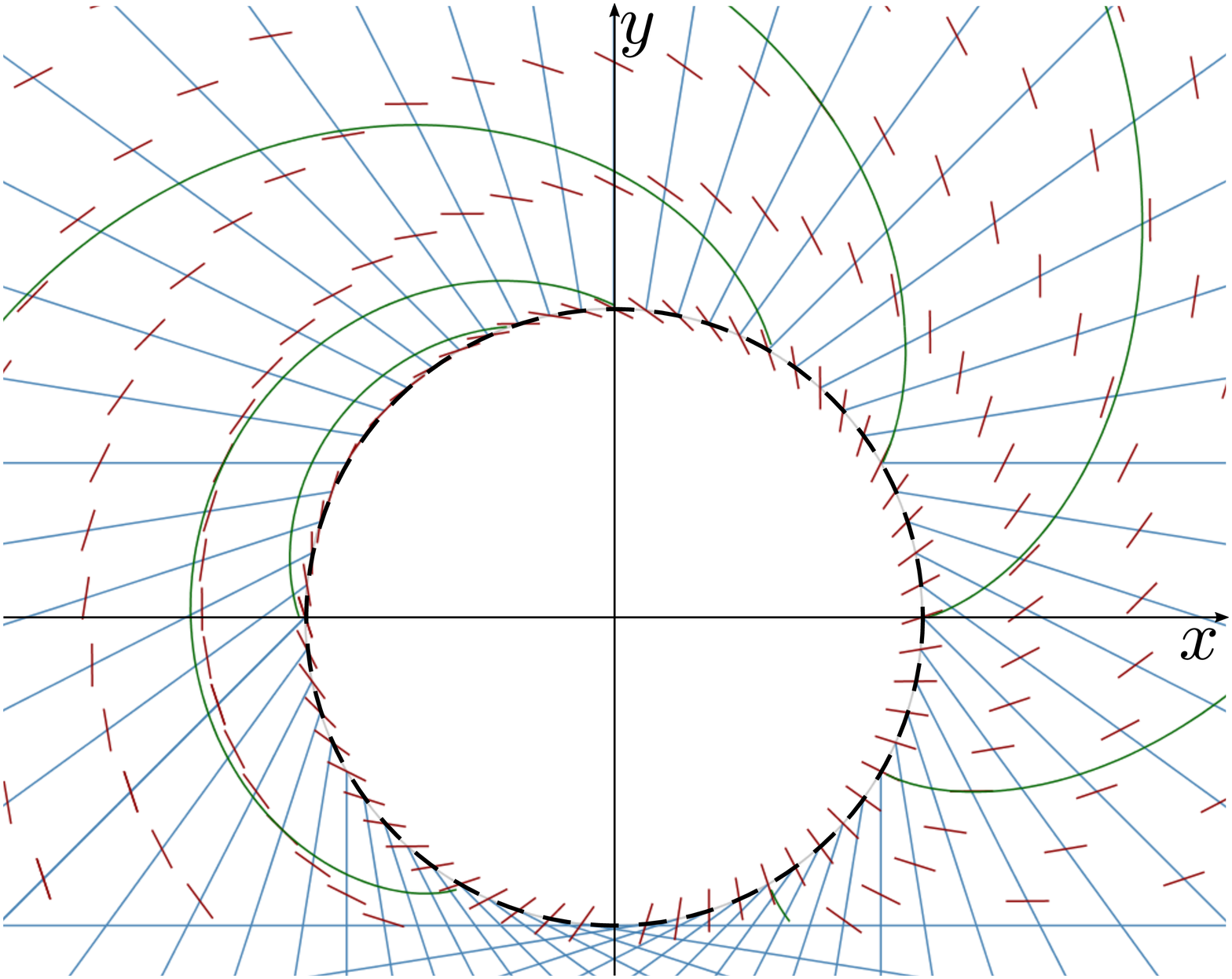}
		\caption{Field lines.}
	\end{subfigure}
	\caption{Planar quasi-uniform distortions as in \eref{eq:defect}, winding outside $\disk$ (black dashed line) with local angle $\alpha$  prescribed  as in \eref{eq:alpha0} with $m=3/2$. Here $\qb=2$ and $c_0=-3\pi(m-1)/2-\arctan(1/\qb)$, so that  $\tanset=\{3\pi/2\}$. The straight lines are the characteristics  where the azimuthal angle $\varphi$ propagates unchanged. In panel (a), the characteristic (gray) lines starting from $\disk$  are extended inside $\disk$ (blue lines), revealing a large number of mutual intersections. In panel (b),   directors are represented by short red segments, while the field lines of $\n$ are drawn in green: they represent a new family of planar quasi-uniform distortions (parameterized in $\qb$) relieving the frustration of a field with topological charge $m=3/2$ on $\disk$.}
	\label{fig:m15}
\end{figure}

\begin{figure}
\centering
 \begin{subfigure}[b]{0.4\textwidth}
 \centering
  \includegraphics[width=\textwidth]{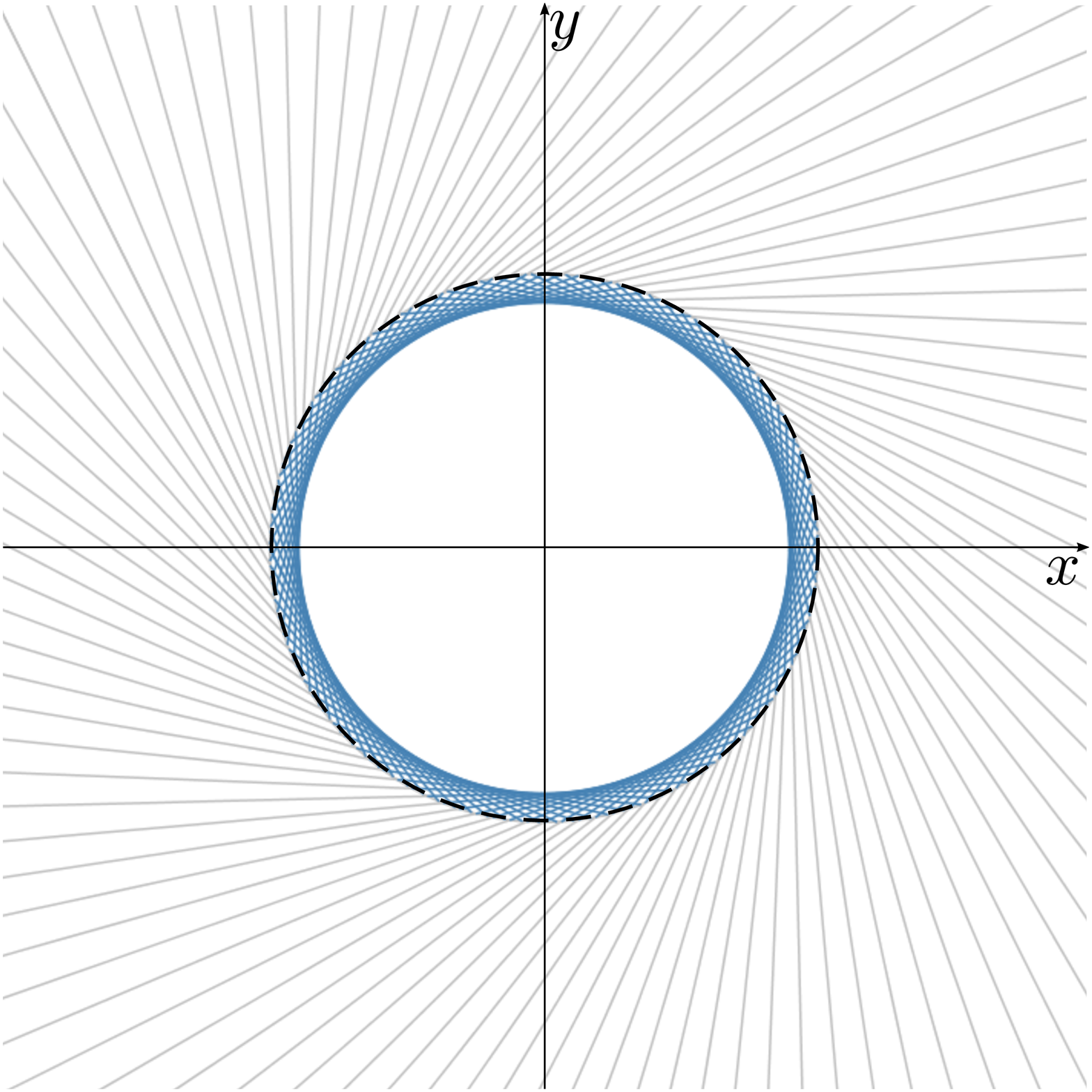}
  \caption{Characteristic lines.}
 \end{subfigure}
 $\qquad$
 \begin{subfigure}[b]{0.4\textwidth}
 \centering
  \includegraphics[width=\textwidth]{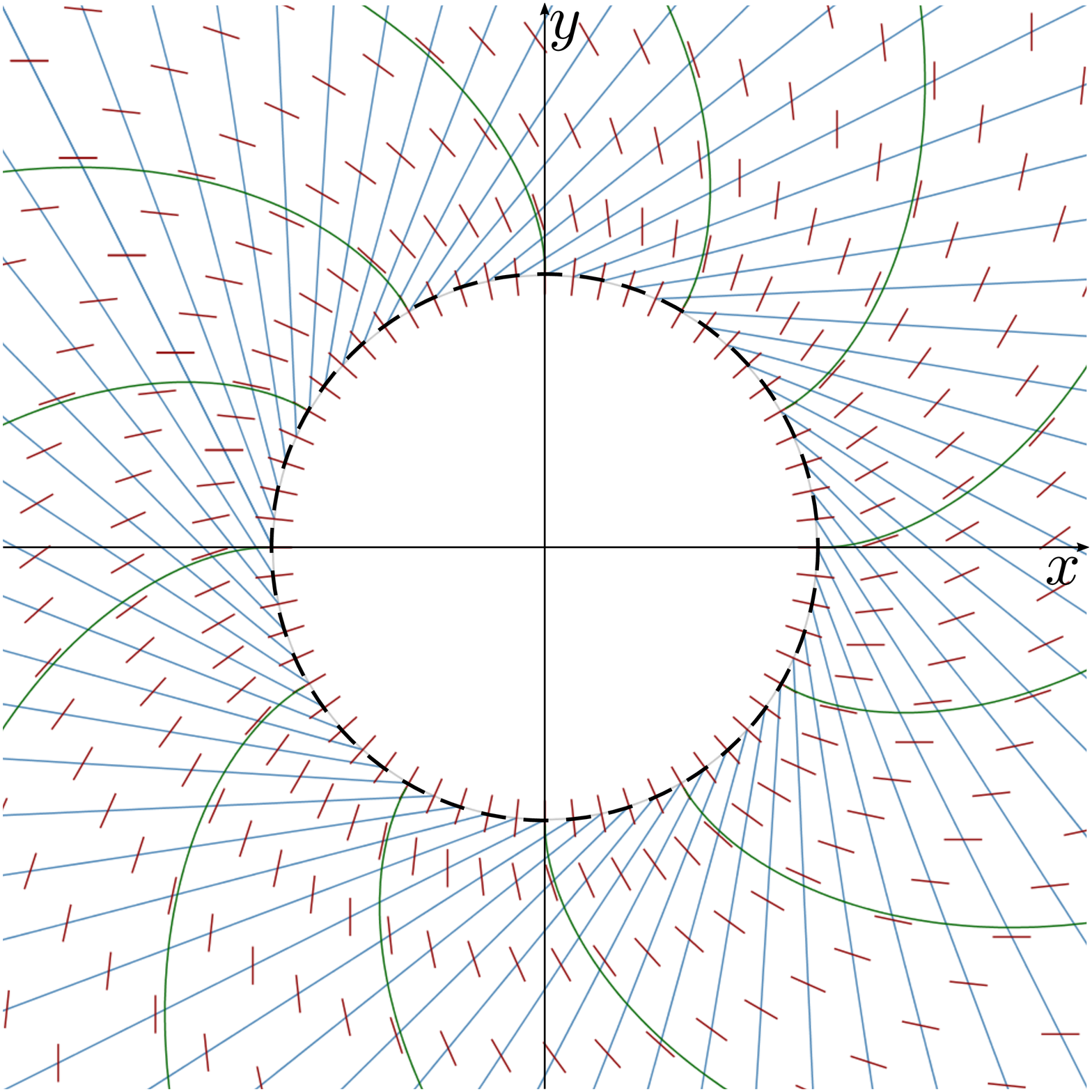}
  \caption{Field lines.}
 \end{subfigure}
 \caption{Planar quasi-uniform distortions as in \eref{eq:defect} winding outside $\disk$ (black dashed line) with local angle $\alpha$ prescribed   as in \eref{eq:alpha0} with $m=1$. Here $\qb=2$ and $c_0=0$. The same conventional representations as in \fref{fig:m15} apply here for characteristics, field lines, and directors.}
\label{fig:m10}
\end{figure}

A different scenario presents itself for $m=3/2$. Inside $\disk$ the characteristic lines have plenty of intersections (see \fref{fig:m15}),
giving rise to a genuinely new family of quasi-uniform distortions (parameterized in $\qb$) relieving in the half-plane $y>-1$ a prescribed field with  charge $m=3/2$ on $\disk$. Similarly, in the case where $m=1$, a new family of quasi-uniform distortions fill the whole plane outside $\disk$, as shown in \fref{fig:m10}.

\begin{figure}[h]
	\centering
	\begin{subfigure}[b]{0.3\textwidth}
		\centering
		\includegraphics[width=\textwidth]{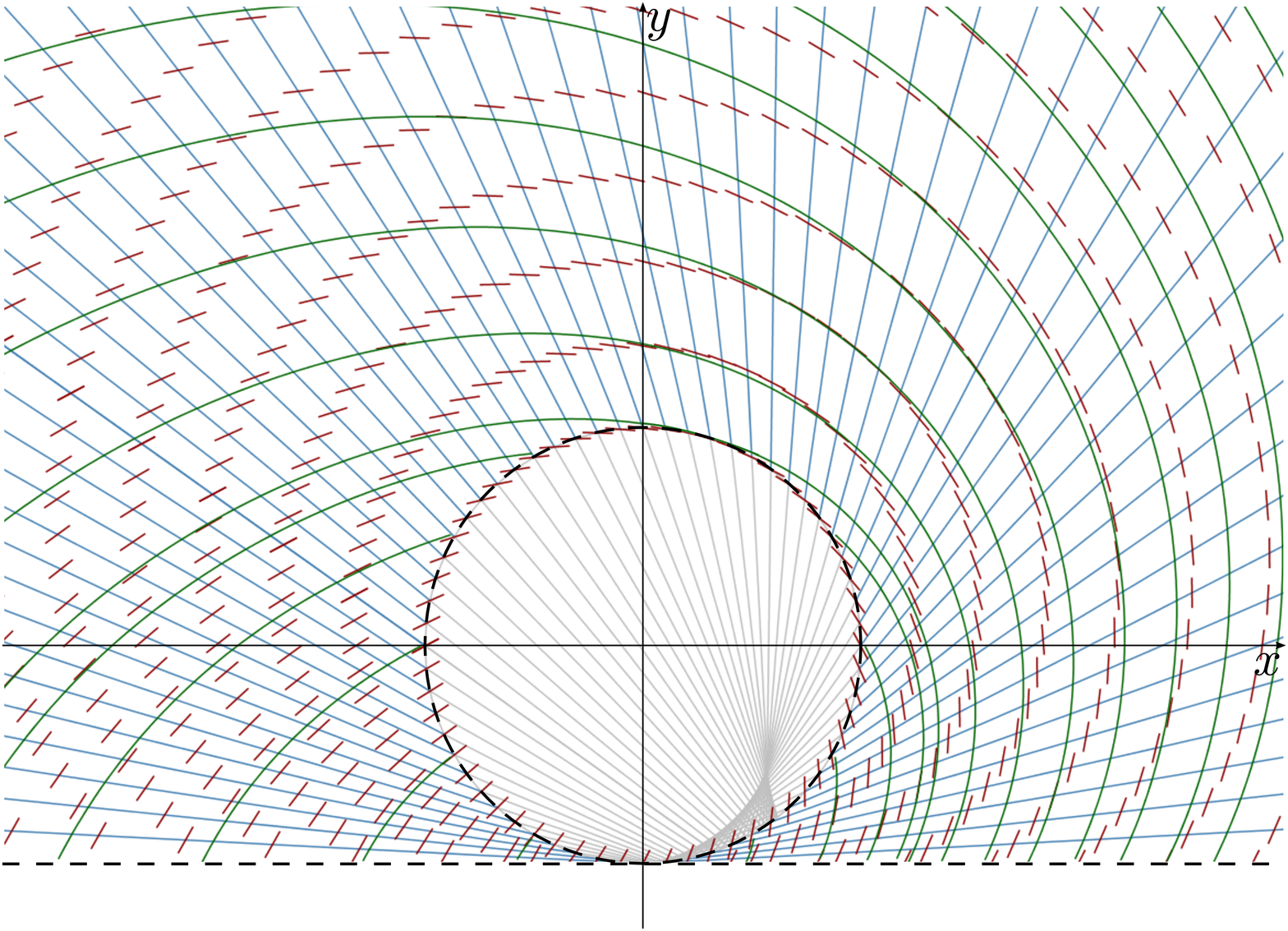}
		\caption{$m=1/2$.}
	\end{subfigure}
	\ 
	\begin{subfigure}[b]{0.3\textwidth}
		\centering
		\includegraphics[width=\textwidth]{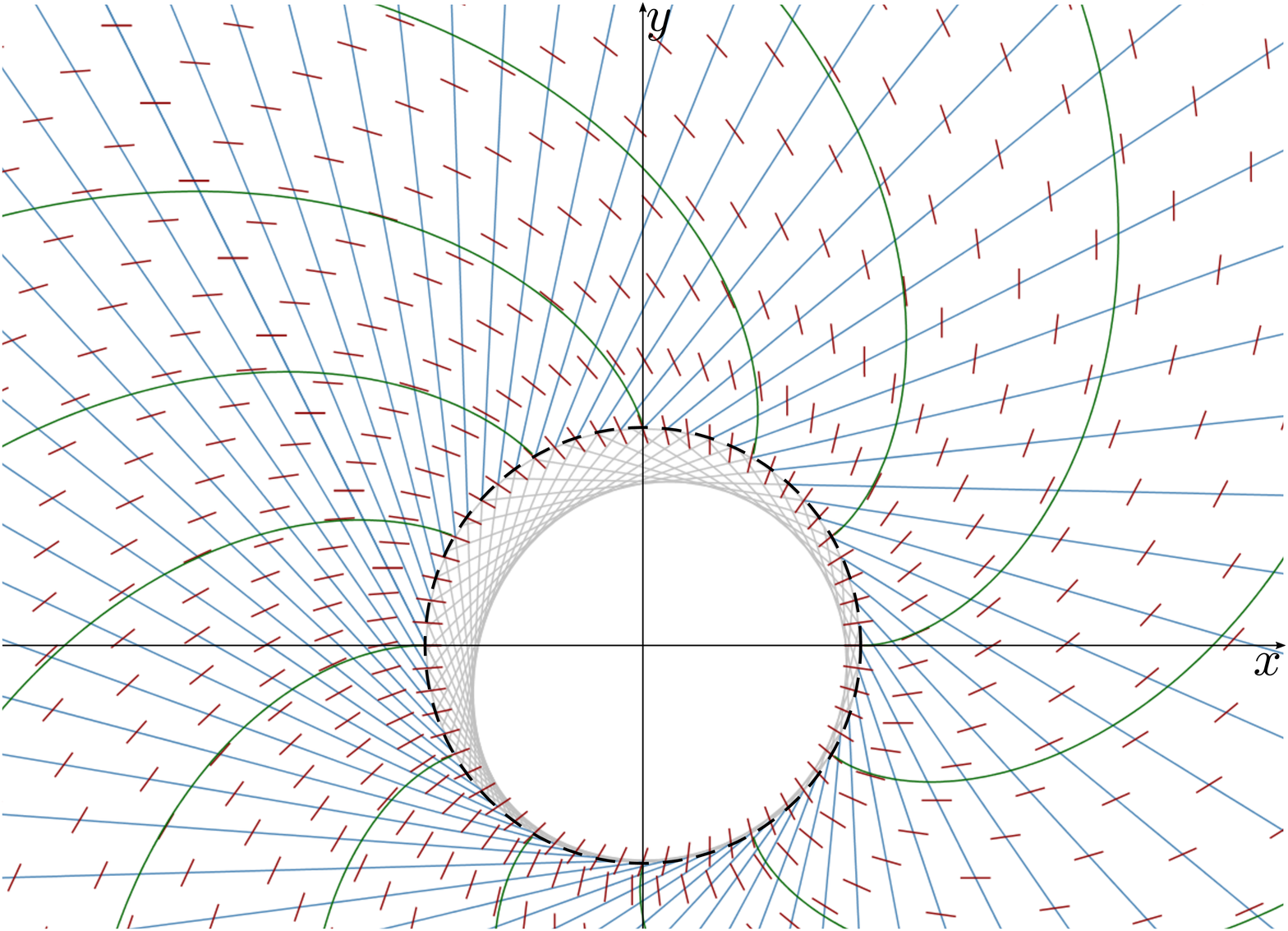}
		\caption{$m=1$.}
	\end{subfigure}
	\ 
	\begin{subfigure}[b]{0.3\textwidth}
		\centering
		\includegraphics[width=\textwidth]{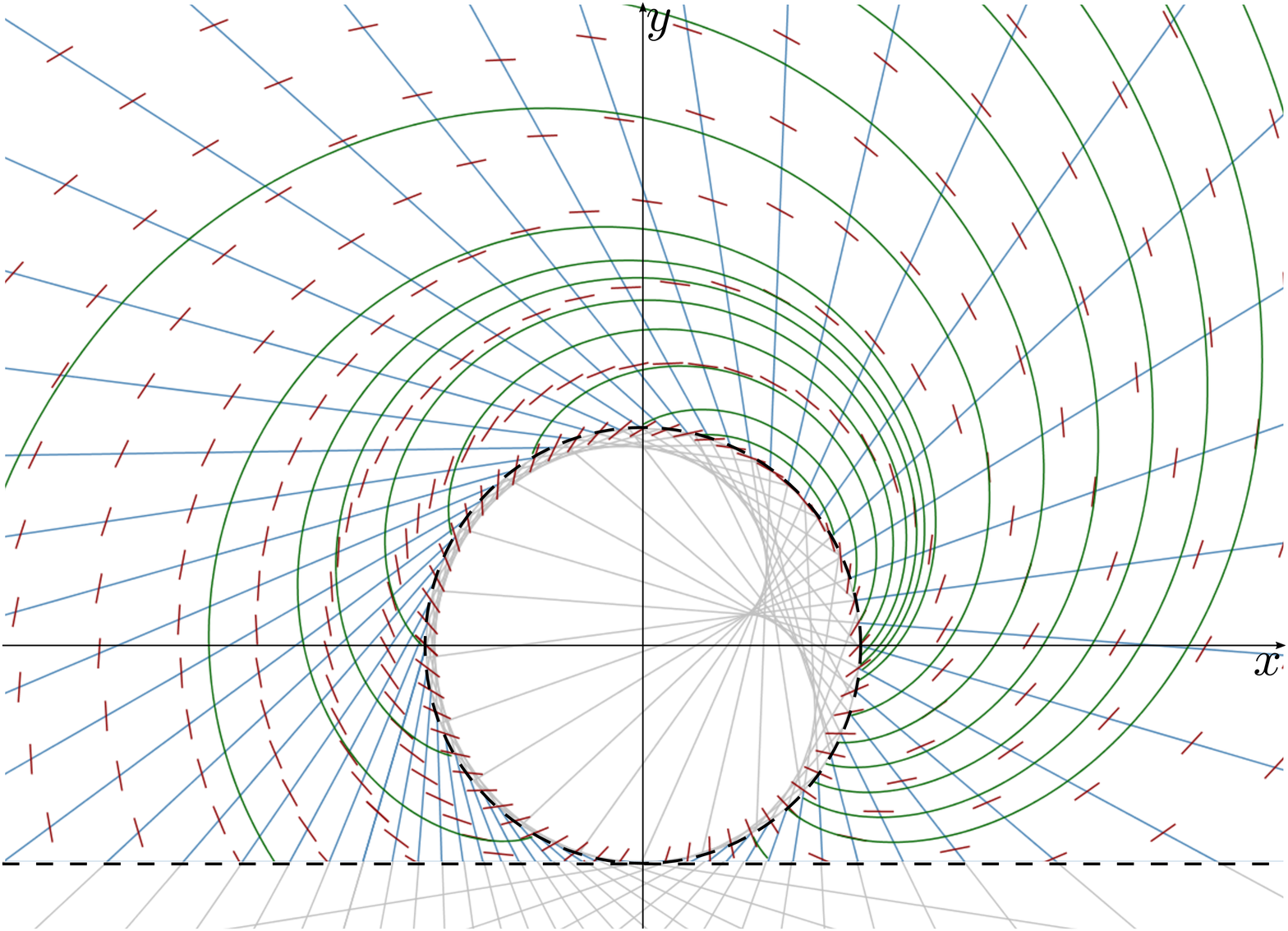}
		\caption{$m=3/2$.}
	\end{subfigure}
	\caption{Planar quasi-uniform distortions as in \eref{eq:defect} winding outside $\disk$ (black dashed line) with local angle $\alpha_0$  as in \eref{eq:alpha2} and different values of the topological charge $m$. Here $\qb=2$ and $c_0=0$ in panel (b), while in both panels  (a) and (c) $c_0$ has been tuned to let $\n$ cover the  whole half-plane $y>-1$. The straight (blue) lines are the characteristics where the azimuthal angle $\varphi$ propagates unchanged; they intersect in many different points inside  $\disk$ (gray lines), thus showing that the corresponding director  fields are different from the spiral in \eref{eq:spirals}.}
	\label{fig:defect_sin}
\end{figure}

Further examples can be produced by slightly modifying the boundary value $\alpha_0(\vartheta_0)$, while leaving its winding number unchanged, so that both the cardinality of $\tanset$ and the qualitative appearance of the   domain $\mathcal{D}$ remain unaltered. 
The idea is to perturb the frustrating boundary condition so as  to cause a \emph{crowding} of the characteristic lines, making them mutually intersect in more than one point.
An example is provided by the function
\begin{equation}\label{eq:alpha2}
	\alpha_0(\vartheta_0)= (m-1)\vartheta_0 + \frac m3\sin\vartheta_0 + c_0
	\quad\mathrm{for}\quad \vartheta_0\in[0,2\pi),
\end{equation}
as illustrated in \fref{fig:defect_sin}.

\section{Common features}\label{sec:universal}
All families of relieving fields $\n$ found in \sref{sec:defect} tend to be asymptotically spiraling as in \eref{eq:spirals}  when they can be extended for  $r\to+\infty$. Here, we present some further evidence for this behaviour  by studying the local angle $\alpha$ on different rays emanating from the origin. In all cases, $\alpha$ tends to a constant determined by $\qb$,  independently of the specific ray. A similar conclusion can be drawn for the families in \sref{sec:halfplane},  thus suggesting a possible universal trait for all frustrations that can be relieved quasi-uniformly in a half-plane.

\begin{figure}[h]
	\centering
	\begin{subfigure}[b]{0.3\textwidth}
		\centering
		\includegraphics[width=\textwidth]{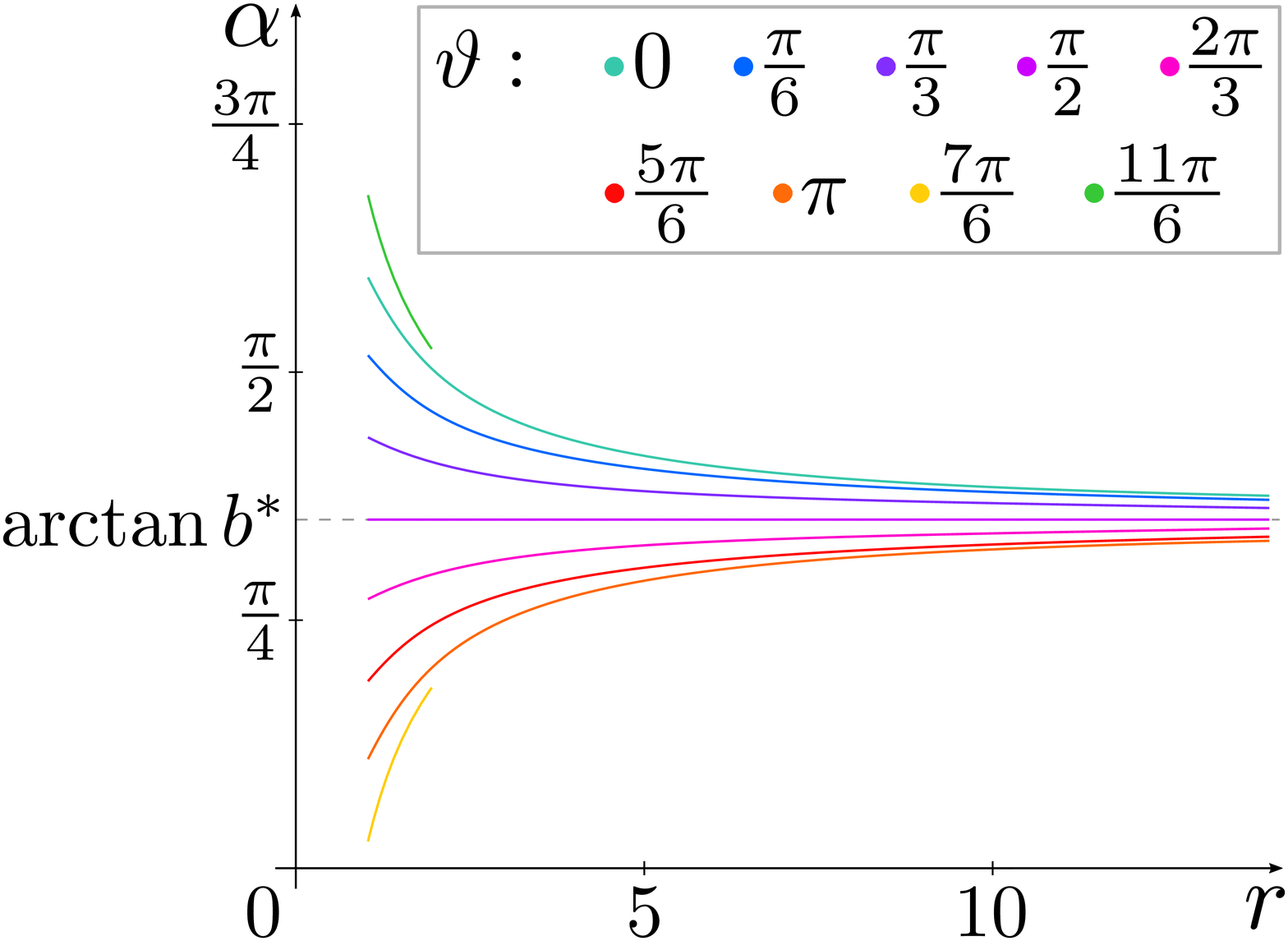}
		\caption{$m=1/2$.}
	\end{subfigure}
	\quad
	\begin{subfigure}[b]{0.3\textwidth}
		\centering
		\includegraphics[width=\textwidth]{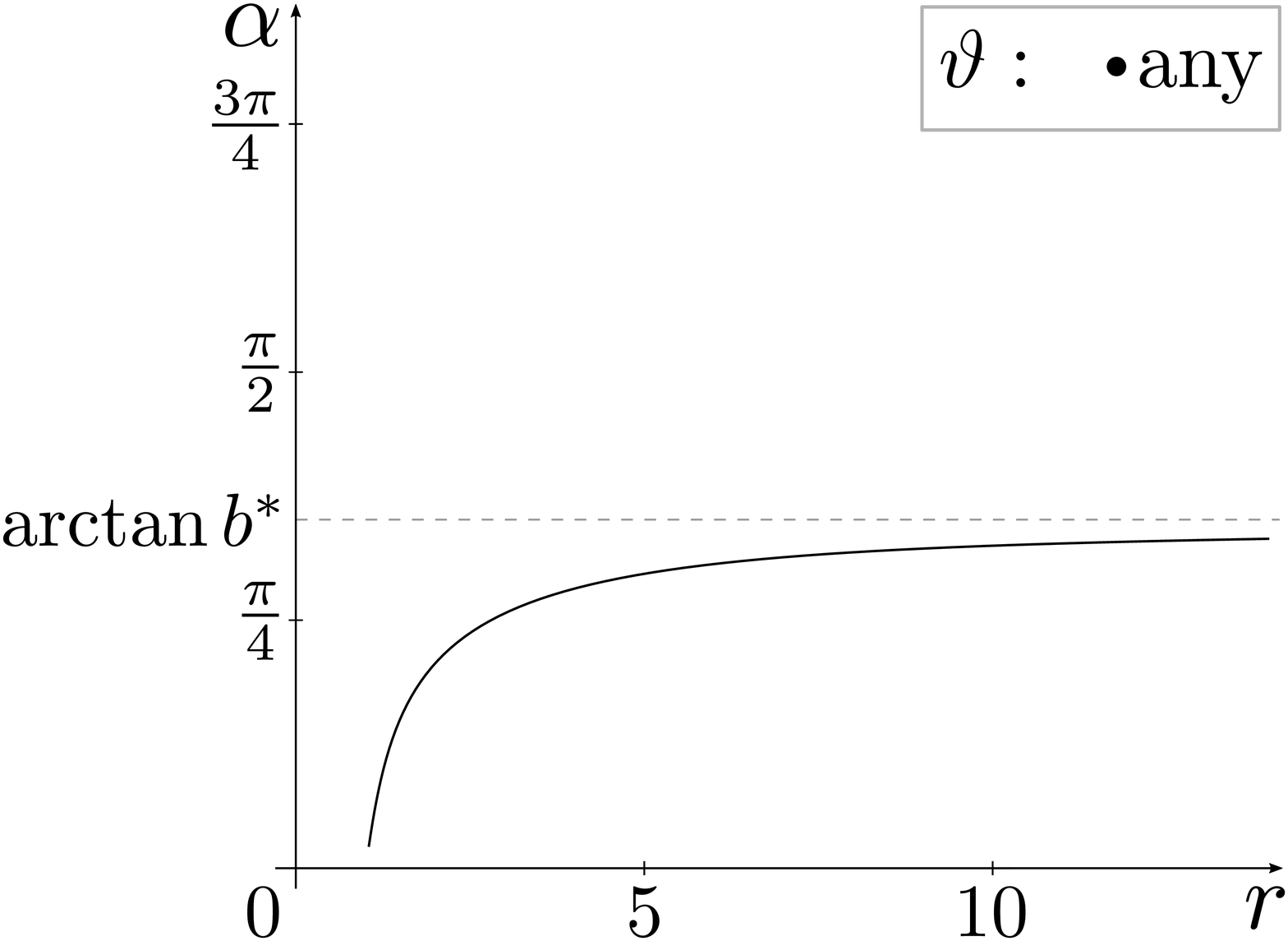}
		\caption{$m=1$.}
	\end{subfigure}
	\quad
	\begin{subfigure}[b]{0.3\textwidth}
		\centering
		\includegraphics[width=\textwidth]{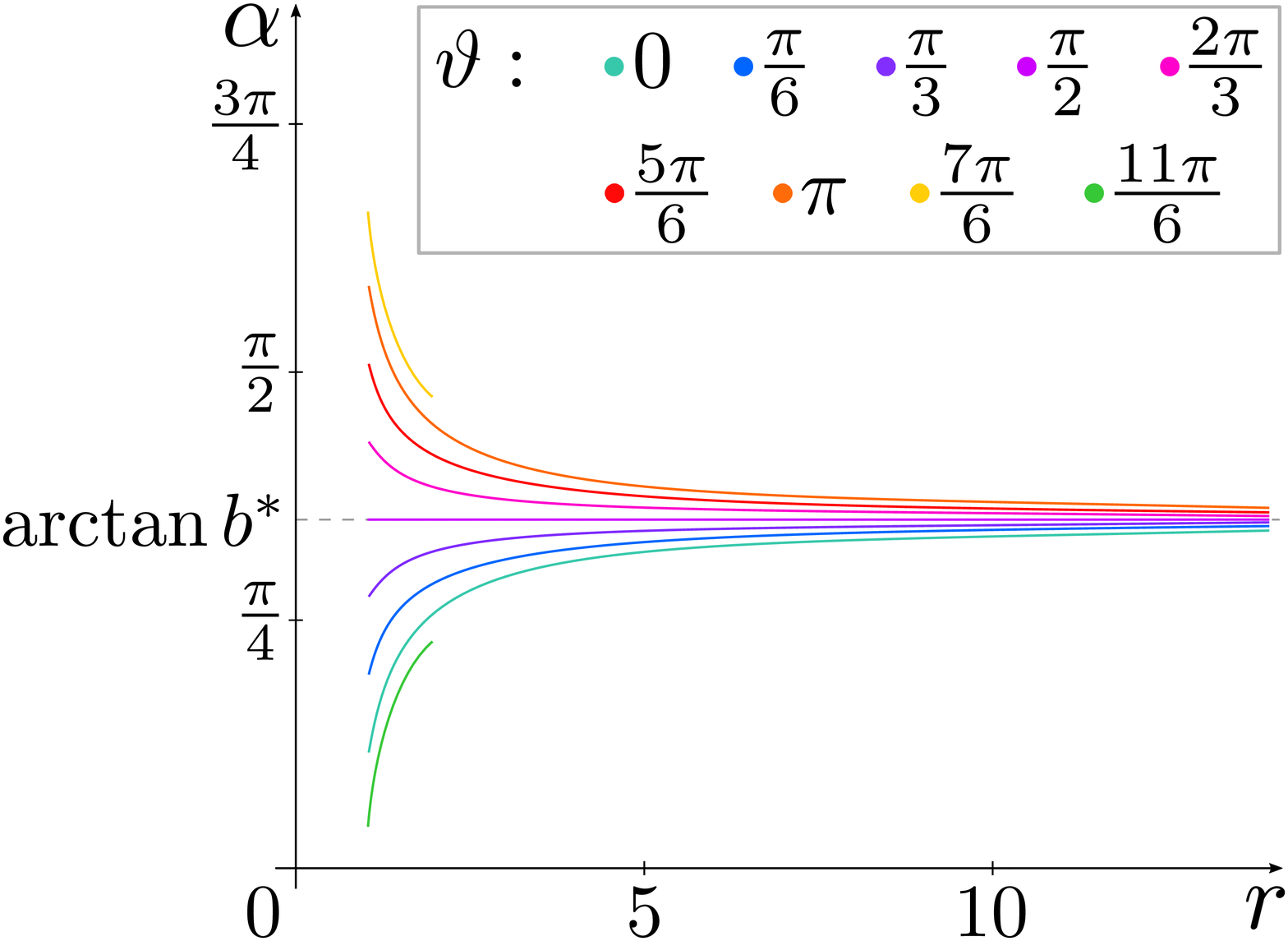}
		\caption{$m=3/2$.}
	\end{subfigure}
	\\ 
	\begin{subfigure}[b]{0.3\textwidth}
		\centering
		\includegraphics[width=\textwidth]{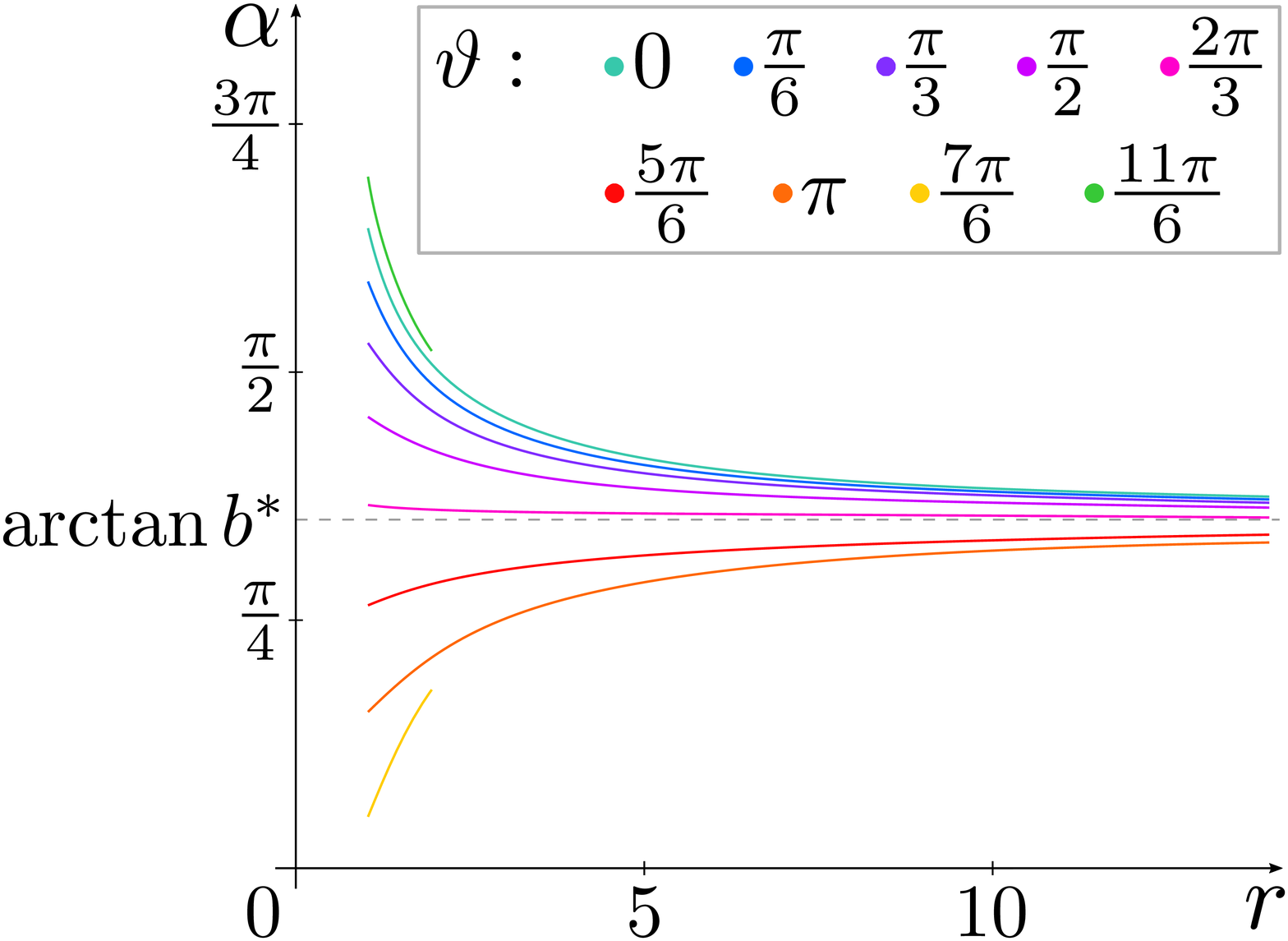}
		\caption{$m=1/2$.}
	\end{subfigure}
	\quad
	\begin{subfigure}[b]{0.3\textwidth}
		\centering
		\includegraphics[width=\textwidth]{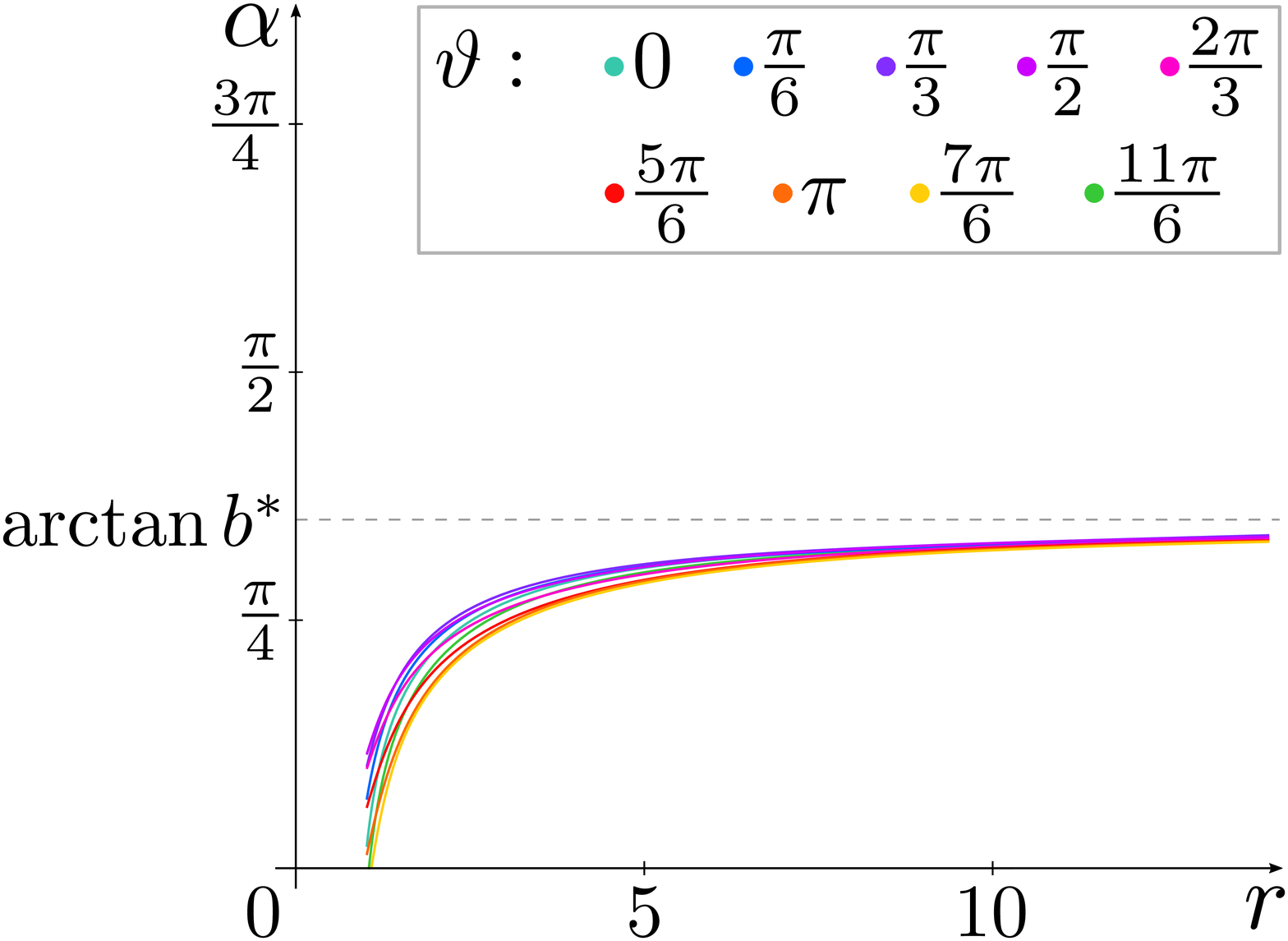}
		\caption{$m=1$.}
	\end{subfigure}
	\quad
	\begin{subfigure}[b]{0.3\textwidth}
		\centering
		\includegraphics[width=\textwidth]{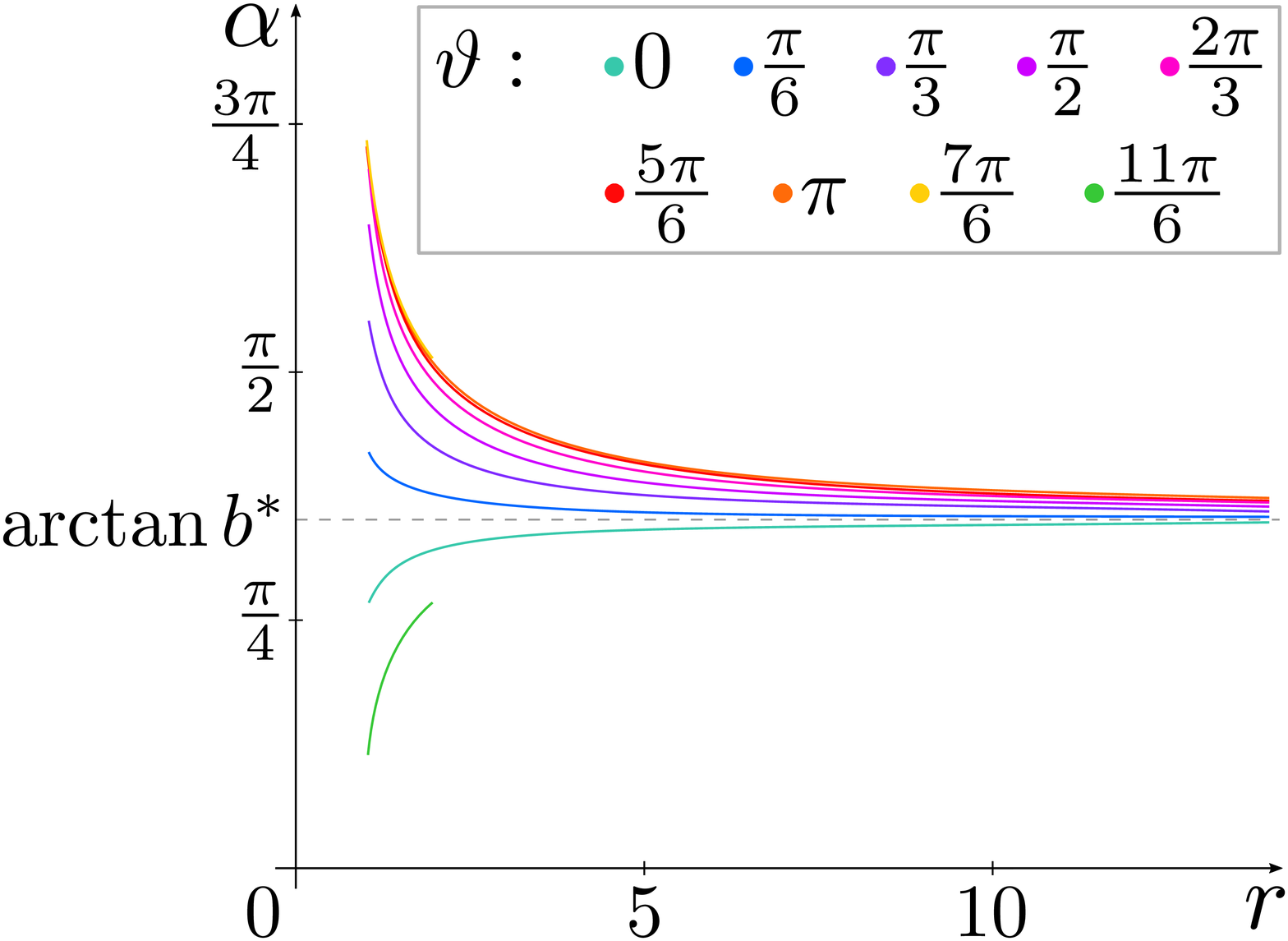}
		\caption{$m=3/2$.}
	\end{subfigure}
	\caption{Behaviour of  local angle $\alpha$  for the planar  quasi-uniform distortions in  \eref{eq:defect} winding outside $\disk$ on different rays emanating from the origin  at different values of $\vartheta$. As the distance $r$ from the origin diverges to $+\infty$, $\alpha$ tends to $\arctan\qb$ (here $\qb=2$). In panels (a), (b), and (c), the trace $\alpha_0$ of $\alpha$ on $\disk$ is given by \eref{eq:alpha0}, while in panels (d), (e), and (f), $\alpha_0$ is prescribed as in \eref{eq:alpha2}. When $m=1/2$ or $m=3/2$, the (yellow and green) lines corresponding to $\vartheta=7\pi/6$ and $\vartheta=11\pi/6$ (respectively) stop where the rays intersect the horizontal line $y=-1$, below which field $\n$ is not defined. In panel (b),  there is only one (black) line representing $\alpha$, as this is the same on all rays.}
	\label{fig:alpha_limit}
\end{figure}

\Fref{fig:alpha_limit}
illustrates the behaviour of the local angle $\alpha$: on each ray emanating from the origin with a fixed angle $\vartheta$ with respect to $\vx$, $\alpha$ tends to $\arctan\qb$, which is precisely the value constantly taken on the planar spiral \eref{eq:spirals}. 

\begin{figure}[h]
	\centering
	\begin{subfigure}[b]{0.45\textwidth}
		\centering
		\includegraphics[width=\textwidth]{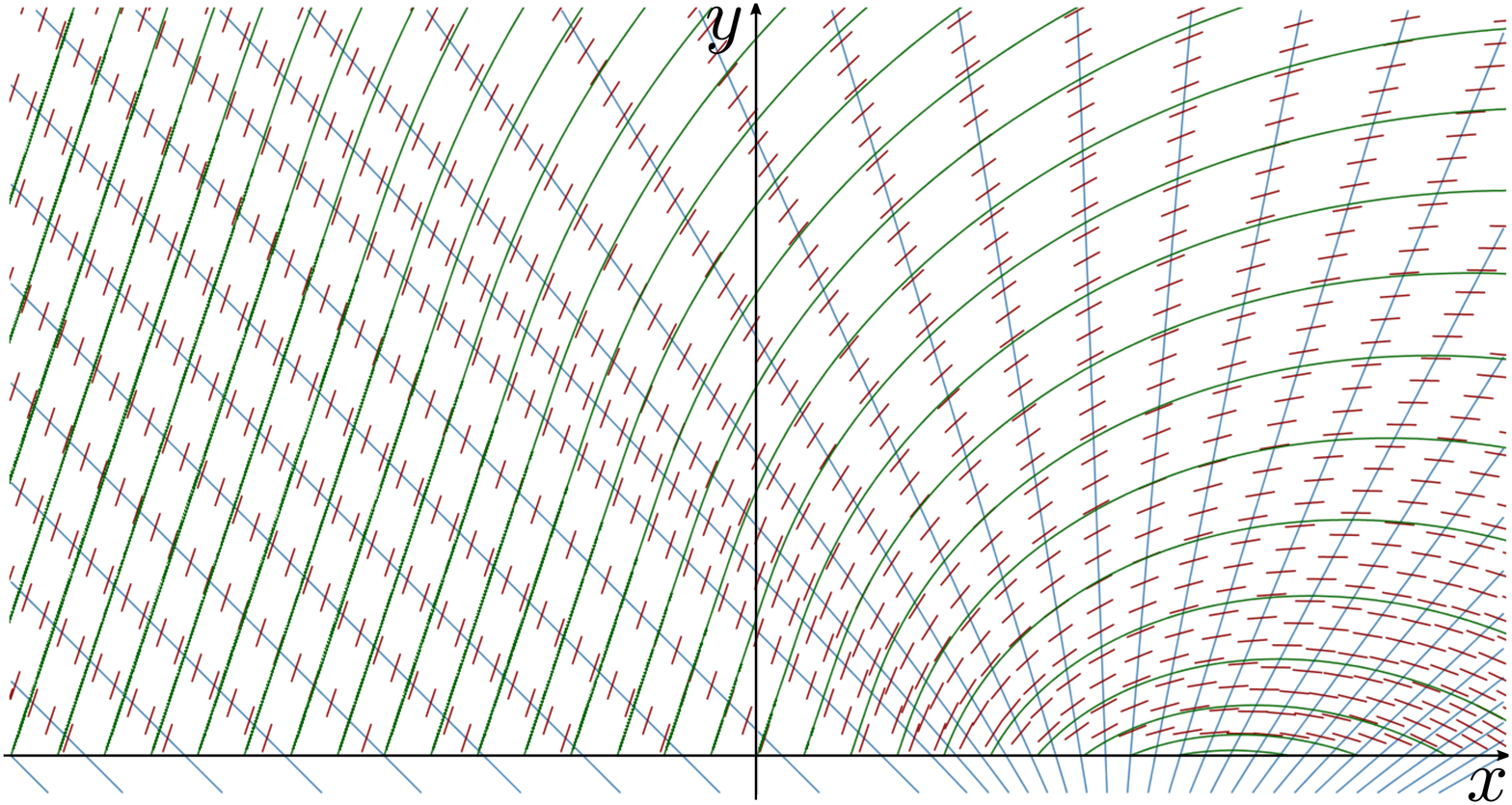}
		\caption{Field lines and characteristics.}
	\end{subfigure}
	$\qquad$
	\begin{subfigure}[b]{0.45\textwidth}
		\centering
		\includegraphics[width=\textwidth]{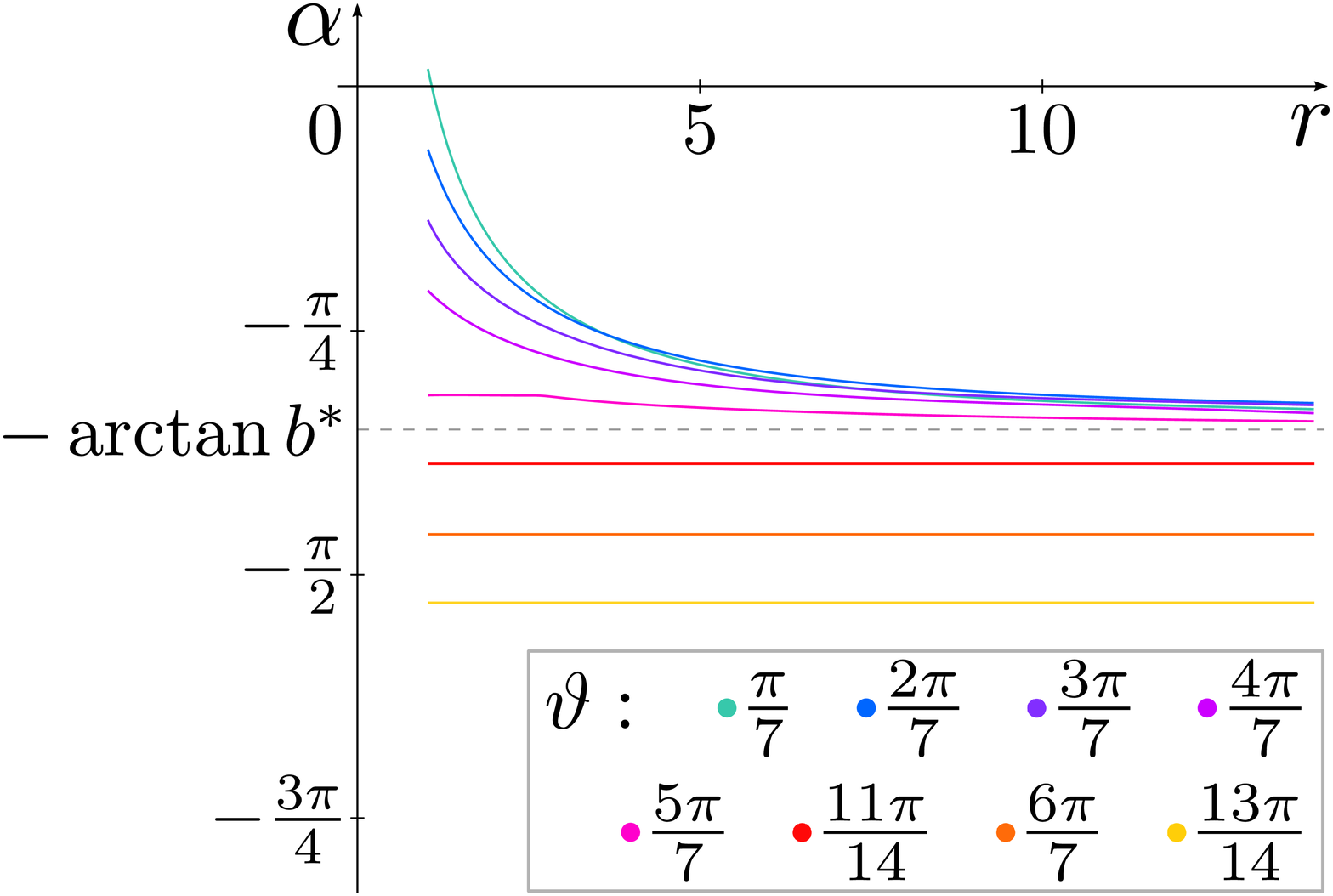}
		\caption{Local angle $\alpha$ along rays.}
	\end{subfigure}
	\caption{Representation of a planar quasi-uniform distortion $\n$ obtained as a solution of \eref{eq:condition1b_half} with frustration  prescribed on the $x$-axis as in \eref{eq:phi0_half2}. The  characteristic lines (where $\n$ propagates unaltered) are parallel to one another for $x_0<0$; their inclination on the line $y=0$ (measured by $|\sloh|$) is first increasing and the decreasing, after crossing the line $x=0$.  As a result, the asymptotic value of the  local angle $\alpha$ depends on the selected ray. In the limit as $r\to+\infty$, $\n$ reveals a hybrid nature: it is either a constant field or a planar spiral. In this  example, $\qb=2$.}
	\label{fig:alpha_halfplane}
\end{figure}

A similar property is exhibited by  the fields constructed in \sref{sec:halfplane}: when the slope $\sloh$ in  \eref{eq:slope_half} is \emph{strictly} monotonic and one characteristic is parallel to the $y$-axis: $\alpha$ tends to $-\arctan\qb$, as in the examples in \fref{fig:halfplane}. However, whenever the slope is everywhere finite or somewhere constant, the asymptotic limit of $\alpha$ on rays depends on the specific ray.  An example of this situation is provided by \fref{fig:alpha_halfplane},
where the field $\n$ is generated by the following frustration prescribed on the $x$-axis,
\begin{equation}\label{eq:phi0_half2}
	\varphi_0(x_0)=\cases{
		-\frac{3\pi}4\exp(-1/x_0^2) - \arctan\qb + \frac{3\pi}4&for $ x_0\geq0$,\\
		-\arctan\qb + \frac{3\pi}4&otherwise.}
\end{equation}

The quasi-uniform distortions constructed in this paper have another important feature in common. Quasi-uniformity in the plane has been proved to require   $\n$ at point $(x,y)$ to make an angle with the characteristic line passing through $(x,y)$ that remains  constant along the characteristic. Such an inclination represents indeed the constant ratio between splay and bend along each characteristic. For $\n$ as in \eref{eq:splay-bend}, denoting by $\bm{e}_0$ the unit vector along the   characteristic through $(x_0,0)$ with slope $\sloh$ given by \eref{eq:slope_half},  we obtain $\bm{e}_0\cdot\n=-(1+{\qb}^2)^{-\frac12}$, while, for $\n$  as in \eref{eq:defect} and  $\bm{e}_0$ tangent to the characteristic with slope $\slod$ given by \eref{eq:defect_slope}, we obtain  $\bm{e}_0\cdot\n=(1+{\qb}^2)^{-\frac12}$. Thus, the  inclination of the director field with respect to each characteristic line depends on $\qb$ only.

\nalert{
\section{One-dimensional uniformity}\label{sec:1d_uniformity}
In this paper, the concept of frustration has been employed with quite an extensive meaning. Since uniformity for a director field $\n$ in two space dimensions is equivalent to have $\n\equiv\n_0$, we regarded any director field prescribed on a curve in the plane so as to be incompatible with a constant extension to the whole plane as a source of frustration. One may wonder whether there is a sensible, restricted notion of frustration, which would further illuminate the role of quasi-uniform distortions as means of frustration relief.

Suppose that $\curve$ is a regular curve in the plane and that a director field $\n$ is prescribed on it as a function of the arch-length parameter $s$ for $\curve$. Letting $\bm{e}_z$ be a unit vector orthogonal to the plane that contains $\curve$, we define $\n_\perp:=\bm{e}_z\times\n$ as a unit vector field everywhere orthogonal to $\n$, so that $(\n,\n_\perp)$ plays the role of a distortion frame in this restricted setting.

Elementary computations show that $\n'=\gamma\n_\perp$, where a prime denotes differentiation with respect to $s$ and $\gamma=\gamma(s)$. The intrinsic definition of one-dimensional uniformity that we introduce requires $\gamma\equiv\gamma_0$.

Let now $\bm{t}=\bm{t}(s)$ be the unit tangent to $\curve$ and set $\bm{t}_\perp:=\bm{e}_z\times\bm{t}$. By representing $\n$ as
\begin{equation}
	\label{eq:n(s)_representation}
	\n(s)=\cos\varphi_0(s)\bm{t}+\sin\varphi_0(s)\bm{t}_\perp,
\end{equation}
we easily obtain that
\begin{equation}
	\label{eq:n_prime}
	\n'=(\varphi_0'+\tau)\n_\perp,
\end{equation}
where $\tau=\tau(s)$ is such that $\bm{t}'=\tau\bm{t}_\perp$. Thus, our notion of one-dimensional uniformity reduces to requiring that 
\begin{equation}
	\label{eq:one-dimensional_uniformity}
	\varphi_0'+\tau\equiv\gamma_0.
\end{equation}

As expected, this condition puts restrictions on the cases studied above. When $\curve$ is the straight line $y=0$, as in \sref{sec:halfplane}, $\bm{t}\equiv\bm{e}_x$ and $\tau\equiv0$, so that \eref{eq:one-dimensional_uniformity} reduces to $\varphi_0'\equiv\gamma_0$. When $\curve$ is $\disk$, as in \sref{sec:defect}, $\bm{t}=\bm{e}_\vartheta$ and $\tau\equiv1$, so that \eref{eq:one-dimensional_uniformity} reduces to $\varphi_0'=\alpha_0'+1\equiv\gamma_0$, showing that whenever \eref{eq:phi} applies the frustration imposed on $\disk$ in \sref{sec:defect} can also be interpreted within the restricted meaning introduced here.

For completeness, we now illustrate  an example of one-dimensional uniformity prescribed on a line segment and the ways it has to relax quasi-uniformly in the plane. We take
\begin{equation}
	\label{eq:linear_phi_0}
	\varphi_0=-\frac{\pi}{4}x_0\quad\mathrm{for}\quad x_0\in[0,1],
\end{equation}
so that the admissible values for $b^\ast$ determined in \sref{sec:halfplane} obey $
b^\ast\geq1$.
Figure~\ref{fig:strip} depicts the outcomes of the analysis performed following the construction described in \sref{sec:halfplane}.
\begin{figure}[h]
	\centering
	\begin{subfigure}[b]{0.45\textwidth}
		\centering
		\includegraphics[width=\textwidth]{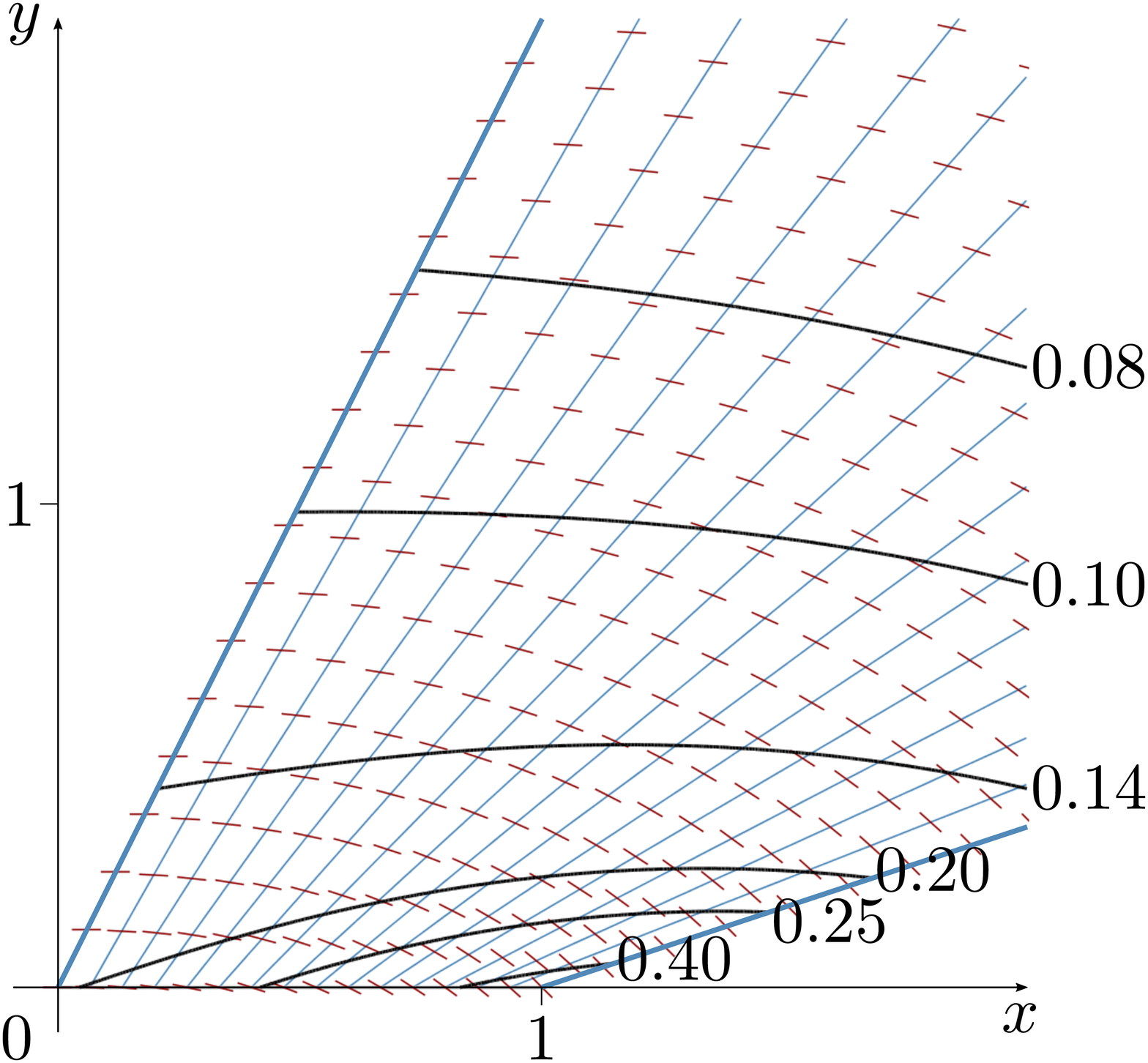}
		\caption{$b^\ast=2$.}
	\end{subfigure}
	$\quad$
	\begin{subfigure}[b]{0.45\textwidth}
		\centering
		\includegraphics[width=\textwidth]{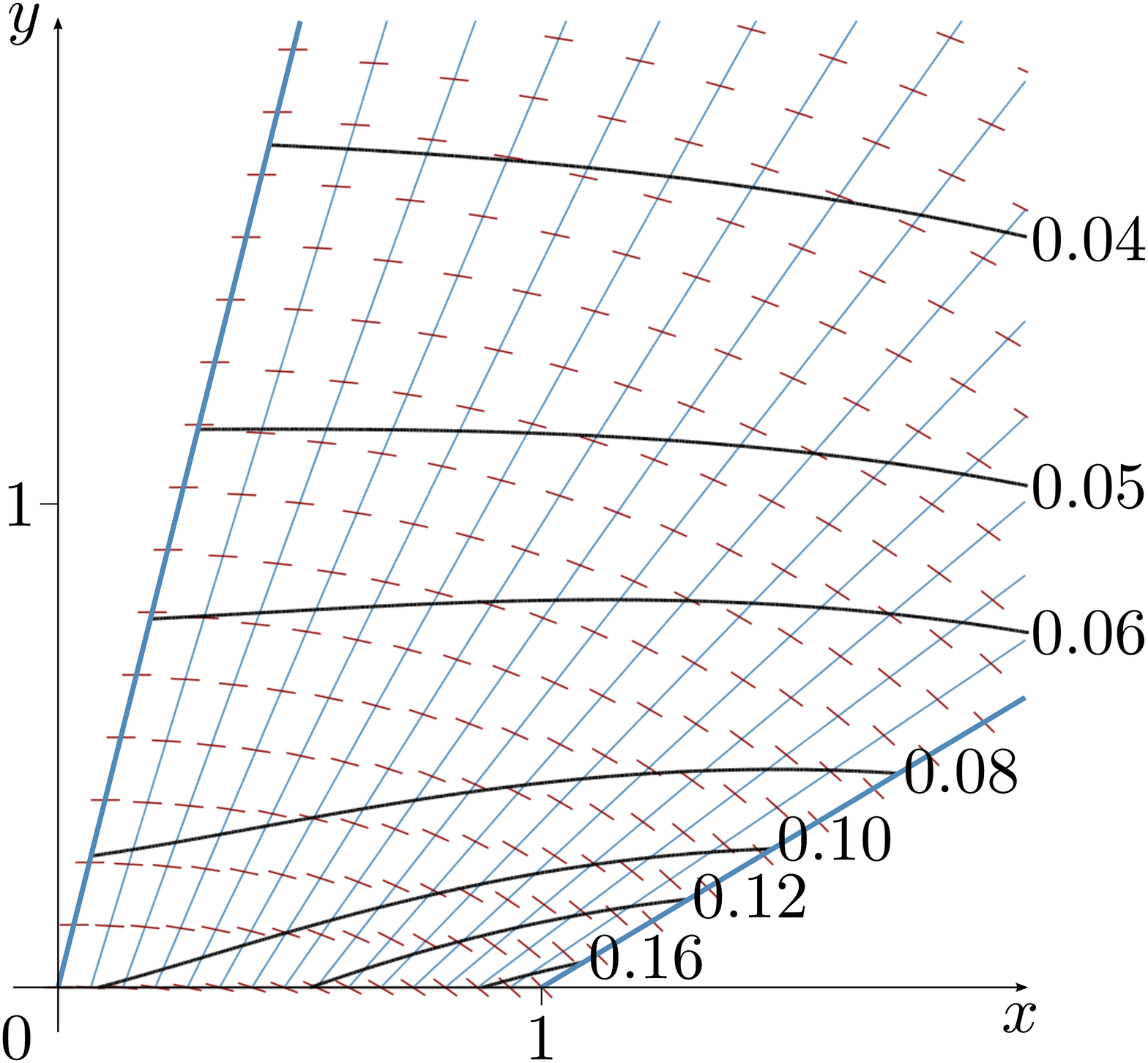}
		\caption{$b^\ast=4$.}
	\end{subfigure}
	\caption{Two instances of one-dimensional uniformity on a line segment relieved in the plane. Characteristics and directors are depicted  as customary in the preceding figures; levels sets of the function $f$ are black curves with the value they carry marked next to them. As expected from \eref{eq:f_y=0}, larger values of $b^\ast$ correspond to smaller values of $f$.}
	\label{fig:strip}
\end{figure}
Alongside with characteristics and directors, here with the aid of \eref{eq:f_y=0} we also show the level sets of the function $f$. The frustrated field relaxes in an unbounded quadrilateral based on the interval $[0,1]$. The level sets of $f$, which represent restricted loci of two-dimensional uniformity, run from one bounding side to the other and straighten while getting away from the frustrated base.  
}

\section{Conclusions}\label{sec:conclusions}
Finding systematic ways for relieving frustration in an ordered medium endowed with a director field as order descriptor is a topic generally tackled in energetic terms. Here, we took instead a completely geometric approach to this problem, identifying quasi-uniformity as a viable notion of frustration relief. Our study was confined to two space dimensions in flat geometry.

Having first realized that, apart from \emph{planar spirals}, no non-trivial quasi-uniform distortion exists in the whole plane, a conclusion that extends the negative results already known for planar uniform distortions \cite{virga:uniform,niv:geometric}, we endeavoured to find planar quasi-uniform distortions in half a plane.

Two general settings were considered: in one, the director field was prescribed along a straight line, in the other it was prescribed on the unit circle. In both cases, we provided a method to construct quasi-uniform distortions obeying the prescribed frustration. We also identified the frustrations that can be relieved in a half-plane and those that cannot, thus turning our method into a selection criterion for relievable frustrations. To show how such a criterion works, we considered the case where a classical Frank's disclination field is prescribed on the unit circle; we proved that such frustrations relax quasi-uniformly  in a half-plane only if the topological charge is either $m=1$ or $m=3/2$ (the case $m=1/2$ being trivially a planar spiral).

Quasi-uniform distortions in a half-plane are plenty, but a universal feature emerged which is common to all the quasi-uniform distortions that we constructed in this paper: away from the generating frustration, they all tend to become a planar spiral (being exactly so only in very selected cases).

A number of issues remained unresolved; we mention just two. First, we have not classified all possible quasi-uniform distortions in half a plane. Second, we have not attempted to characterize quasi-uniform distortions on curved surfaces. Accomplishing  these tasks would possibly extend to the realm of quasi-uniformity in two space dimensions results already achieved for uniformity in dimensions three   \cite{virga:uniform,dasilva:moving,pollard:intrinsic} and two \cite{niv:geometric}, respectively.


\ack{This research was performed under the auspices of the Italian \emph{INdAM (GNFM)}. A.P. wishes to acknowledge financial support from the Italian \emph{MIUR} through  European Programmes  \emph{REACT EU 2014-2020} and \emph{PON 2014-2020 CCI2014IT16M2OP005}. \nalert{Both authors are grateful to the anonymous Reviewers of this paper for their stimulating insights and suggestions.}}

\appendix
\section{The Oseen-Frank energy}\label{app:oseen_frank}
For ordinary nematic liquid crystals, the undistorted ground state coincides with a constant director field $\n$: the molecular orientation is the same at each point occupied by the material, with no distortion.

The elastic free energy penalizes any distortion that would make $\gradn\neq\mathbf{0}$. In the classical elastic theory formulated by Oseen \cite{oseen:theory} and Frank \cite{frank:theory}, the cost for the deviation from the ground state is measured by the free energy density 
\begin{equation}\label{eq:frank_density}
	\eqalign{
		\Fdensity(\n,\gradn) &:= \frac12K_{11}(\dv\n)^2 + \frac12K_{22}(\n\cdot\curl\n)^2 \\&\qquad+ \frac12K_{33}\big|\n\times\curl\n\big|^2 +
		 K_{24}\Big[\tr(\gradn)^2-(\dv\n)^2\Big],}
\end{equation}
where each term corresponds to a specific elastic distortion and the nonnegative coefficients $K_{11}$, $K_{22}$, $K_{33}$ and $K_{24}$ are Frank's elastic constants of \emph{splay}, \emph{bend}, \emph{twist} and \emph{saddle-splay}, respectively. Equation \eref{eq:frank_density} delivers the most general frame-indifferent function $W(\n,\gradn)$, at most quadratic in $\gradn$, which enjoys the nematic symmetry (see, for example, Chapt.\,3 of \cite{virga:variational}).

By use of \eref{eq:q_identity} and
denoting $B:=|\vb|$, $\Fdensity$ takes  the following equivalent form (see \cite{selinger:interpretation}),
\begin{equation}\label{eq:Sdensity}
	\Fdensity= \frac{1}{2}(K_{11}-K_{24})S^{2} + \frac{1}{2}(K_{22}-K_{24})T^{2} + \frac{1}{2}K_{33}B^2 + K_{24}(2q^2),
\end{equation}
which is semipositive definite whenever the elastic constants satisfy \emph{Ericksen's inequalities} \cite{ericksen:inequalities},
\begin{equation}\label{eq:ericksen}
	K_{11}\geq K_{24}\geq0,\quad
	K_{22}\geq K_{24}\geq0,\quad
	K_{33}\geq0.
\end{equation}

\section{Compatibility conditions}\label{app:compatibility}
In this Appendix we provide a set of compatibility conditions that a function $\f$ must satisfy in order to be the common factor in \eref{eq:qu_gradn} for a quasi-uniform distortion. We first arrive at a set  of  nine conditions for a general field $\n$ in three-dimensional space, then we specialize them to the case of a planar quasi-uniform distortion.

\subsection{General three-dimensional case}\label{app:three_dimensional_case}
The orthonormality of the distortion frame $\dframe$ requires  the gradients $\gradn_1$, $\gradn_2$, and $\gradn$ to be represented by  three vectors $\cc_1$, $\cc_2$, and $\cd$ (called the \emph{connectors} in \cite{virga:uniform}) through the equations\footnote{\nalert{The method of connectors is in a way intermediate between Cartan's method of moving frames \cite{clelland:frenet} employed in \cite{dasilva:moving,pollard:intrinsic} and classical vector calculus employed in \cite{dasilva:compatible}, being perhaps closer in spirit to the latter.}} 
\begin{equation}\label{eq:connectors}
\fl
 \gradn = \n_1\tpr\cc_1 + \n_2\tpr\cc_2,\ 
 \gradn_1 = -\n\tpr\cc_1 + \n_2\tpr\cd,\ 
 \gradn_2 = -\n\tpr\cc_2 - \n_1\tpr\cd.
\end{equation}
Connectors $\cc_1$ and $\cc_2$ are completely determined by demanding that $\gradn$ be given as in \eref{eq:qu_gradn},
\begin{equation}\label{eq:c1c2}
\fl
 \cc_1 = \f\Bigg[\Bigg(\frac{\qS}{2}+\qq\Bigg)\n_1 - \frac{\qT}{2}\n_2 - \qb_{1}\n\Bigg],\quad
 \cc_2 = \f\Bigg[\frac{\qT}{2}\n_1 + \Bigg(\frac{\qS}{2}-\qq\Bigg)\n_2 - \qb_{2}\n\Bigg],
\end{equation}
while the third connector, which we decompose as $\cd=d_1\n_1+d_2\n_2+d_3\n$, remains undetermined.

The existence of a quasi-uniform distortion filling the whole space requires that both fields $\n$ and $\n_1$ (and so also $\n_2=\n\times\n_1$) are uniquely defined everywhere, which, under smoothness assumptions, amounts to require that both gradients $\gradn$ and $\gradn_1$ be integrable. In a flat space, the latter request is equivalent to the symmetry in the last two legs of the   second gradients 
\begin{eqnarray}\label{second_gradients}
\eqalign{
 \grad^2\n &= \n_1\tpr\grad\cc_1 - \n\tpr\cc_1\tpr\cc_1 + \n_2\tpr\cc_1\tpr\cd + \n_2\tpr\grad\cc_2 \\
 &\qquad- \n\tpr\cc_2\tpr\cc_2 - \n_1\tpr\cc_2\tpr\cd,} \\
 \eqalign{
 \grad^2\n_1 &= - \n\tpr\grad\cc_1 - \n_1\tpr\cc_1\tpr\cc_1 - \n_2\tpr\cc_1\tpr\cc_2 + \n_2\tpr\grad\cd \\
 &\qquad- \n\tpr\cd\tpr\cc_2 - \n_1\tpr\cd\tpr\cd.}
\end{eqnarray} 
Such  integrability conditions reduce to the symmetry of the following six second-order tensors obtained by contraction of the first leg of the third-order tensors in \eref{second_gradients} with the vectors of the frame $\dframe$,
\begin{eqnarray}\label{eq:symmetry_conditions} 
  \n_1\cdot\grad^2\n &= \grad\cc_1  - \cc_2\tpr\cd, \\
  \n_2\cdot\grad^2\n &= \cc_1\tpr\cd + \grad\cc_2, \\
  \n\cdot\grad^2\n &= - \cc_1\tpr\cc_1 - \cc_2\tpr\cc_2,\label{eq:third} \\
  \n_1\cdot\grad^2\n_1 &= - \cc_1\tpr\cc_1 - \cd\tpr\cd, \label{eq:fourth}\\
  \n_2\cdot\grad^2\n_1 &= - \cc_1\tpr\cc_2 + \grad\cd, \\ 
  \n\cdot\grad^2\n_1 &= - \grad\cc_1 - \cd\tpr\cc_2.
\end{eqnarray}
Both tensors in \eref{eq:third} and \eref{eq:fourth} are already symmetric; the symmetry requirements for the remaining tensors amount to $12$ scalar equations, not all of which turn out to be independent.

Keeping in mind that $f$ depends on position $\x$, we write $\gradf=f_1\n_1+f_2\n_2+f_3\n$  and set $\grad d_1=d_{11}\n_1+d_{12}\n_2+d_{13}\n$, $\grad d_2=d_{21}\n_1+d_{22}\n_2+d_{23}\n$, and $\grad d_3=d_{31}\n_1+d_{32}\n_2+d_{33}\n$, finally arriving at the following set of nine compatibility conditions,
\begin{eqnarray}
\fl
 \frac{T}{2}f_1 + \Bigg(\frac{S}{2}+q\Bigg)f_2 = -\f ^2b_{1}T + 2\f  q d_1, \label{eq:symmetry1}\\
 \fl
 b_1f_1 + \Bigg(\frac{S}{2}+q\Bigg)f_3 = \f ^2\Bigg[\frac{T^2}{4}-\Bigg(\frac{S}{2}+q\Bigg)^2-b_1^2\Bigg] + \f  b_2d_1, \label{eq:symmetry2}\\
 \fl
 b_1f_2 - \frac{T}{2}f_3 = \f ^2\Bigg(\frac{ST}{2}-b_1b_2\Bigg) + \f (b_{2}d_2-2q d_3), \label{eq:symmetry3}\\
 \fl
  \Bigg(\frac{S}{2}-q\Bigg)f_1 - \frac{T}{2}f_2 = \f ^2b_{2}T + 2\f q d_2, \label{eq:symmetry4}\\
  \fl
 b_2f_1 + \frac{T}{2}f_3 = - \f ^2\Bigg(\frac{ST}{2}+b_{1}b_2\Bigg) - \f (b_1d_1+2q d_3), \label{eq:symmetry5}\\
 \fl
 b_2f_2 + \Bigg(\frac{S}{2}-q\Bigg)f_3 = \f ^2\Bigg[\frac{T^2}{4}-\Bigg(\frac{S}{2}-q\Bigg)^2-b_2^2\Bigg] - \f b_1d_2, \label{eq:symmetry6} \\
 \fl
 \f ^2\Bigg(\frac{S^2}{4}-q^2+\frac{T^2}{4}\Bigg) = - \f T d_3 - d_1^2 - d_2^2 + d_{12} - d_{21}, \label{eq:symmetry7}\\
 \fl
 \f ^2\Bigg[b_{1}\frac{T}{2}-b_{2}\Bigg(\frac{S}{2}+q\Bigg)\Bigg] = \f \Bigg(\frac{S}{2}+q\Bigg)d_1 + \f \frac{T}{2}d_2 - \f b_{1}d_3 - d_2d_3 + d_{13} - d_{31}, \label{eq:symmetry8}\\
 \fl
\f ^2\Bigg[b_{1}\Bigg(\frac{S}{2}-q\Bigg)+b_{2}\frac{T}{2}\Bigg] = \f \Bigg(\frac{S}{2}-q\Bigg)d_2 - \f \frac{T}{2}d_1 - \f b_{2}d_3 + d_1d_3 + d_{23} - d_{32}, \label{eq:symmetry9}
\end{eqnarray}
where $\fchar$, with $\dchar$ all \emph{constants}, are the corresponding distortion characteristics. These equations are equivalent versions, fit to the present purpose, of the compatibility conditions considered in \cite{dasilva:compatible}. As for using them as sufficient conditions to determine quasi-uniform distortions, the same caveats issued in \cite{dasilva:compatible} also apply here, as witnessed by the limited use we make of them below.

\subsection{Spatial twist-bend}\label{sec:spatial_twist_bend}
Here we show that 
conditions \eref{eq:symmetry1}--\eref{eq:symmetry9} are consistent with the family of \emph{heliconical} quasi-uniform distortions constructed in \cite{pollard:intrinsic}, which in a Cartesian frame $\cframe$ can be described as 
\begin{equation}\label{eq:pa}
 \n = \sin\alpha\cos g\vx \pm \sin\alpha\sin g\vy + \cos\alpha\vz,
\end{equation}
with $g=g(z)$ and $\alpha\in\R$ a constant.
By writing
\begin{equation}\label{eq:gradn_pa}
 \grad\n = g_{,z}\sin\alpha[-\sin g\vx\tpr\vz \pm \cos g\vy\tpr\vz]
\end{equation}
and by identifying  the distortion frame $\dframe$ with
\begin{equation}\label{eq:frame_pa}
\fl
\eqalign{
 \n_1 &= \frac1{\sqrt2}[(\cos\alpha\cos g \pm \sin g)\vx + (\cos\alpha\sin g \mp \cos g)\vy - \sin\alpha\vz], \\
 \n_2 &= \frac1{\sqrt2}[(\cos\alpha\cos g \mp \sin g)\vx + (\cos\alpha\sin g \pm \cos g)\vy - \sin\alpha\vz],}
\end{equation}
it becomes a simple matter  to check by direct inspection that the field in \eref{eq:pa} is quasi-uniform, as its distortion characteristics can be written as $\dstar$ with
\begin{equation}\label{eq:modes_pa}
\fl
 \qS = 0,
 \quad
 \qT = \mp \sin^2\alpha,
 \quad
 \qq = \frac{\sin^2\alpha}2 = \mp\frac \qT2,
 \quad
 \qb_1 = - \qb_2 = \frac{\sin\alpha\cos\alpha}{\sqrt2}, f=g_{,z}.
\end{equation}
Since $S=0$ and $T=\pm2q$, the field in \eref{eq:pa} is a twist-bend quasi-uniform distortion. By extracting the connector $\cd$ from \eref{eq:frame_pa} and projecting $\gradf$ on the distortion frame $\dframe$, we arrive at the following equations,
\begin{eqnarray}
\gradf = g_{,zz}\vz = \frac{g_{,zz}}{\sqrt2}[-\sin\alpha(\n_1 + \n_2) + {\sqrt2}\cos\alpha\n],\\
f_1 = f_2 = -\frac{g_{,zz}}{\sqrt2}\sin\alpha,
\quad
f_3 = g_{,zz}\cos\alpha, \\
\cd = g_{,z}\cos\alpha\vz = \frac{g_{,z}}{\sqrt2}\cos\alpha[-\sin\alpha(\n_1 + \n_2) + {\sqrt2}\cos\alpha\n], \\
\grad d_1 = \grad d_2 = -\frac{\sin\alpha\cos\alpha}{\sqrt2}\gradf, 
\quad
\grad d_3 = \cos^2\alpha\gradf,
\end{eqnarray}
whose use in the above compatibility conditions shows that they are identically satisfied.
\subsection{Planar splay-bend}
In general, a planar distortion  is a splay-bend (with $S=0$ and $T=\pm2q$). For it to be quasi-uniform, it must also satisfy the compatibility conditions \eref{eq:symmetry1}--\eref{eq:symmetry9}, which reduce to
\begin{eqnarray}
 \Bigg(\frac{S}{2}+q\Bigg)f_2 = 2\f  q d_1, \label{eq:sb1}\\
 b_1f_1 + \Bigg(\frac{S}{2}+q\Bigg)f_3 = - \f ^2\Bigg[\Bigg(\frac{S}{2}+q\Bigg)^2 + b_1^2\Bigg] + \f  b_2d_1, \label{eq:sb2}\\
 b_1f_2 = - \f ^2b_1b_2 + \f (b_2d_2-2q d_3), \label{eq:sb3}\\
  \Bigg(\frac{S}{2}-q\Bigg)f_1 = 2\f q d_2, \label{eq:sb4}\\
 b_2f_1 = - \f ^2b_1b_2 - \f (b_1d_1+2q d_3), \label{eq:sb5}\\
 b_2f_2 + \Bigg(\frac{S}{2}-q\Bigg)f_3 = - \f ^2\Bigg[\Bigg(\frac{S}{2}-q\Bigg)^2 + b_2^2\Bigg] - \f b_1d_2, \label{eq:sb6} \\
 \f ^2\Bigg(\frac{S^2}{4} - q^2\Bigg) = - d_1^2 - d_2^2 + d_{12} - d_{21}, \label{eq:sb7}\\
 - \f ^2b_2\Bigg(\frac{S}{2}+q\Bigg) = \f \Bigg(\frac{S}{2}+q\Bigg)d_1 - \f b_{1}d_3 - d_2d_3 + d_{13} - d_{31}, \label{eq:sb8}\\
\f ^2b_1\Bigg(\frac{S}{2}-q\Bigg) = \f \Bigg(\frac{S}{2}-q\Bigg)d_2 - \f b_{2}d_3 + d_1d_3 + d_{23} - d_{32}. \label{eq:sb9}
\end{eqnarray}

Since for a planar field either $\n_1$ or $\n_2$ is constant, by \eref{eq:connectors},  the connector $\cd$ vanishes identically. Furthermore, letting  $\f$  be different from zero, we see from \eref{eq:splay_bend_definition} that  \eref{eq:sb7} is identically satisfied, while \eref{eq:sb8} and \eref{eq:sb9} require that either $\qb_2$ or $\qb_1$ vanishes.
More specifically, if for $\n_1\equiv\vz$ then by \eref{eq:connectors}   $\cc_1=\bm{0}$, which by  \eref{eq:c1c2} implies that $\qb_1=0$ and $\qS=-2\qq$, coherently with \eref{eq:splay_bend_definition}. Conditions \eref{eq:sb4} and \eref{eq:sb5} then read as
\begin{equation}
 2\qq f_1 = 0
 \quad\mathrm{and}\quad
 \qb_2f_1 = 0,
\end{equation}
respectively,
whence $f_1=0$ (otherwise $\n$ is constant), and by \eref{eq:sb6}
\begin{equation}\label{eq:f1_zero} 
 \qb_2f_2 - 2\qq f_3 = - \f ^2\big[4(\qq)^2 + (\qb_2)^2\big].
\end{equation}
On the other hand, if $\n_2\equiv\vz$ then  $\gradn_2=\bm{0}$ and $\cc_2=\bm{0}$, again  by \eref{eq:connectors}, whence it follows from \eref{eq:c1c2} that $\qb_2=0$ and $\qS=2\qq$. In this case, the nontrivial compatibility conditions are \eref{eq:sb1} and \eref{eq:sb3}, which imply $f_2=0$, and also \eref{eq:sb2}, which reads as
\begin{equation}\label{eq:f2_zero} 
 \qb_1f_1 + 2\qq f_3 = - \f ^2\big[4(\qq)^2 + (\qb_1)^2\big].
\end{equation}
The difference between \eref{eq:f1_zero} and \eref{eq:f2_zero} only lies  in the conventional  (but useful) requirement that $\n_1$ is the eigenvector of $\tD$ with \emph{positive} eigenvalue $q$; we shall consider these  two cases as  essentially the same: in both of them, the quasi-uniformity requirement translates into basically the same equation, of which we study now two simple consequences. 

For $q=0$, by \eref{eq:splay_bend_definition},  the field $\n$ is a pure bend distortion and, by letting (without loss of generality) $\vb = b\n_1$ and setting $\f=b$ (whence $\qb=1$), we obtain from  \eref{eq:f2_zero} that  $f_1 = - \f ^2$, together with $f_2=0$.

For $q>0$, we can let $\f=q$, so that $\qq=1$. Then, assuming that  $\vb\neq\bm{0}$, we arrive at the following dichotomy,
\begin{eqnarray}
 \mathrm{either}
 \quad
 \qS = -2, \quad
 \qb_1 = f_1 = 0, \quad
 \qb_2f_2 - 2f_3 = - \f ^2\big[4 + (\qb_2)^2\big] \label{eq:sb-} 
 \\
 \mathrm{or}
 \quad
 \qS = 2, \quad
 \qb_2 = f_2 = 0, \quad
 \qb_1f_1 + 2f_3 = - \f ^2\big[4 + (\qb_1)^2\big]. \label{eq:sb+} 
\end{eqnarray}
Spirals as in \eref{eq:spirals} fall into \eref{eq:sb-} when $\cos\alpha>0$ and into \eref{eq:sb+} when $\cos\alpha<0$. 

\section{Planar distortion characteristics}\label{app:sb_cartesian}
In this Appendix, we collect a number of details concerning the distortion characteristics of the planar fields considered in  sections~\ref{sec:halfplane} and \ref{sec:defect} of the main text.

\subsection{Frustrated line}\label{app:frustrated_line}
We start from the gradient decomposition in \eref{eq:grad_n} for the director
field $\n$ represented by \eref{eq:splay-bend}. It readily follows from the  latter that
\begin{equation}\label{eq:P_cartesian}
	\eqalign{
		\tP(\n) = \tI - \n\tpr\n
		&= \sin^2\varphi\vx\tpr\vx - \sin\varphi\cos\varphi(\vx\tpr\vy+\vy\tpr\vx) \\
		&\qquad+ \cos^2\varphi\vy\tpr\vy + \vz\tpr\vz}
\end{equation}
and from \eref{eq:gradn_half} that  
\begin{equation}\label{eq:S_cartesian}
 S = \dv\n = -(\varphi_{,x}\sin\varphi - \varphi_{,y}\cos\varphi)
\end{equation}
and
\begin{equation}\label{eq:b_cartesian}
	\vb = -(\grad\n)\n = -(\varphi_{,x}\cos\varphi + \varphi_{,y}\sin\varphi)(-\sin\varphi\vx + \cos\varphi\vy).
\end{equation}
Hence
\begin{equation}\label{eq:bn_cartesian}
\eqalign{
 \vb\tpr\n
 = (\varphi_{,x}\cos\varphi + \varphi_{,y}\sin\varphi)&\big[\sin\varphi\cos\varphi(\vx\tpr\vx - \vy\tpr\vy) \\
 &\quad+ \sin^2\varphi\vx\tpr\vy - \cos^2\varphi\vy\tpr\vx\big].}
\end{equation}
Moreover, the skew-symmetric part $\gradn_{\rm skw}$ of $\gradn$ is given by
\begin{equation}\label{eq:skw_cartesian}
 2\gradn_{\rm skw} := \gradn - \gradn^{\top}
 = (\varphi_{,x}\cos\varphi + \varphi_{,y}\sin\varphi)(\vy\tpr\vx - \vx\tpr\vy),
\end{equation}
which delivers
$\curl\n = (\varphi_{,x}\cos\varphi + \varphi_{,y}\sin\varphi)\vz$ (and $T=0$).

We can now use  \eref{eq:grad_n} to derive 
\begin{equation}\label{eq:D_cartesian}
\eqalign{
 \tD &= \gradn + \bm{b}\tpr\n -\frac{S}{2}\tP(\n)\\
 &=\frac{\varphi_{,x}\sin\varphi - \varphi_{,y}\cos\varphi}2
 \big[\sin\varphi\cos\varphi(\vx\tpr\vy + \vy\tpr\vx) \\
 &\qquad- \sin^2\varphi\vx\tpr\vx 
 - \cos^2\varphi\vy\tpr\vy 
 + \vz\tpr\vz\big],}
\end{equation}
whose eigenvalues are $\pm\frac12(\varphi_{,x}\sin\varphi - \varphi_{,y}\cos\varphi)$. Therefore, when $\varphi_{,x}\sin\varphi > \varphi_{,y}\cos\varphi$
\begin{equation}\label{eq:q+_cartesian}
 q = -\frac S2 = \frac{\varphi_{,x}\sin\varphi - \varphi_{,y}\cos\varphi}2
\end{equation}
and the distortion frame $\dframe$ is given by
\begin{equation}\label{eq:frame+_cartesian}
 \n_1 := \vz,
 \quad
 \n_2 := \sin\varphi\vx - \cos\varphi\vy,
 \quad
 \n := \cos\varphi\vx + \sin\varphi\vy,
\end{equation}
while for $\varphi_{,x}\sin\varphi < \varphi_{,y}\cos\varphi$ 
\begin{equation}\label{eq:q-_cartesian}
 q = \frac S2 = -\frac{\varphi_{,x}\sin\varphi - \varphi_{,y}\cos\varphi}2
\end{equation}
and 
\begin{equation}\label{eq:frame-_cartesian}
 \n_1 := - \sin\varphi\vx + \cos\varphi\vy,
 \quad
 \n_2 := \vz,
 \quad
 \n := \cos\varphi\vx + \sin\varphi\vy.
\end{equation}
In the very special case where $\varphi_{,x}\sin\varphi = \varphi_{,y}\cos\varphi$ both $S$ and $q$ vanish, and the field carries a pure bend distortion (which is of course quasi-uniform) with
\begin{equation}\label{eq:frame0_cartesian}
 b_1 = -(\varphi_{,x}\cos\varphi + \varphi_{,y}\sin\varphi),
 \qquad
 b_2 = 0
\end{equation}
and distortion frame as in \eref{eq:frame-_cartesian}.

\nalert{To obtain $f$ from \eref{eq:f_line}, we first parameterize the characteristic described by \eref{eq:chareq_nonpar_half} as follows\footnote{It should be noted that neither here nor in \eref{eq:characteristic_circle_s} $s$ represents the arc-length along characteristics.}
\begin{equation}
	\label{eq:charactistic_line_s}
	\cases{x=x_0+s=:\xi(x_0,s),\\
	y=\frac{\sin\varphi_0+b^\ast\cos\varphi_0}{\cos\varphi_0-b^\ast\sin\varphi_0}s=:\eta(x_0,s),}
\end{equation}
and then use the identity $\varphi(\xi(x_0,s),\eta(x_0,s))=\varphi_0(x_0)$ to derive the linear system
\begin{equation}
	\label{eq:linear_system_line}
	\cases{\varphi_{,x}\xi_{,s}+\varphi_{,y}\eta_{,s}=0,\\
	\varphi_{,x}\xi_{,x_0}+\varphi_{,y}\eta_{,x_0}=\varphi_0',}
\end{equation}
in $(\varphi_{,x},\varphi_{,y})$. Solving \eref{eq:linear_system_line}, we readily arrive at
\begin{equation}
	\label{eq:f_line_appendix}
	f(x_0,s)=\frac12\frac{\varphi_0'}{\frac{1+(b^\ast)^2}{\cos\varphi_0-b^\ast\sin\varphi_0}s\varphi_0'-(\sin\varphi_0+b^\ast\cos\varphi_0)},
\end{equation}
from which both \eref{eq:f_y=0} and \eref{eq:f_line_infinity} in the main text follow.
}

\subsection{Frustrated circle}\label{app:sb_defect}
For the director field in \eref{eq:defect}, we easily see that
\begin{equation}\label{eq:P_defect}
	\fl
	\tP(\n) 
	= \sin^2\alpha\er\tpr\er - \sin\alpha\cos\alpha(\er\tpr\et+\et\tpr\er) + \cos^2\alpha\et\tpr\et + \vz\tpr\vz.  
\end{equation}
It also follows from \eref{eq:gradn_defect} that
\begin{equation}\label{eq:bend_defect}
 \bm{b} = \Bigg(\alpha_{,r}\cos\alpha+\frac{1+\alpha_{,\vartheta}}{r}\sin\alpha\Bigg)(\sin\alpha\er-\cos\alpha\et)
\end{equation}
and 
\begin{equation}\label{eq:bendn_defect}
	\fl
	\eqalign{
		\bm{b}\tpr\n  
		= \Bigg(\alpha_{,r}\cos\alpha+\frac{1+\alpha_{,\vartheta}}{r}\sin\alpha\Bigg)
		&\big(\sin\alpha\cos\alpha\er\tpr\er - \cos^2\alpha\et\tpr\er\\
		&\quad + \sin^2\alpha\er\tpr\et - \sin\alpha\cos\alpha\et\tpr\et\big). }
\end{equation}
Moreover,
\begin{equation}
 2\gradn_{\rm skw} = \Bigg(\alpha_{,r}\cos\alpha+\frac{1+\alpha_{,\vartheta}}{r}\sin\alpha\Bigg)(\et\tpr\er-\er\tpr\et) 
\end{equation}
 and
 \begin{equation}
 \curl\n = \Bigg(\alpha_{,r}\cos\alpha+\frac{1+\alpha_{,\vartheta}}{r}\sin\alpha\Bigg)\vz. \label{eq:curl_defect}
\end{equation}

Bu use of \eref{eq:grad_n}, we then arrive at
\begin{equation}\label{eq:D_defect}
\fl
\eqalign{
 \tD
 = \frac{1}{2}\Bigg(\frac{1+\alpha_{,\vartheta}}{r}\cos\alpha &- \alpha_{,r}\sin\alpha\Bigg)\big(
 \sin^2\alpha\er\tpr\er - \sin\alpha\cos\alpha\er\tpr\et \\
 &\quad  - \sin\alpha\cos\alpha\et\tpr\er+ \cos^2\alpha\et\tpr\et - \vz\tpr\vz\big),}
\end{equation}
whose eigenvalues are
\begin{equation}\label{eq:q_defect}
 \pm\frac S2 = \pm\frac{1}{2}\Bigg(\frac{1+\alpha_{,\vartheta}}{r}\cos\alpha - \alpha_{,r}\sin\alpha\Bigg).
\end{equation}
Thus,  for $S\geq0$, the distortion frame $\dframe$ has
\begin{equation}\label{eq:frame1_defect}
 \n_1 =\sin\alpha\er - \cos\alpha\et,
 \qquad
 \n_2 = -\vz,
\end{equation}
while for $S<0$
\begin{equation}\label{eq:frame2_defect}
 \n_1 = \vz,
 \qquad
 \n_2 = \sin\alpha\er - \cos\alpha\et.
\end{equation}
A quick comparison with \eref{eq:bend_defect} shows that $\bm{b}$ and $\n_1$ are collinear in the former case, while $\bm{b}$ and $\n_2$ are collinear in the latter.

\nalert{It follows from \eref{eq:characteristic_circle} and \eref{eq:defect_slope} that coordinates $(\vartheta_0,s)$ along characteristics can be related to the Cartesian coordinates $(x,y)$ through the change of variables
\begin{equation}\label{eq:characteristic_circle_s}
	\cases{x=\cos\vartheta_0+s=:\xi(\vartheta_0,s),\\
	y=\sin\vartheta_0-\frac{b^\ast\cos\varphi_0-\sin\varphi_0}{b^\ast\sin\varphi_0+\cos\varphi_0}s=:\eta(\vartheta_0,s),}
\end{equation}
which here replaces \eref{eq:charactistic_line_s}. Use of \eref{eq:characteristic_circle_s} in the identity $\varphi(\xi(\vartheta_0,s),\eta(\vartheta_0,s))=\varphi_0(\vartheta_0)$ and resort to the relations
\begin{equation}
	\label{eq:linear_system_circle}
	\cases{\frac1r\varphi_{,\vartheta}=-\sin\vartheta\varphi_{,x}+\cos\varphi_{,y},\\
	\varphi_{,r}=\cos\vartheta\varphi_{,x}+\sin\vartheta\varphi_{,y},}
\end{equation}
where $\vartheta$ is to be related to $\vartheta_0$ through \eref{eq:characteristic_circle_s} and the identities $\cos\vartheta=x/r$ and $\sin\vartheta=y/r$, lead us from \eref{eq:f_circle} to $f$ expressed as a function of $(\vartheta_0,s)$. The resulting expression is rather complicated and inexpressive, but it easily reduces to \eref{eq:f_disk_s=0}, as can also be checked directly by letting $\vartheta=\vartheta_0$ in \eref{eq:linear_system_circle}. Also, for large $s$, $f=O(1/s^2)$, which justifies \eref{eq:f_disk_infinity}.
}

\section{Characteristic lines}\label{app:sb_char_curves}
We collect here details of computations needed in \sref{sec:defect}.

Equation~\eref{eq:defect_drdtheta} follows from \eref{eq:defect_chareq_nonpar}, as
\begin{equation}\label{eq:defect_drdtheta1}
 \int \frac{1}{r}\dd r = -\int \frac{\qb\sin(\varphi_0 - \vartheta) + \cos(\varphi_0 - \vartheta)}{\qb\cos(\varphi_0 - \vartheta)-\sin(\varphi_0 - \vartheta)}\dd\vartheta 
\end{equation}
and then 
\begin{equation}\label{eq:defect_drdtheta_app}
\fl
 r = R_0|\qb\cos(\varphi_0 - \vartheta)-\sin(\varphi_0 - \vartheta)|^{-1},
 \quad\mathrm{with}\quad R_0:=r_0|\qb\cos\alpha_0-\sin\alpha_0|>0 .
\end{equation}

Both $\varphi:=\alpha+\vartheta$ and $R:=r|\qb\cos\alpha-\sin\alpha|$ are constant along the characteristic curves since
\begin{equation}
\cases{
\frac{\dd\varphi}{\dd t} = \frac{\dd\alpha}{\dd t} + \frac{\dd\vartheta}{\dd t} = 0,& \\
\eqalign{
\frac{\dd R}{\dd t} &= \frac{\dd r}{\dd t}|\qb\cos\alpha-\sin\alpha| + r\frac{\dd}{\dd t}|\qb\cos\alpha-\sin\alpha|\\
&=  r(\qb\sin\alpha+\cos\alpha)|\qb\cos\alpha-\sin\alpha| \\
&\qquad- r\frac{(\qb\cos\alpha-\sin\alpha)^2}{|\qb\cos\alpha-\sin\alpha|}(\qb\sin\alpha+\cos\alpha) = 0.}& \\
}
\end{equation}
Outside $\disk$, that is for $r>1$, we thus obtain that 
\begin{equation}
\fl
\cases{
r = \Bigg|\frac{\qb \cos\alpha_0-\sin\alpha_0}{\qb \cos(\alpha_0+\vartheta_0-\vartheta)-\sin(\alpha_0+\vartheta_0-\vartheta)}\Bigg|
= \Bigg|\frac{\qb \cos\alpha_0-\sin\alpha_0}{\qb \cos\alpha-\sin\alpha}\Bigg|,&\\
\alpha = \alpha_0 + \vartheta_0 - \vartheta.&\\
}
\end{equation}
Letting in the $(x,y)$ plane be $x=r\cos\vartheta$ and $y=r\sin\vartheta$, since the function
\begin{equation}
\eqalign{
R &= |\qb r\cos\alpha - r\sin\alpha| \\
&= |\qb r[\cos(\alpha_0 + \vartheta_0)\cos\vartheta + \sin(\alpha_0 + \vartheta_0)\sin\vartheta] \\
&\qquad- r[\sin(\alpha_0 + \vartheta_0)\cos\vartheta - \cos(\alpha_0 + \vartheta_0) \sin\vartheta]| \\
&= |(\qb \cos\varphi_0 - \sin\varphi_0)x + (\qb \sin\varphi_0 + \cos\varphi_0)y| 
}
\end{equation}
is also constant, we derive  the expression in \eref{eq:defect_slope} for the slope $\slod$ of the characteristic lines.

By remarking that $(\qb \cos\varphi_0 - \sin\varphi_0)^2 + (\qb \sin\varphi_0 + \cos\varphi_0)^2 = (\qb) ^2 + 1$ and requiring $x(s)^2+y(s)^2>1$, where $s$ is the arc-length parameter along characteristics, we arrive at
\begin{equation}\label{eq:char_arc1} 
\cases{
x(s) = \frac{|\qb \sin\varphi_0 + \cos\varphi_0|}{\sqrt{(\qb)^2+1}}s + \cos\vartheta_0,&\\
y(s) = -\sgn(\qb \sin\varphi_0 + \cos\varphi_0)\frac{\qb \cos\varphi_0 - \sin\varphi_0}{\sqrt{(\qb)^2+1}}s + \sin\vartheta_0, &\\
}
\end{equation}
or
\begin{equation}\label{eq:char_arc2} 
\cases{
x(s) =\cos\vartheta_0,&\\
y(s) = s+\sin\vartheta_0,  &\\
}
\qquad \mathrm{if}\ \tan(\varphi_0)=-\frac{1}{\qb}
\end{equation}
for
\begin{equation}\label{eq:char_sign} 
\fl
\cases{
 \eqalign{
s\in[0,+\infty) &\quad\mathrm{if}\ (\qb \sin\varphi_0 + \cos\varphi_0)(\qb \sin\alpha_0 + \cos\alpha_0)>0 \\
  &\qquad\mathrm{or}\ \qb \sin\alpha_0 + \cos\alpha_0 \neq  \qb \sin\varphi_0 + \cos\varphi_0 = 0\ \mathrm{and}\ \vartheta_0\in(0,\pi),}\\
 \eqalign{
s\in(-\infty,0] &\quad\mathrm{if}\ (\qb \sin\varphi_0 + \cos\varphi_0)(\qb \sin\alpha_0 + \cos\alpha_0)<0 \\
  &\qquad\mathrm{or}\ \qb \sin\alpha_0 + \cos\alpha_0 \neq  \qb \sin\varphi_0 + \cos\varphi_0 = 0\ \mathrm{and}\ \vartheta_0\in(\pi, 2\pi),} \\
\eqalign{
	s\in\R &\quad\quad\quad\ \ \mathrm{if}\ \qb \sin\alpha_0 + \cos\alpha_0 = 0.}
}
\end{equation}
The last case in \eref{eq:char_sign} represents the characteristic lines  tangent to $\disk$ that delimit the  domain $\mathcal{D}$ in \eref{eq:defect_domain}.


\section*{References}


\providecommand{\newblock}{}

\end{document}